\documentclass{aa}  
\usepackage{graphicx}
\usepackage{txfonts}
\begin{document}

\title{Expanding shells around young clusters - S 171/Be 59
  \thanks{Based on observations collected at the Nordic Optical Telescope, La Palma, Spain; and Onsala Space Observatory, Sweden }}

   \author{G. F. Gahm\inst{1}\thanks{Deceased}
             \and M. J. C. Wilhelm\inst{2}  
              \and C. M. Persson\inst{3}
             \and A. A. Djupvik\inst{4, 5}
             \and S. F. Portegies Zwart\inst{2}}

   \institute{Stockholm Observatory, AlbaNova University Center, Stockholm University,
              SE-106 91 Stockholm, Sweden 
              \and Leiden University, PO Box 9500, NL-2300 RA Leiden, Holland \\
              email:  \mbox{wilhelm@strw.leidenuniv.nl}
              \and Chalmers University of Technology, Department of Space, Earth and Environment, Onsala Space Observatory,\\  
              SE-439 92 Onsala, Sweden.
              \and Nordic Optical Telescope, Rambla Jos\'{e} Ana Fern\'{a}ndez P\'{e}rez 7, ES-38711 Bre\~{n}a Baja, Spain 
              \and Department of Physics and Astronomy, Aarhus University, Ny Munkegade 120, DK-8000 Aarhus C, Denmark 
               }
   \date{}

   %\abstract{}{}{}{}{} 
   
   % 5 {} token are mandatory
\abstract
   {Some \ion{H}{ii} regions that surround young stellar clusters are bordered by molecular shells that appear to expand at a rate inconsistent with our current model simulations. In this study we focus on the dynamics of Sharpless 171 (including NGC 7822), which surrounds the cluster Berkeley 59.} 
   %\object
   {We aim to compare the velocity pattern over the molecular shell with the mean radial velocity of the cluster for estimates of the expansion velocities of different shell structures, and to match the observed properties with model simulations. }
   {Optical spectra of 27 stars located in Berkeley 59 were collected at the Nordic Optical Telescope, and a number of molecular structures scattered over the entire region were mapped in $^{13}$CO(1-0) at Onsala Space Observatory.  }
   {We obtained radial velocities and MK classes for the cluster's stars. At least four of the O stars are found to be spectroscopic binaries, in addition to one triplet system. From these data we obtain the mean radial velocity of the cluster. From the $^{13}$CO spectra we identify three shell structures, expanding relative to the cluster at moderate velocity (4 km/s), high velocity (12 km/s), and in between. The high-velocity cloudlets extend over a larger radius and are less massive than the low-velocity cloudlets. We performed a model simulation to understand the evolution of this complex. }
  {Our simulation of the Sharpless 171 complex and Berkeley 59 cluster demonstrates that the individual components can be explained as a shell driven by stellar winds from the massive cluster members. However, our relatively simple model produces a single component. Modelling of the propagation of shell fragments through a uniform interstellar medium demonstrates that dense cloudlets detached from the shell are decelerated less efficiently than the shell itself. They can reach greater distances and retain higher velocities than the shell. }

%\keywords{ISM: \ion{H}{ii} regions - ISM: molecules - ISM: kinematics and dynamics - ISM: evolution - Stars: clusters - Stars: kinematics/dynamics - Individual: Berkeley 59, S 171, NGC 7822 }
\keywords{ISM: \ion{H}{ii} regions - ISM: molecules - ISM: kinematics and dynamics - ISM: evolution - ISM: individual objects: Berkeley 59, S 171, NGC 7822 - Stars: kinematics/dynamics}

 \maketitle
%
%________________________________________________________________

\section{Introduction}
\label{sec:intro}

There are in our galaxy numerous \ion{H}{ii} regions surrounding young stellar clusters that contain several massive O and B stars. Far- and extreme-UV radiation and stellar wind from O and B stars heat, ionise, and shape the surrounding gas, and these expanding bubbles drive compressed molecular gas outwards. The evolution of such regions has been subject to a large number of theoretical studies, starting with the pioneering work by \citet{str39} (and e.g. the reviews in \citealt{hen07}, \citealt{kru14}, \citealt{dal15}, and \citealt{haw18}). 

Once the first O stars have formed, the surrounding molecular cloud is compressed into a shell that is accelerated outwards. The shell breaks up into filaments, elephant trunks, and globules, and eventually it slows down and disperses. The expansion velocity of the shell obtained from simulations depends on the physical processes assumed, the initial conditions, and the source driving the expansion. Numerical studies show that the shells surrounding single stars rarely expand with velocities greater than 5 km/s \citep[e.g.][]{hos06,art11,bis15}, although shell velocities of the order $\sim$10 km/s can be attained for very massive stars \citep{gre20} or a low-density ambient interstellar medium \citep[ISM;][]{fre03}. A compact cluster of massive stars can drive a single, common shell, and these shells can reach velocities in excess of 10 km/s \citep[e.g.][]{dal14,lan21a}, but this requires a large population of high-mass stars.

From existing estimates of the central velocities of various \ion{H}{ii} regions in comparison to those measured for molecular line
emission in the surrounding shells, it appears that the expansion velocities as a rule are of the order of a few km~s$^{-1}$, in
congruence with the predictions from some of the above-mentioned theoretical model simulations. However, we have found several regions
for which one could suspect much higher velocities. A closer inspection shows that published estimates of central velocities and
shell velocities are in fact rather uncertain. In this paper we focus on the Sharpless~171 (S 171) complex, which has the young cluster Berkeley~59 (Be 59)
at its centre.

\begin{figure*}[t]
\centering
\includegraphics[width=8.30cm]{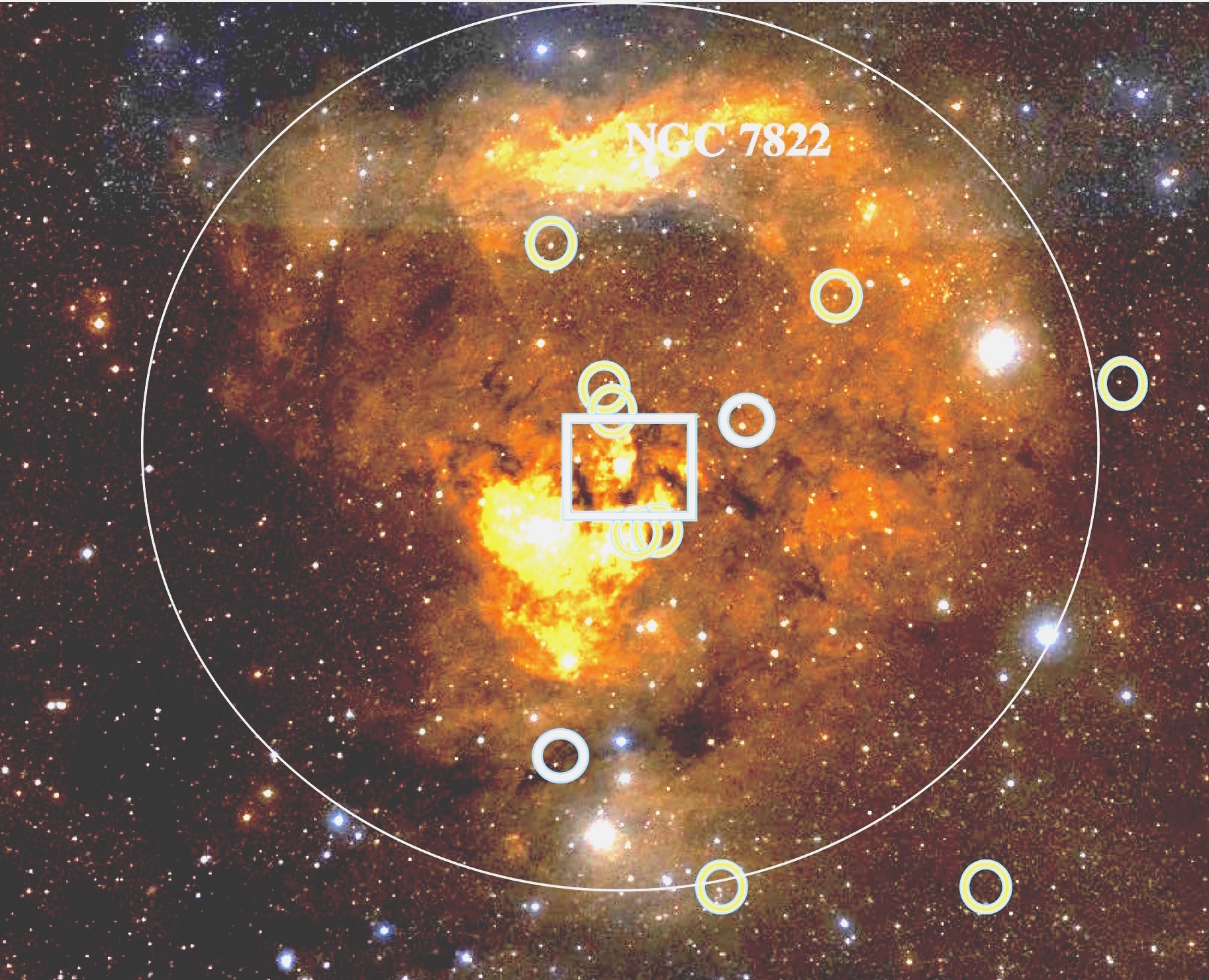}
\includegraphics[width=9.00cm]{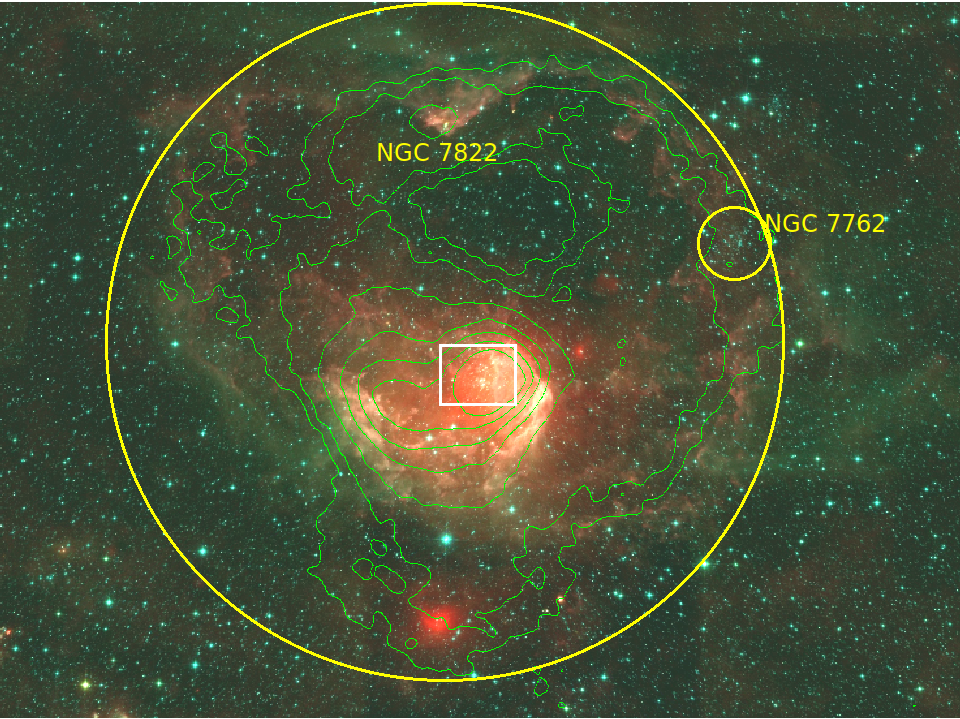}
\caption{Optical image from the Digitized Sky Survey (left) and mid-IR image from Wide-field Infrared Survey Explorer (WISE, right), featuring the \ion{H}{ii} region S~171 in relation to the cluster Berkeley 59, which is located in the central white box. The northern part of the nebula is called NGC 7822. The large circle has a radius of 1.6$\degr$ and shows the approximate extent of the \ion{H}{ii} region. The green contours show the 4 mm (70 GHz) continuum map from Planck. The smaller circles in the optical image mark the positions of the hottest members of the extended Cep OB 4 association; light blue circles represent spectral types O9 and B0, and yellow circles types B1 and B2. North is up, and east to the left. The WISE image is a colour composition of red (22 $\mu$m), green (4.6 $\mu$m), and blue (3.4 $\mu$m). Figure~\ref{fig:field} is a zoomed-in view of the area in the white box. }
\label{fig:OBass}
\end{figure*}

Literature data on the kinematics of the stars, ionised gas, and shell structures in this complex suggest that the molecular shell expands at a very high velocity. However, existing maps of the molecular line emission from shell structures are rather coarse, some line profiles are complex, and only a few stars have so far been measured for radial velocity (RV). We therefore decided to more closely inspect the kinematics of stars and nebular features in the complex. We collected high resolution spectra, at the Nordic Optical Telescope (NOT), of a number of stars in an area at the central cluster to obtain a mean central velocity of the cluster and information on spectroscopic binarity and membership. Moreover, we mapped the molecular line emission from the associated shell surrounding the cluster at Onsala Space Observatory (OSO). The molecular line velocities observed from different parts of the shell in comparison to the cluster velocity should provide an estimate of the expansion velocity of the complex.

The S 171 complex is described in Sect.~\ref{sec:S171} with an account of existing information on the velocities of the stars, the ionised gas, and shell structures. The observations and the data reduction are described in Sect.~\ref{sec:obs}. The general results are given in Sect.~\ref{sec:results} and discussed in more detail in Sect.~\ref{sec:discussion}. We compare these results with numerical simulations in Sect.~\ref{Sect:Simulations} and end with a summary in Sect.~\ref{sec:conclusions}.

\section{The S 171 complex}
\label{sec:S171}

\begin{table*}
\centering 
\caption{Basic data.}
\begin{tabular} {lccccccc}
\hline
     \noalign{\smallskip}
Designations:  &  Position & Distance & Radius & Age & $V_{*}$ &  $V_{*}$(range) & $\bar V_{\ion{H}{ii}}$  \\
 Cluster / Nebula & RA, Dec. (2000) & kpc & pc  & Myr &  km/s (helio)  & km/s (lsr) & km/s (lsr) \\
     \noalign{\smallskip}
\hline
     \noalign{\smallskip}
Be 59 / S 171$^{1}$ & 00:02:10 +67:25:10 & 1.1$^{2}$ & 31$^{2}$ & ~2$^{3,4,5,6}$ & -14.5$^{7}$-6.5$^{8}$ & -5.2 to +4.1 & -19$^{10}$ -9.2$^{11}$ -7.5$^{12}$  \\ 
(NGC 7822, W 1) &  & & &  & -5.2$^{9}$ &   & -8.1$^{13}$ -10.1$^{14}$ -12.0$^{15}$ \\ 
 \noalign{\smallskip}
\hline
\end{tabular}
\label{tab:Be59}
\tablefoot{
\tablefoottext{1}{ \citet{sha59};}
\tablefoottext{2}{This work, Sect.~\ref{sec:pm};}
\tablefoottext{3}{ \citet{maj08};}
\tablefoottext{4}{ \citet{pan08};}
\tablefoottext{5}{ \citet{panw18};}
\tablefoottext{6}{ \citet{get18a,get18b};}
\tablefoottext{7}{ \citet{liu89};}
\tablefoottext{8}{\citet{kha05};}
\tablefoottext{9}{ \citet{dia02}: combined data from various sources;}
\tablefoottext{10}{ \citet{dow74};}
\tablefoottext{11}{ \citet{ped80};}
\tablefoottext{12}{\citet{loz87}: H$\alpha$, [\ion{N}{ii}];}
\tablefoottext{13}{\citet{ros80}: mean for the area just west of the cluster;}
\tablefoottext{14}{\citet{geo70}: H$\alpha$.}
\tablefoottext{15}{\citet{fic90}: H$\alpha$. }}
\end{table*}

\begin{figure}
\centering
\includegraphics[angle=00, width=8.9cm]{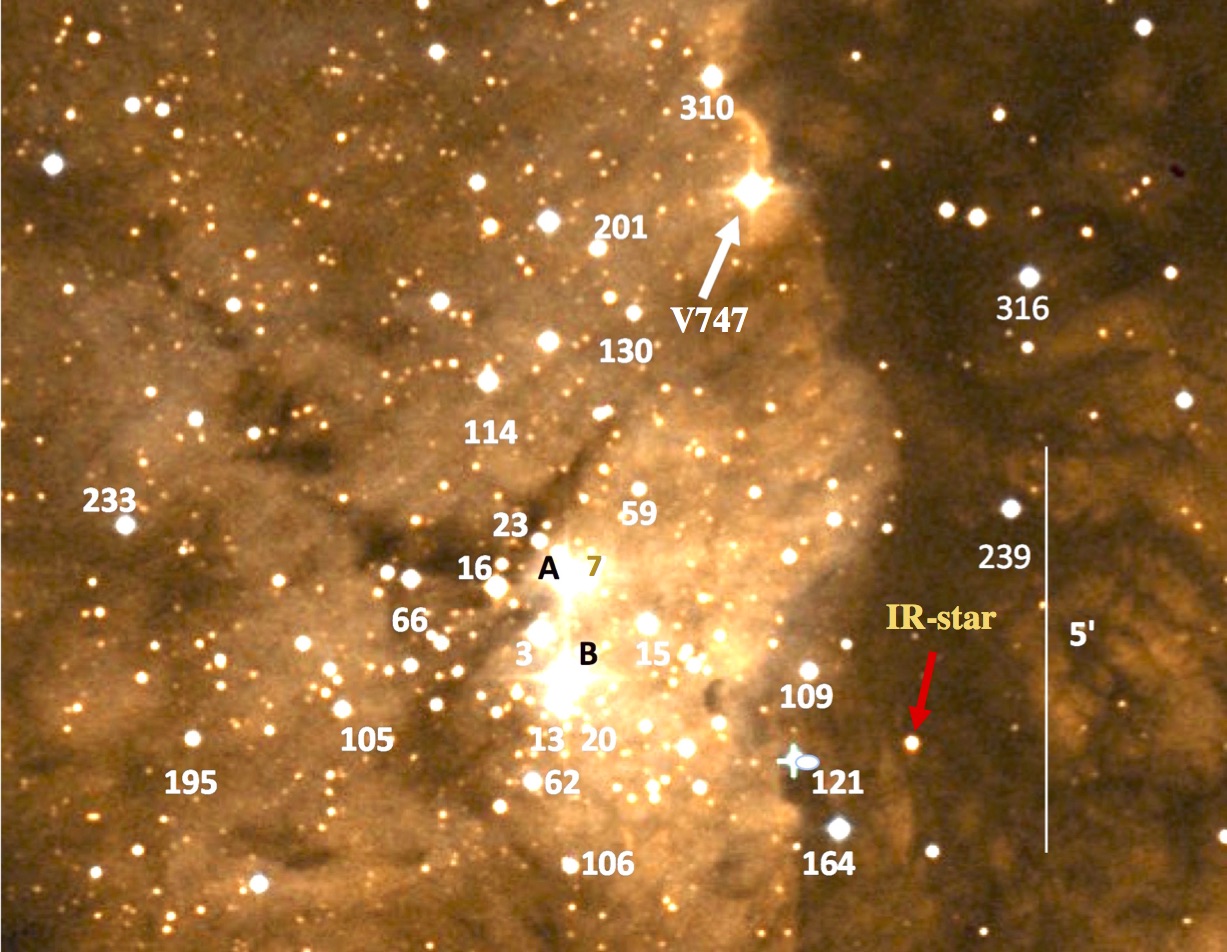}
\caption{Locations of program stars in the Berkeley~59 region according to the designations in Table~\ref{tab:stars} on an image from the Digitized Sky Survey. The cluster centre is put between the bright stars {\it A} (BD+66$\degr$1674) and {\it B}  (BD+66$\degr$1675). The white arrow points at BD+66$\degr$1673, V747 Cep. The red arrow marks the position of a bright IR source. North is up, and east is to the left. }
\label{fig:field}
\end{figure}

Figure~\ref{fig:OBass} shows large-scale optical and infrared (IR) images of the S~171 nebula surrounding Berkeley~59  (galactic coordinates $l^{\rm II} = 118.2^{\circ}, b^{\rm II}= 5.0^{\circ}$). The nebula has a rather circular shape with an approximate extent as depicted by the large circle. The emission is less intense in the south-eastern part, and foreground obscuring clouds are scattered over the nebula. The arc of enhanced emission to the north and north-west have been listed as separate nebulae, NGC 7822 and the Cepheus Loop, but are clearly part of the same \ion{H}{ii} region. In the right panel of Fig.~\ref{fig:OBass} we have marked the location of the cluster NGC~7762, which lies at the north-west border of the \ion{H}{ii} region and seems to have a similar distance as Berkeley~59 \citep{liu19}, but according to \citet{pat95} it is an intermediate-age open cluster of 1.8 Gyr, and has a very different RV \citep{cas16}, for this reason we regard it as unrelated to Berkeley~59.
A more widespread association of early type stars, called Cep~OB4, was recognised by \citet{mac68}, and members of spectral types earlier than B3 are marked in the left image of Fig.~\ref{fig:OBass} (reviewed in \citealt{kun08}). 
Berkeley~59 is well confined within the area outlined by the central box, which is enlarged in Fig.~\ref{fig:field}.
Many fainter pre-main-sequence stars are associated with the cluster \citep{coh76,pan08,rosv13,get17,panw18}. After submitting this paper we became aware of the work by \citet{min21}, a major IR study of the young stellar population and its spatial distribution, where they find the youngest sources located in the surrounding shells and pillars and the older ones in the bubble surrounding the OB association, from which they suggest that the expanding HII region may have triggered star formation in Cep OB4.

Maps of the continuous radio emission show that S~171 has developed into a more or less spherical \ion{H}{ii} region of thermal emission but with a void in the south-eastern part. Fig.~\ref{fig:OBass} shows the 4 mm Planck map as contours overlaid on the IR image, and these are morphologically very similar to the 9 cm radio continuum map from \citet{ros80}. NGC 7822 and the bright patch seen close to the south-east from the centre in the optical image in Fig.~\ref{fig:OBass} show enhanced radio emission \citep{chu70}. The central, most intense emission comes from an area offset by about 8$\arcmin$ to the south-west of the cluster and was mapped in more detail in \citet{ang77} and \citet{har81}. Isophotes of forbidden line emission are congruent with the radio maps \citep{loz87}. A review of works related to the stellar content and the nebula is given in \citet{kun08}.    

The general properties of the S~171 complex are given in Table~\ref{tab:Be59}, where we have adopted a central position for the complex as between the two bright stars in Be~59 (A and B in Fig.~\ref{fig:field}). The radius of the \ion{H}{ii} region given in Column 4 is based on the distance to the complex cited in Column 3 and corresponds to the circle in Fig.~\ref{fig:OBass} with a radius of 1.6$\degr$.  Column 6 gives different estimates of the mean heliocentric RV of the stellar cluster, and the corresponding range in velocity is expressed in local standard of rest (lsr) in Column 7. The last column gives published central velocities of the ionised gas, mostly from radio recombination lines but also from H$\alpha$ and [\ion{N}{ii}] lines. 

Published cluster velocities are not very accurate and are based on only two or three stars, and they differ considerably from the central velocities obtained for the ionised gas. The latter are also discordant, and very different velocities were obtained for different positions over the nebula \citep[e.g.][]{ped80,ros80,loz87}. The lines are broad, full width at half maximum (FWHM) $>$ 30 km~s$^{-1}$, with complex line profiles in some directions, which makes it difficult to assign a precise value of the central velocity. 

The northern molecular cloud related to NGC 7822 was mapped in CO emission in \citet{elm78} and \citet{woo83}, and the shell structures related to the S 171 nebula were mapped in CO by \citet{lei89,yan92,yon97}, and in H$_{2}$CO by \citet{ros83}. These maps cover most of the complex with grid spacings of several arcminutes \citep[2$\arcmin$ for the central part in]{yan92}. The associated velocity patterns turned out to be rather complex with multiple velocity components at several positions, ranging in RVs from --1 km~s$^{-1}$ to --19 km~s$^{-1}$ (lsr) in some cloudlets. Considering also the viewing aspect angle of these cloudlets it appears from some of the central velocities listed in Table~\ref{tab:Be59} as parts of the shell may expand at very high velocities.

We can assume that the present structure of the S~171 complex is the result mainly of the interaction between the hot members in Be~59 with the surrounding molecular shell. Since the central velocity of the complex is not well established, and existing CO maps are rather coarse, we found it motivated to obtain new RVs of several stars in Be 59 and of the molecular line emission from the shell at a higher spatial resolution than obtained before.   

\section{Observations, reductions, and measurements}
\label{sec:obs}

\subsection{Optical spectroscopy}
\label{sec:spectra}

We aimed at including all OB stars in the Berkeley~59 cluster in our sample and furthermore reaching as low a mass as possible, but due to extinction and the signal-to-noise ratio (S/N) needed in the spectra to determine RVs, a practical limit was set around V $\sim$ 14.5 mag. A few stars had determined spectral types from previous studies, and for the rest we made rough estimates based on IR and optical photometry \citep{pan08} to select the program stars  (Sect.~\ref{sec:results}).
Optical spectra were collected for 27 stars in the Be~59 region at the 2.6 m NOT during two observing runs in October 2015 and August 2016 plus spectra obtained as filler programs in service nights. We used the high-resolution FIber-fed Echelle Spectrograph \citep[FIES;][]{tel14}, with a resolution of R $\sim$ 25000 covering the wavelength range from 3700--7300 \AA\ when used with Charge Capture Device 13 (CCD13) and 3620--8580 \AA\  when used with CCD15, which was installed in October 2016. Our program stars are marked in Fig.~\ref{fig:field} with numbers as defined in the subsequent section.

%\subsection{Reductions and measurements}
%\label{sec:reduction} 

\begin{table*}
\centering
\caption{Program stars, parallaxes, and proper motions.}
\begin{tabular} {llccc}
 \hline
\noalign{\smallskip}
Star ID$^a$  & RA \ \ \ \ \ \ \ Dec. & $\pi$  & $\mu$(RA) & $\mu$(Dec.)   \\
           & \ \ \ \ \ \ (2000.0)  &  mas & mas/yr & mas/yr       \\
\noalign{\smallskip}
\hline
\noalign{\smallskip}

P316, TYC 4026-151-1       & 00 01 11.6  +67 29 22 & 1.67 (0.009) & 7.40 (0.04) & -3.75 (0.01) \\
P239                       & 00 01 14.2  +67 26 34 & -3.00 (0.33) & -3.03 (0.35) & 7.47 (0.36) \\
IR star$^{1}$               & 00 01 26.7  +67 23 46 & 1.09 (0.05) & -1.09 (0.16) & -2.51 (0.15)  \\
P164                       & 00 01 36.1  +67 22 43 & 3.36 (0.03) & 10.54 (0.04) & -9.44 (0.04) \\
P109                       & 00 01 39.7  +67 24 37 & 1.53 (0.02) & -17.53 (0.03) & 1.78 (0.03)\\
P121                       & 00 01 41.5  +67 23 33 & 2.93 (0.02) & -6.83 (0.03) & -0.54 (0.03) \\
V747 Cep$^{2}$              & 00 01 46.9  +67 30 25 & 1.02 (0.03) & -1.57 (0.04) & -1.77 (0.05) \\
P310, SOP 7                & 00 01 51.9  +67 31 47 & & & \\
P15, SOP 11                & 00 02 00.1  +67 25 11 & 1.10 (0.05) & -2.49 (0.08) & -1.77 (0.08) \\
P59                        & 00 02 01.2  +67 26 49 & 0.89 (0.02) & -1.43 (0.03) & -1.78 (0.03) \\
P7                         & 00 02 07.7  +67 25 42 & 0.94 (0.03) & -1.54 (0.03) & -1.69 (0.05) \\
P106                       & 00 02 09.7  +67 22 16 & 4.28 (0.02) & -6.75 (0.04) & -33.92 (0.03) \\
BD+66$\degr$1674           & 00 02 10.2  +67 25 45 & 0.82 (0.03) & -1.54 (0.05) & -2.76 (0.05) \\
BD+66$\degr$1675           & 00 02 10.3  +67 24 32 & 0.93 (0.04) & -1.61( 0.05) & -2.04 (0.05) \\
P20, SOP 14                & 00 02 10.6  +67 24 09 & 0.93 (0.03) & -1.46 (0.05) & -1.98 (0.04) \\
P13,                       & 00 02 12.2  +67 24 14 & 0.95 (0.02) & -1.41 (0.03) & -2.00 (0.03) \\
P130, SOP 4                & 00 02 12.5  +67 28 36 & 0.92 (0.03)  & -1.99 (0.04) &  -1.87 (0.04)  \\
P201, TYC 4294-330-1, SOP 16  & 00 02 12.5  +67 30 03 & 3.15 (0.03) & & \\
P3, MacC15, SOP 6          & 00 02 13.6  +67 25 04 & 0.94 (0.03) & -1.71 (0.04) & -1.94 (0.04) \\ 
P23, SOP 13                & 00 02 13.7  +67 26 11 & 0.87 (0.04) & -1.55 (0.06) & -1.73 (0.05) \\
P62                        & 00 02 14.7  +67 23 17 & 0.91 (0.02) & -1.99 (0.03) & -1.99 (0.03) \\
P16, MacC29, SOP 2         & 00 02 19.1  +67 25 38 & 0.83 (0.03) & -1.88 (0.09) & -2.63 (0.08) \\
P114, SOP 9                & 00 02 20.1  +67 28 07 & 0.68 (0.04) & -1.99 (0.06) & -2.22 (0.05) \\
P66, SOP 8                 & 00 02 29.9  +67 25 44 & 0.95 (0.03) & -1.49 (0.05) & -2.22 (0.06) \\
P105                       & 00 02 38.6  +67 24 09 & 0.98 (0.03) & -1.16( 0.04) & -1.96 (0.04) \\
P195                       & 00 02 57.5  +67 23 48 & 0.89 (0.02) & -1.78 (0.03) & -1.66 (0.03) \\
P233                       & 00 03 06.0  +67 26 22 & 1.47 (0.03) & -4.26 (0.04) &  0.27 (0.05) \\
\noalign{\smallskip}
\hline

\end{tabular}
\label{tab:astrometric} 

\tablefoot{
  \tablefoottext{1}{2MASS J00012663+6723461, WISE J000126.63+672346.1, UCAC4 787-000033}
  \tablefoottext{2}{BD+66$\degr$1673}}
\end{table*}

Most targets were observed on at least two different dates (separated by a minimum of four days) to check for spectroscopic binaries. Because our targets are mainly early type stars with few available lines, often strongly rotationally broadened, we aimed for a S/N of $\sim$ 50 in the final spectra. However, our targets are heavily reddened (typically by 5-7 visual magnitudes). This lowers the S/N ratio significantly in the blue part of the spectra, and thereby the number of useful lines. Exposure times were up to 1800 s, and for the faintest targets two consecutive exposures were taken. From the list in \citet{fek99} we selected as early-type RV standards HR 1810 (B2.5~V), HR 7512 (B8  III), HR 8404 (B9.5 V), and HR 7903 (A0 III), stars not found to vary in velocity and with accuracies in the range 0.05 to 0.4~km~s$^{-1}$. We also observed Vega (A0 V), and for later spectral types we used the RV standard HD 31253 (F8). In addition, we collected spectra of a number of MK standards used for spectral classification of our targets.   

Standard calibrations (21 halogen flats, seven biases, and one ThAr) were obtained each afternoon. Data reduction was made using the FIEStool pipeline \citep{ste05}. Over the course of our observations, FIES was upgraded with a new CCD, in October 2016, and new octagonal fibres, in June 2017. The merging of the orders in FIEStool was particularly troublesome in the spectra of blue stars with the old setup but was enhanced with the new one.\ Thus, the merging was improved manually by cutting the noisy edges of individual orders. We then normalised the spectra by fitting a continuum.

We measured the stellar RVs using both cross-correlation techniques (`fxcor' in IRAF) as well as direct measurements of line centres by fitting a Gaussian profile to each line (`rvidlines' in IRAF). For most OB-type stars we mainly used the \ion{He}{i} and \ion{He}{ii} lines when available, while for later B and early A stars, and for targets where few lines were available, we also included the hydrogen lines.
  When both methods were applicable, we compared the results and typically found consistent values within the errors. The instrument drift is practically negligible compared to the typical measurement errors in the RVs of our targets. Multi-epoch RV results for the program stars are listed in Tables~\ref{tab:RVstars1} and~\ref{tab:RVstars2}.

\subsection{Molecular line observations}
\label{sec:COobs}

We observed $^{13}$CO \mbox{$J$\,=\,1\,--\,0}  at 110.201~GHz from the region surrounding Be 59 during several runs in 2016 and 2017 using a 3        mm SIS receiver at the 20 m telescope at OSO.  Frequency-switching mode was used with a 1600-channel hybrid digital autocorrelation spectrometer with a channel spacing of 25 kHz (0.07~km~s$^{-1}$), a bandwidth of 40~MHz, and the velocity resolution is about 0.2~km~s$^{-1}$ after smoothing. The single-sideband system temperature was typically around \mbox{150\,--\,400~K}. The pointing was checked regularly, and we estimate the pointing error to be less than a few arcseconds. At 110~GHz, the FWHM beam size of the 20 m antenna is 34$\arcsec$, and the main beam efficiency $\sim$0.30 (for an average elevation of approximately 30$^\circ$). The chopper-wheel method was used for the intensity calibration. The data reduction was performed with the spectral line software package  {\tt xs}\footnote{http://www.chalmers.se/rss/oso-en/observations/data-reduction-software}.

\begin{figure*}[t]
\centering
\includegraphics[angle=00, width=9.0cm]{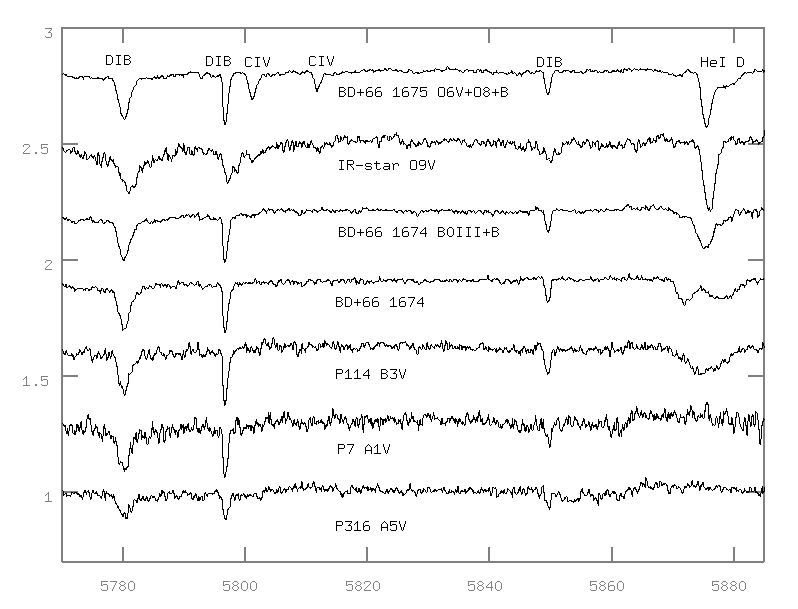}
\includegraphics[angle=00, width=8.0cm]{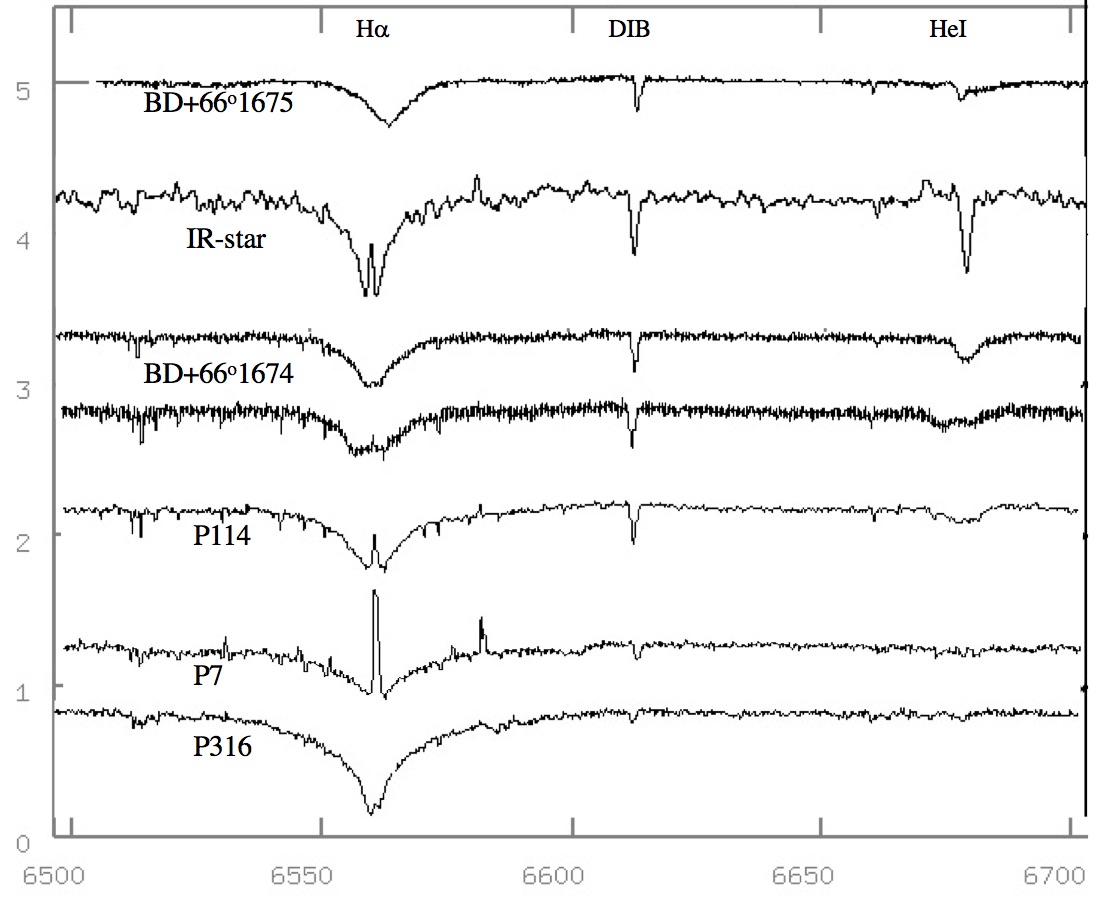}
\caption{Examples of normalised spectra obtained for some program stars over two selected wavelength regions. Certain spectral features discussed in the text are marked. The spectra are ordered in steps of 0.3 (intensity scale) in the left panel and 1.0 in the right panel and with a smaller spacing between the third and fourth spectra; the double-lined spectroscopic binary BD+66$\degr$1674 is shown at two different phases. To improve the S/N for the faint IR star, its left spectrum is a combination of many observations, Doppler-corrected for the binary motion, which smears out the DIBs.}  
\label{fig:spectra}
\end{figure*}

\begin{table*}
\centering
\caption{Spectral properties and heliocentric RVs.}
\begin{tabular} {llllcclccl}
 \hline
\noalign{\smallskip}
Star ID & V/(B-V)$^a$ & MK$_{lit.}$$^b$ & MK$_{NOT}$$^c$ & DIB$^d$ &  A$_V$ & V(Lit.)$^e$ & V(NOT)$^f$ & Number$^g$ & Notes$^h$ \\
           &  &  &  &  & &   km/s &   km/s  & nights &   \\
\noalign{\smallskip}
\hline
\noalign{\smallskip}

V747 Cep$^*$     &  10.07/1.31   & O5.5 V  & O5.5 V      & s & 6.1 &   & EB, SB1  & 12  & br  \\
BD+66$\degr$1675 &   9.08/1.08   & O7.5 V  & O6 V/O8 V/B & s & 4.6 & -9.9$^1$-9.3$^2$ & SB3 & 13 & \\ 
P16              &  11.81/1.91 v & O9      & O7 V        & s & 9.4 &   & SB1 & 7 &  \\ 
IR star$^{**}$    &               &         & O9 V        & s & 11.9 &   & SB1 & 18 &  \\     
BD+66$\degr$1674 &   9.60/1.07   & O9.7 IV & B0 III/B    & s &  3.4 & -9.1$^1$+5.2$^2$ & SB2, SB3?  & 12 &   \\
P66              &  12.95/1.54   &         & B0.5 V      & s &  6.9 &   & -20.4$\pm3.4$ & 2 &   \\
P3               &  11.30/1.14 v & B0.5 V  & B0.5 V      & s &   3.7 & -24.5$^2$ & -14.4$\pm$2.8 & 12 & br \\
P310             &  12.48/1.32   &         & B1 V(n)     & s &   6.1 &   & -21.1$\pm2.4$ & &   \\
P130             &  12.61/1.63   &         & B1 V(n)     & s &   7.6 &   & -22.0$\pm4.2$ & 2 & br  \\
P23              &  13.84/1.64   &         & B2.5Vn      & s &   6.1 &   & SB1 & 2 & br \\ 
P62              &  13.48/1.19   &         & B3 Vn       & s &   4.4 &   & -16.2$\pm2.2$ & & \\ 
P114             &  12.43/1.35   &         & B3 V(n)     & s &   5.3 &   & -18.7$\pm5.4$ & 2 & br \\
P15              &  12.78/1.25 v & B3 V    & B5 Vn       & s &   5.1 &   & -17.5$\pm3.4$ & 2 &  \\
P13              &  13.64/1.30   &         & B5 Vn       & s &   4.2 &   & -24.5$\pm1.3$ & 2 & br \\
P105             &  13.65/1.43   &         & B5 Vn       & s &   6.8 &   & -21.2$\pm4.6$ & & br  \\
P20              &  13.40/1.43 v & B8 III  & B8 Vn       & s &   4.5 &   & -16.5$\pm6.6$ & 2&  \\
P59              &  14.88/1.42   &         & B9 Vn       & s &   4.6 &   & -21.8$\pm5.8$ & &  \\
P195             &  14.22/1.47   &         & B9 Vn       & s &   4.1 &   & -14.2$\pm9.5$ & & \\
P7               &  13.87/1.24   &         & A1 Vn       & s &   3.9 &   & -15.3$\pm1.2$ & 2 &  \\
P316             &  12.47/0.72   &         & A5 V        & w &   1.2 &   & -20.4$\pm3.8$ & & nm \\
P233             &  13.00/0.82   &         & F8 IVn      & w &   0.3 &   & -12.0$\pm0.9$ & & nm \\
P164             &  12.27/0.60   &         & G2 V        & n &   0.3 &   & -6.0$\pm0.2$ & & nm \\
P239             &  12.95/1.03   &         & G3 V        & w &   0.5 &   & -30.0$\pm1.0$ & & nm  \\
P121             &  13.42/0.85 v &         & G5 Vn       & n &   1.1 &   & -7.2$\pm0.3$ & & nm \\
P201             &  12.43/0.86 v &         & G5 V        & n &   0.6 & -50.3$^3$ & SB2 & & nm  \\
P109             &  13.50/1.12 v &         & G8 Vn       & w &   1.0 & -29.2$^3$ & -30.5$\pm0.7$ & & nm  \\
P106             &  14.32/1.19 v &         & K5 Vn       & n &   0.0 &   & -12.9$\pm1.5$ & & nm  \\ 
\noalign{\smallskip}
\hline
\end{tabular}
\label{tab:stars} 

\tablefoot{
\tablefoottext{*}{BD+66$\degr$1673} 
\tablefoottext{**}{2MASS J00012663+6723461 } 
\tablefoottext{a}{Photometric data from \citet{pan08} except for BD+66$\degr$1673, 1674, 1675 taken from \citet{maj08}; stars marked v: variable in brightness according to \citet{lat11} and/or as listed in SIMBAD. }
\tablefoottext{b}{Spectral class from SIMBAD. } 
\tablefoottext{c}{Spectral class from NOT spectra. Components in binaries are separated. } 
\tablefoottext{d}{Diffuse interstellar bands:  strong (s), weak (w), or not present (n) }
\tablefoottext{e}{Heliocentric RV from the literature: 1. \citet{cra74}; 2. \citet{liu89}; 3. Gaia Data Release no 2. }
\tablefoottext{f}{Heliocentric RV derived from NOT spectra; close binaries are noted as single-lined $SB1$, double-lined $SB2$, triple-lined SB3, and eclipsing binary $EB$.}
\tablefoottext{g}{Number of nights of observations if more than one.}   
\tablefoottext{h}{br: broad lines with v sin $i$:s $>$~150~km~s$^{-1}$; nm: not member.}}

\end{table*}

Since the complex extends over a very large area in the sky, we had to select primarily areas where distinct shell structures are seen as dark foreground features in optical images. Some of these features were listed in the catalogue of dark nebulae in \citet{lyn62}. A total of 489 positions were observed including both observations over grids and at single positions, and some areas have a denser spatial coverage than others. In addition, observations of positions located outside obscuring clouds provided reference spectra.

\section{Results}
\label{sec:results}

\subsection{Program stars and parallaxes plus proper motions }
\label{sec:pm}

Most of our program stars in Be~59 are common to the stars observed by \citet{pan08}, and we chose their sequential number as prime designation (P). The stars are marked accordingly on the image in Fig.~\ref{fig:field}. The program stars were selected with the purpose of determining the stellar RVs and are therefore a brightness limited sample. However, we searched the mid-IR WISE point source catalogue \citep{cut13} for early type stars extinguished behind dark clouds within a radius of 8 arc minutes of the cluster centre. The star we name IR star (identical with  WISE J000126.63+672346.1) was located in the direction of very opaque shell structures obscuring the bright nebulosity and although faint in the visual, its brightness at mid-IR wavelengths was at the same level as for the well known O~stars in the cluster, for this reason
we included it in our program, and our multi-epoch spectroscopy will reveal it to be a binary O9\,V~star.

The program stars are listed according to increasing right ascension (RA) in Table~\ref{tab:astrometric} and include designations from \citet[][MaxC]{mac68}, Sharma et al. (\cite {sha07}, SOP), and the Tycho mission \cite[][TYC]{hog00} with the parallaxes and proper motions, when available, from the Gaia Data Release 2 (proper motion) and 3 \citep[parallaxes;][]{gai16,gai18,gai20}. Numbers are in milliarcseconds (mas) and quoted errors in parentheses. The parallax of the IR star is consistent with membership in Be~59. The parallaxes are well confined around a mean of 0.9~$\pm$~0.1~mas, corresponding to a distance of 1.1 kpc excluding six stars, which appear to be foreground stars. We adopt this distance for the present study.
The corresponding radius of the \ion{H}{ii} region is 31 pc as listed in Table~\ref{tab:Be59}. 

From the proper motions, excluding the six stars as above, we obtain a mean value for the cluster of $\Delta$$\alpha$ = --1.7 and $\Delta$$\delta$ = --2.0 mas/yr. The spread around this mean is small. Our cluster mean values, also regarding the parallax, are in good agreement with the values obtained by \citet{kuh19} based on 225 stars in Be~59.

\subsection{Stellar properties and radial velocities}
\label{sec:stellar}

We found a number of program stars to be spectroscopic binaries, including the IR star, and for most of these we extended the observations to get a hand on orbital data. Examples of normalised spectra collected at NOT of some program stars for two spectral regions are shown in Fig.~\ref{fig:spectra} with some spectral features discussed in the text marked. The third and fourth spectra from top show the double-lined spectroscopic binary BD+66$\degr$1674 at two phases when the stellar lines from the two components merge and split. For stars considered as members the diffuse interstellar bands (DIBs) are of about the same strength but weaker in foreground stars as evident for the star P316 at bottom of the panels. As seen in the bottom panel narrow H$\alpha$ line emission, flanked by [\ion{N}{ii}] emission in some cases, are superimposed on the broad H$\alpha$ absorption lines. These emission lines originate in the surrounding emission nebula.

Most program stars had not been classified on the MK system before, which we did by comparisons to standard stars based on traditional line ratios in the blue spectral region \citep{gra09}, also for the secondary components in binaries observed at large line splittings. For OB stars \citep{wal09} we used \ion{He}{ii} to \ion{He}{i} equivalent width ratios in the blue whenever possible, and, despite multiple epoch spectra, there are uncertainties in the line ratios due to blending of the binary or triple components.
The blue part of the spectrum of the reddened IR star is noisy, but using ratios based on equivalent widths of lines in the yellow and red parts we conclude that it is of spectral type O9~V (lines of e.g. \ion{He}{i}, \ion{He}{ii}, \ion{O}{iii} 5592~\AA, \ion{C}{iv} 5801, and 5812 \AA).
These line ratios were measured also for all early type stars for comparison. The most luminous star is the eclipsing and spectroscopic binary V747~Cep of spectral type O5.5~V.

In Table~\ref{tab:stars} we present the results from our spectroscopic survey, where the stars are ordered according to spectral type. Photometric data from the literature are in Column~2.  MK spectral classes as listed in the Set of Identifications, Measurements, and Bibliography for Astronomical Data (SIMBAD) Astronomical Database are in Column~3 and our estimates in Column~4. Column~5 gives information on the strength of the DIBs. Column~6 gives the visual extinction derived from the Two Micron All Sky Survey (2MASS) photometric data combined with intrinsic $J-H$ colours inferred from the spectral types obtained from \citet{koo83}. We assume that the excess in the $J-H$ colour is a measure of the extinction and use the relation $A_V = R_V \times E(J-H)/R(J-H) = 11.9 \times E(J-H)$, for R$_V$= 3.1 and the empirical 2MASS extinction coefficient $R(J-H)$ = 0.26 \citep{yua13}.
Column~7 lists heliocentric RVs from the literature and in Column~8 are our measurements with corresponding rms errors. For stars observed in only one night, the errors are from the determination of the RV from that single spectrum. Spectroscopic binaries are denoted as single-lined (SB1), double-lined (SB2), and triple-lined (SB3), and these will be discussed in more detail in the subsequent section. In Column~9 are the number of nights when FIES spectra were taken if more than one night. Individual measurements are listed in Tables~\ref{tab:RVstars1},~\ref{tab:RVstars2}, and \ref{tab:RVstars3}.

Based on the distance derived from the Gaia parallaxes and, if not available, from the photometry combined with spectral type and extinction, the strength of the diffuse interstellar absorption bands, and RVs, we find that all stars of spectral type A1 and earlier are likely members of Be~59. Those that we regard as non-members, are marked in Column 10, which also contains notes on stars with broad He or metal lines (br), for which we measured v~sin $i$ ~$> 150$\,km s$^{-1}$.

Fig.~\ref{fig:field} might give the impression that some members in Be~59 are not located behind foreground shell structures. Yet, they suffer from relatively large extinctions. The optical image indicates fragments of foreground clouds seen in absorption against a cavity illuminated by the H$\alpha$ nebulosity, which is even more clearly displayed in the IR in Figs.~\ref{fig:OBass} and \ref{fig:RVpositions}, where the contour overlay shows radio continuum emission enhanced at the location of the central cluster. Towards the west there is a pronounced ridge of dust behind which the IR star is located, and as shown in Table~\ref{tab:stars} the measured extinction towards individual stars is highly variable over the cluster area, but A$_V >$ 3 mag for all cluster members.

\subsection{Completeness of OB stars}
\label{sec:OBsample}

Next we wanted to determine how complete our sample of OB stars was. B3\,V stars have absolute magnitudes $M_V~=~-1.6~\pm~1.3$ mag \citep{weg06} and apparent magnitudes m$_V$ = M$_V$ + dm + A$_V$, where dm = $10.21$ mag for Be~59. Our magnitude limit for spectroscopy was set at V $\sim$ 14.5 mag, which means we can reach completeness for B3\,V stars only for regions with A$_V$~$\leq$~6~mag, and completeness for O7~V stars for A$_V$~$\leq$~9 mag. Among the most reddened sources in our sample is P16 and the IR star with visual extinctions slightly $>$ 9 mag while the median A$_V$ value is 5.1 mag. From the WISE catalogue search within a radius of 8$\arcmin$ (2.6~pc from the cluster centre) we find 38 sources brighter than 9.7 mag in the W1 (4.3 $\mu$m) band. This is the apparent magnitude of a B3\,V star if extinguished by A$_V$ = 5.1 mag, equivalent to A$_{W1}  = 0.30$ mag \citep{yua13}. Twelve of these are among our optically selected program stars (all of which are later found to be cluster members), and the brightest of the remaining 26 sources is the IR star ($W1 = 7.15 \pm 0.04$ mag). The rest constitute a mix of intermediate-mass stars and background giants, but a few could be reddened B~stars belonging to the cluster, although all are fainter than the IR star by 1 mag or more at 3.4 $\mu$m. Based on this we conclude that our sample may not be complete all the way to B3 stars, but most likely includes all O~stars in the area we studied.

\subsection{Spectroscopic binaries}
\label{sec:binaries}   

The frequency of spectroscopic binaries can be expected to be high for the most massive stars. According to \citet{chi12} stars with masses $>$16 $M_{\sun}$ as many as 82\% are close binary systems, while this fraction drops to about 20\% for 3 $M_{\sun}$  stars. In order to check for binarity we obtained repeated observations for the earliest type stars (Table~\ref{tab:stars}).

BD+66$\degr$1675 and BD+66$\degr$1674 were regarded as spectroscopic binaries in \citet{liu89} and \citet{cra74}. We found that BD+66$\degr$1675 is a triple-lined spectroscopic binary (SB3) and our data points suggest a period of 74 days and a systemic velocity of -10 km s$^{-1}$, as well as semi-amplitudes of 6, 150, and 280 km~s$^{-1}$ for components 1, 2, and 3, respectively. While components 1 and 2 have \ion{He}{ii} lines, only the \ion{He}{i} lines are detected in the faintest component, indicating spectral type B (Fig.~\ref{fig:multiepoch2}).

BD+66$\degr$1674 is a double-lined spectroscopic binary (SB2).\ However, individual lines are often heavily blended, and we could not extract associated periods and amplitudes.\ There are perhaps three components.

\begin{figure}[t]
\centering
\includegraphics[angle=00, width=8.7cm]{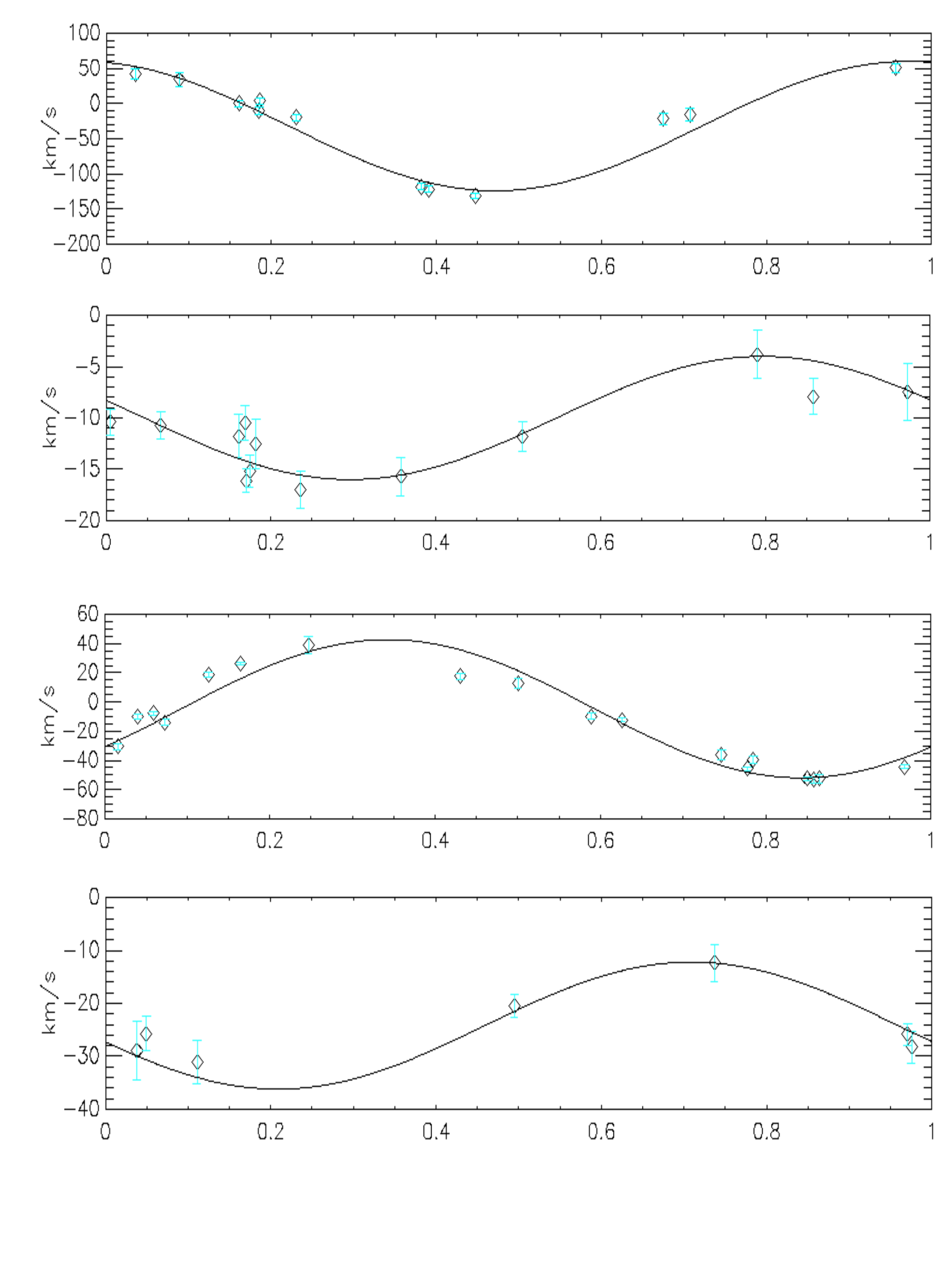}
\caption{Phase diagrams for (from the top): V747 Cep for a period of 5.33146 days; BD+66$\degr$1675, period of 74 days; the IR star, period of 9.51 days; and P16, period of 13 days. }
\label{fig:Phase}
\end{figure}

V747 Cep (BD+66$\degr$1673) is an eclipsing binary with a photometric period of P = 5.33146 days, an orbital eccentricity of $e$ = 0.3, and an inclination $i \sim $ 75$^{\circ}$ according to \citet{maj08}. \citet{MAp} assigned it a spectral type of O5.5~V(n)((f)) and found that it is also a spectroscopic binary. We can confirm that V747 Cep is a spectroscopic binary. In the data from two nights we note an indication of line splitting, but more observations are needed to confirm whether it is of type SB2 (Table~\ref{tab:RVstars1} and Fig.~\ref{fig:multiepoch1}). The RV data fit well with the photometric period and suggest a semi-amplitude of K$_1$ = 90 km~s$^{-1}$ around a systemic velocity of -26 km~s$^{-1}$ (Fig.~\ref{fig:Phase}, top panel), giving a close orbit with a semi-major axis of only 0.04~AU or $\sim$ 10 R$_\odot$. At the time of submission, we became aware of a more detailed study of this target in \citet{tri21}.

Multiple spectra of the hidden O~star we call the IR star (2MASS J00012663+6723461) demonstrates that it is a spectroscopic binary, for which we estimate a period of 9.51 days and K$_1$ = 45 km~s$^{-1}$, assuming a circular orbit, giving a projected semi-major axis $a~\sin i$ around 0.04~AU.

For P16, with only seven observations, only preliminary values can be assigned: a period of 13 days with a semi-amplitude of 12 km~s$^{-1}$. Examples of phase diagrams are shown in Fig~\ref{fig:Phase}. Finally, the late type foreground star, P201 (G5 V), show two velocity components and is likely an SB2.

\subsection{Mean cluster velocity}
\label{sec:MeanVel} 

Excluding spectroscopic binaries and non-members from Table~\ref{tab:stars}, we obtain a mean RV of the cluster of --18.75 $\pm$ 3.3  km s$^{-1}$ (--9.5 km s$^{-1}$ in lsr), which implies a standard error on the mean of 0.9 km s$^{-1}$ if we assume a Gaussian distribution. Hence, our derived mean cluster velocity is close to that derived for the central velocity of the ionised gas by \citet{ped80} as quoted in Table~\ref{tab:Be59}.

\subsection{The molecular line emission}
\label{sec:COvel}

\begin{figure*}[t]
\centering
\includegraphics[angle=00, width=6.9cm]{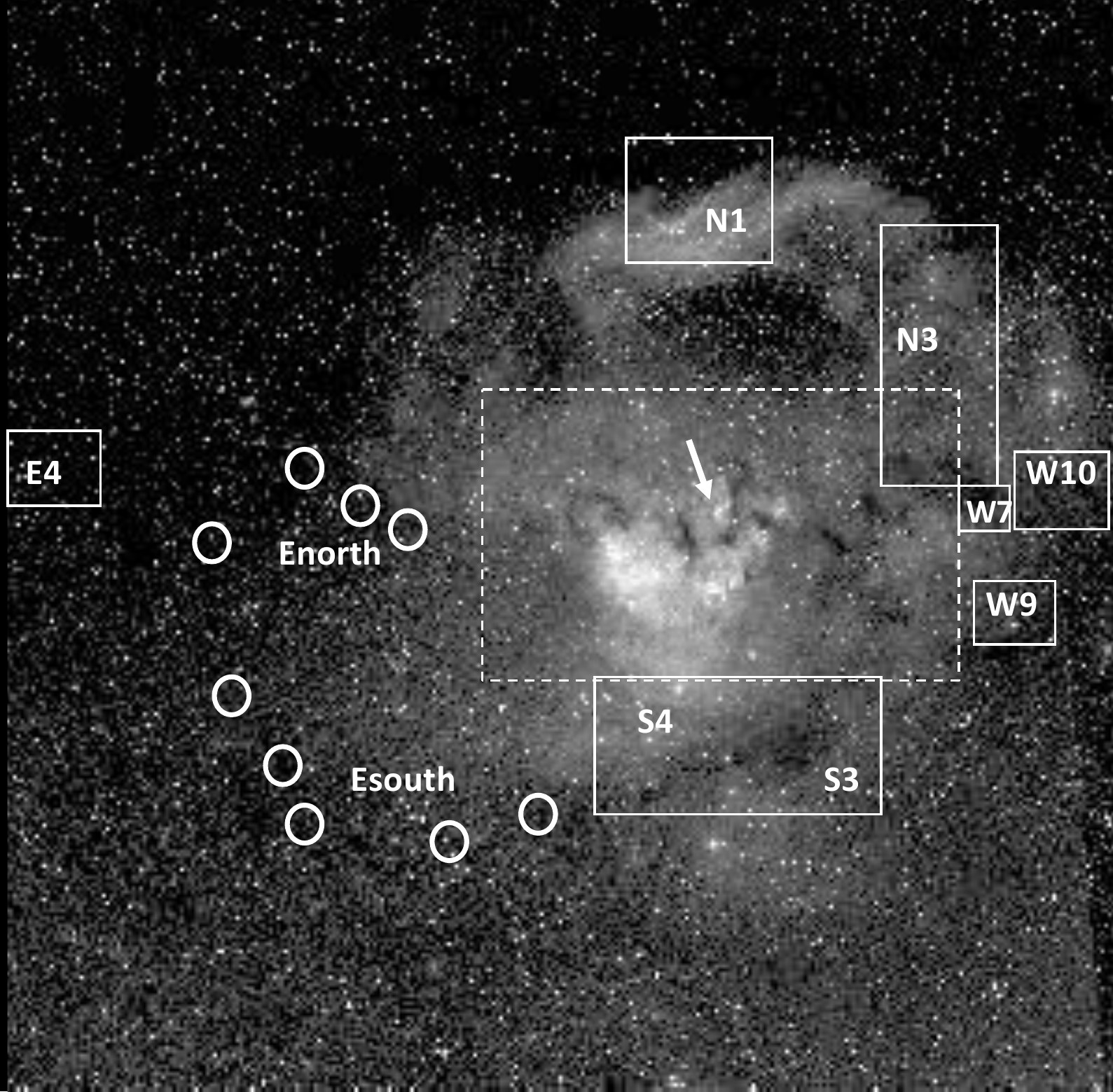}
\includegraphics[angle=00, width=10.65cm]{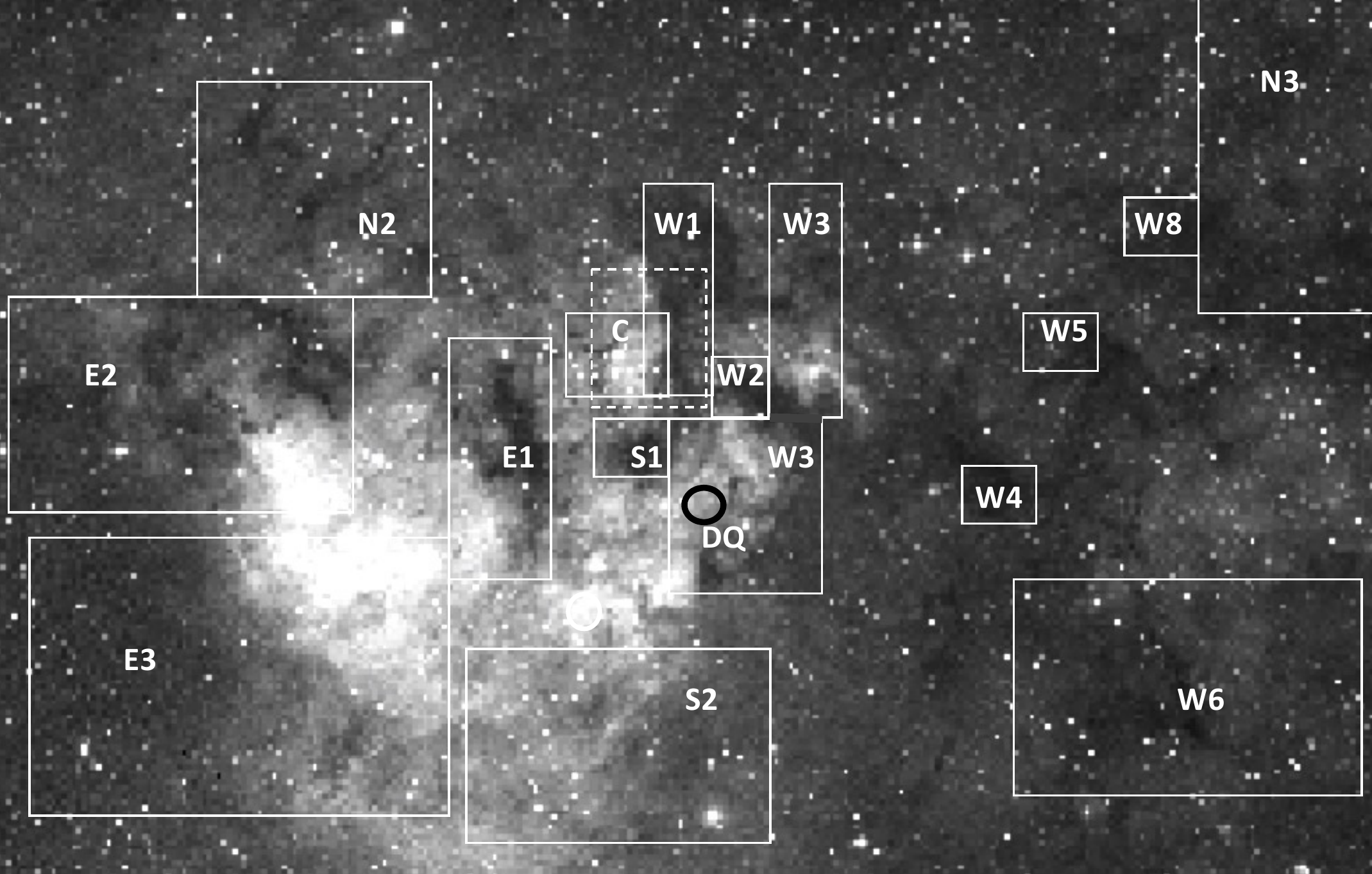}
\caption{Map of the fields we obtained CO spectra in. Left: Large region, 4.4$\times$4.4 degrees, surrounding the emission nebula S 171.\ The arrow points at the position of the central cluster. The northern arc of bright nebulosity is NGC 7822. In this panel the more peripheral areas observed for $^{13}$CO emission are marked and labelled. The circles (Enorth and Esouth) mark observations at selected positions along two dusty striations that extend to the east from the centre. The dashed box marks the central region shown in the right panel. Right: Image spanning 1.9$\times$1.2 degrees.\  The core of the stellar cluster is inside the dashed box. The position of an elephant trunk, called the `Dancing Queen' is marked with a black circle. North is up and east to the left in both images, which are from the red Deep Sky Survey.}
\label{fig:171fields} 
\end{figure*}

\begin{figure}[t]
\centering
\includegraphics[angle=00, width=8cm]{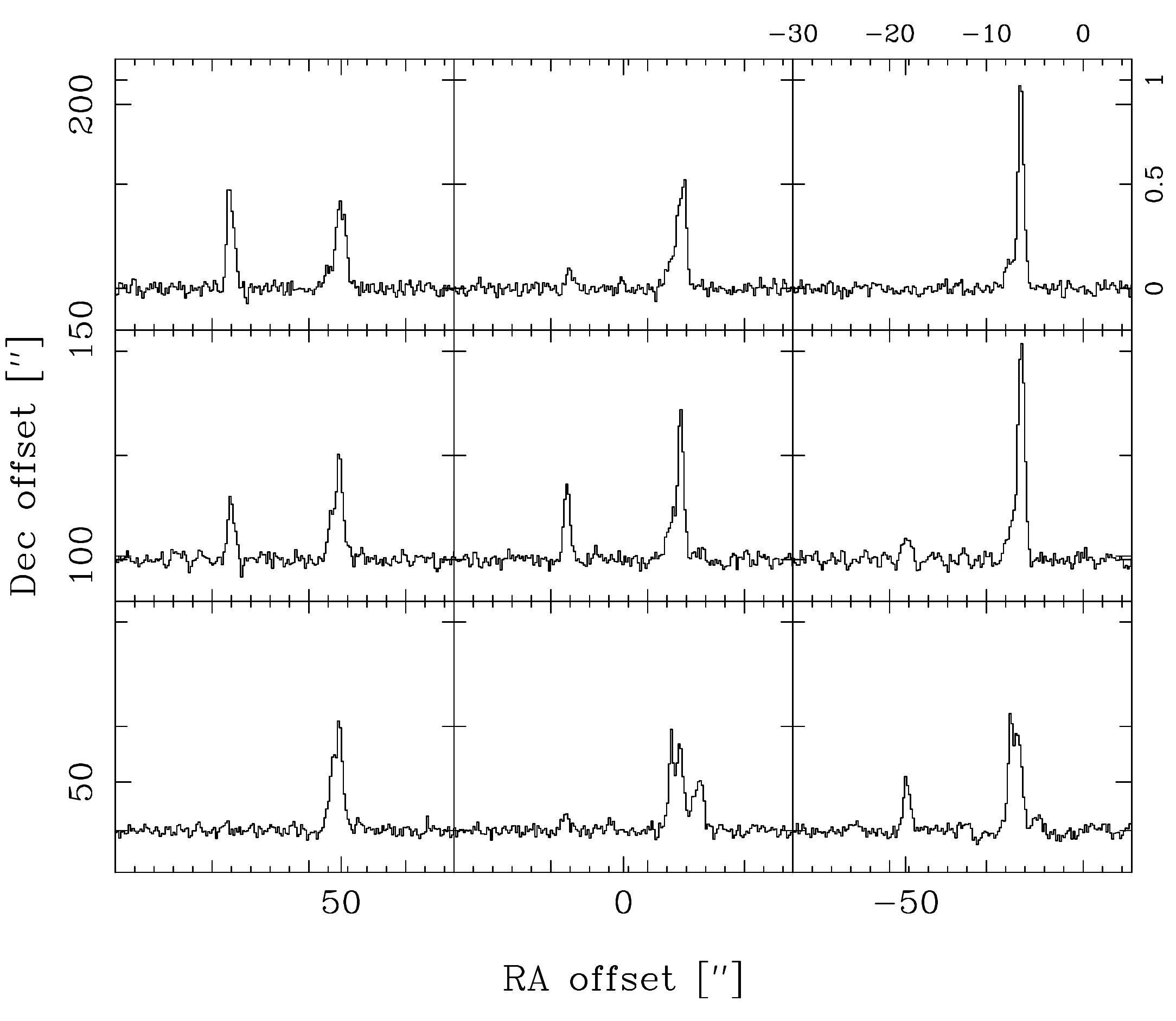}
\caption{Part of the map of $^{13}$CO spectra obtained for area W8. The offsets in RA and Dec. from the central position are in steps of 60$\arcsec$. The velocity interval, in km s$^{-1}$, and the antenna temperature, $T_{A}$ in Kelvin, are indicated in the upper-right panel. The component at around --18~km~s$^{-1}$, present in most panels, is related to the foreground shell, whereas the component at around --7~km~s$^{-1}$ has a local origin.}
\label{fig:W8map} 
\end{figure}

\begin{figure}[t]
\centering
\includegraphics[angle=00, width=7.5cm]{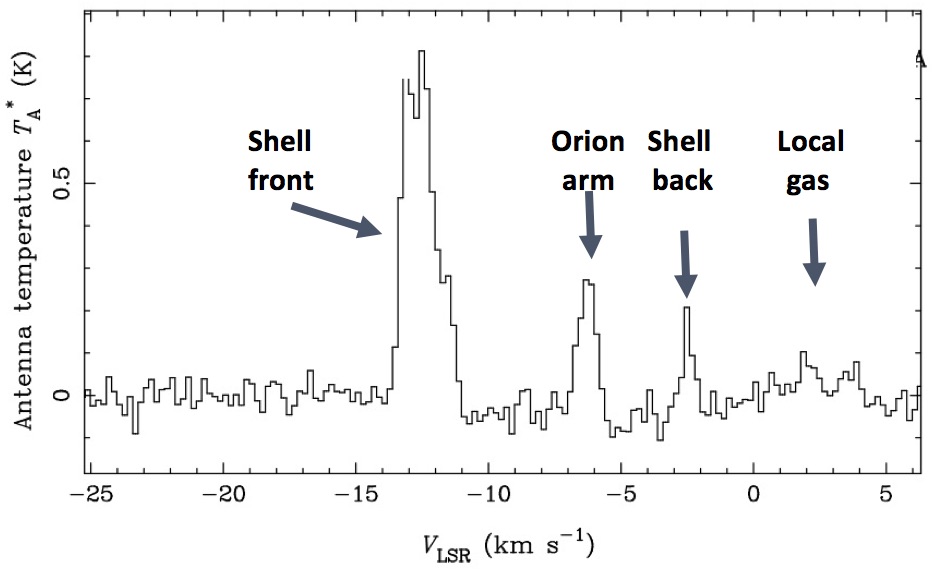}
\caption{$^{13}$CO spectra from one position in area E2. The strongest signal at around --13~km~s$^{-1}$ comes from the foreground shell, while the components at --6 and  +2~km~s$^{-1}$ are related to gas connected to the Orion arm and Gould's Belt. The component at around --2~km~s$^{-1}$ may trace the remote part of the molecular shell.}
\label{fig:E2E} 
\end{figure}

We obtained $^{13}$CO spectra from a total of 489 positions in the areas labelled in Fig.~\ref{fig:171fields}. The left panel shows a large area with the complex at the centre. The circles mark single positions observed along two dusty striations of dust that extend from the central parts to the east, and which are seen as distinct features, especially on the photographic version of the Palomar Sky-Atlas. Field E4 is located about 3$\degr$ east of the cluster and was originally included as a reference position outside the complex. Since we discovered a rather strong signal at high negative velocity here, we extended the observations over what appears to be isolated dust clouds. The right panel shows the areas observed in the central region of S 171. The circle at `DQ' marks the position of an elephant trunk, called the Dancing Queen, mapped in detail in $^{12}$CO and $^{13}$CO in \citet{gah06}.   

Examples of spectra are shown in Figs.~\ref{fig:W8map} and \ref{fig:E2E}. Signals from the molecular gas related to the foreground shell is seen in most panels for W8 in Fig.~\ref{fig:W8map} at $v_{\rm lsr}$ $\sim$ --18~km~s$^{-1}$ and at $v_{\rm lsr}$ $\sim$ --13~km~s$^{-1}$ for E2 in Fig.~\ref{fig:E2E}. These components are blue-shifted relative to the mean cluster velocity as derived in Sect.~\ref{sec:MeanVel} and are present in the bulk of our spectra. In the following, we refer to components at velocities $\leq$~--10~km~s$^{-1}$ as $V_{b}$. Especially in the central fields these lines can be split into two or three components well separated in velocity or blended as in Fig.~\ref{fig:E2E} (shell front). There is another component present at certain positions at velocities around --4~km~s$^{-1}$ (as in Fig.~\ref{fig:E2E}), slightly red-shifted relative to the cluster, which we refer to as $V_{r}$. As will be discussed in Sect.~\ref{sec:discussion} these could be related to the remote side of the complex. Representative spectra for each region are presented in the Appendix, Fig.~\ref{fig:COsp1} and Fig.~\ref{fig:COsp2}.

Our spectra also contain additional components, which are not related to the S 171 complex. In most directions we got signals at velocities around --2 to +4 ~km~s$^{-1}$, which we identify with the local gas, called feature A, related to the Gould's Belt \citep[e.g.][]{lin73}. Components at $v_{\rm lsr}$ $\sim$ --6~km~s$^{-1}$ are related to gas in the local spiral arm, the Orion arm, sometimes referred to as  `the other local feature'. As expected this component declines statistically in intensity with increasing galactic latitude. In our optical spectra of cluster members, interstellar absorption lines of  \ion{Na}{i} D are present in practically all stars. Also, these lines share the velocity of this local gas. Both local components are present in the majority of our $^{13}$CO spectra (Fig.~\ref{fig:E2E}). Emission from the background Perseus arm, at expected velocities of around $\leq$ --30~km~s$^{-1}$, does not enter our spectra since the S 171 complex is at a galactic latitude of $\sim$ +5$\degr$.

\begin{table*}
\centering
\caption{Velocities and peak intensities for the blueshifted and redshifted components in the fields marked in Fig.~\ref{fig:171fields}. The V$_{b}$ refers to velocities $\leq$~--10~km~s$^{-1}$, blueshifted with respect to the velocity of the central cluster of --9.5~km~s$^{-1}$, and V$_{r}$ refers to the interval --2 to --6~km~s$^{-1}$.  Column 2 gives the number of positions mapped, while the number of positions for which a given velocity component was found is given in parentheses after the V$_{b}$ and V$_{r}$ velocity ranges. $T_\mathrm{A}^*$(max) refers to the peak antenna temperature for $^{13}$CO. Details are provided in Sect.~\ref{sec:COvel}.}
\begin{tabular} {lclclc}

\hline
\noalign{\smallskip}
Area &  Number of  & $V_{b}$      &  $T_\mathrm{A}^*$(max)  & $V_{r}$    &   $T_\mathrm{A}^*$(max) \\
     &  positions  & (km/s lsr)  &                       & (km/s lsr) &                       \\ 
\noalign{\smallskip}     
\hline
\noalign{\smallskip}

W10    &  3 & -14.5 to -13.3 (2)       & 1.7     &  \ldots           & \ldots  \\
W9     & 13 & \ldots                   & \ldots  & -6.8 to -5.1 (11) & 0.8  \\   
W8     & 19 & -18.3 to -18.0 (7)       & 0.5     &  -5.5 to -4.8 (3) & 0.5   \\   
W7     & 13 & -12.5 to -10.5 (8)       & 1.4     &  \ldots           & \ldots  \\   
W6     & 25 & -12.1 to -12.0 (3)       & 0.9     &  \ldots           & \ldots  \\
W5     &  7 & -14.6 to -13.9 (7)       & 3.7     &  \ldots           & \ldots  \\
W4     &  9 & -13.2 to -11.9 (3)       & 1.2     &  \ldots           & \ldots  \\
W3     & 73 & -17.3 to -11.2 (57)$^{*}$ & 5.5     & -6.5 to -5.3 (15) & 0.3   \\
W2     &  6 & -16.4 to -12.5 (5)$^{*}$  & 4.3     &  \ldots           & \ldots  \\
W1     & 47 & -17.5 to -12.0 (47)$^{*}$ & 7.5     & -5.6 to -5.5 (3)  & 0.5    \\
C      &  6 & -21.0 to -13.5 (3)$^{*}$  & 2.9     & -6.5 to -6.3 (2)  & 0.6 \\
N3     &  9 & \ldots                   & \ldots  & \ldots            & \ldots  \\
N2     &  5 & -21.5 to -11.8 (2)$^{*}$  & 3.5     & -4.5 to -3.8 (2)  & 0.6  \\
N1     &  2 & -11.1 to -10.0 (2)       & 4.1     & \ldots            & \ldots  \\
S1     &  5 & -14.0 to -13.2 (5)       & 1.0     & \ldots            & \ldots  \\
S2     & 27 & -13.8 to -11.3 (8)       & 0.2     & \ldots            & \ldots  \\
S3     & 35 & \ldots                   & \ldots  & -3.1 to -2.9 (15) & 1.0 \\
S4     &  3 & -12.4 (1)                & 0.2     & -3.0 (1)          & 0.6 \\
E1     & 54 & -13.7 to -12.2 (2)       & 0.5     & \ldots            & \ldots \\
E2     & 21 & -13.7 to -12.2 (14)$^{*}$ & 1.4     & -4.3 to -2.5 (6)  & 0.3  \\
E3     & 76 & -13.3 to -10.2 (8)$^{*}$  & 0.7     & \ldots            & \ldots  \\
Esouth &  5 & -14.3 (1)                & 1.7     & \ldots            & \ldots  \\
Enorth &  4 & -15.6 to -12.7 (3)       & 0.4     & \ldots            & \ldots  \\
E4     & 29 & -14.5 to -10.3 (22)$^{*}$ & 0.6     & \ldots            & \ldots  \\

\noalign{\smallskip} 
\hline
\end{tabular}

\tablefoot{
  \tablefoottext{*}{Two or three separate $V_{b}$ components present at one or several positions.}
}

\label{tab:OSOvel} 
\end{table*} 

We made Gaussian fits to the line profiles, from which we extracted peak antenna temperature $T_\mathrm{A}^*$ (outside the earth atmosphere), central velocity, FWHM line widths, and for more complex profiles we decomposed the signals into separate components. The 1 $\sigma$ noise level at 0.2~km~s$^{-1}$ resolution is normally $<$ 40 mK, but reaches above 200 mK at several offset positions in W1, 6, 7,  and E2. The $V_{b}$ and $V_{r}$ components are as rule rather narrow, and in most cases the line widths are $<$ 1.5~km~s$^{-1}$ (FWHM).    

For most areas in Fig.~\ref{fig:171fields} we defined several central positions from which offset observations were made. The details of observed central positions and offsets can be found in Tables~\ref{tab:OSOpos} in the Appendix, where Column 3 gives the total number of observed positions for each sub-area. In Column 4 all offsets in RA are given in arcminutes, and for each of these the observed offsets in declination (Dec.) are given in Column 5. In Columns 6 and 7 we list those offsets in Dec. (for each offset in RA)  where we have detected a $V_{b}$ component ($\leq$ --10~km~s$^{-1}$) and a $V_{r}$ component (velocity interval --2 to --6~km~s$^{-1}$). Non-detections are marked with a dots. 

The amount of data so obtained is extensive since $V_{b}$ components are present in the majority of the positions, and for an overview we have chosen to present the ranges of different parameters for each area in Table~\ref{tab:OSOvel}. Here, the total number of observed positions is given in Column 2. The range in velocity for the $V_{b}$ component is given in Column 3 with the number of positions where it was detected in parentheses. Column 4 gives the maximum peak antenna temperature for the $V_{b}$ component observed in the area. Within each area the range in $T_\mathrm{A}^*$ can be large. Columns 5 and 6 gives the corresponding values for the $V_{r}$ component. 

Finally, another circumstance to consider when interpreting the data is that there is a system of molecular clouds that move at high negative velocities extending several degrees along the galactic plane. This system was mapped in $^{13}$CO by \citet{yon97}, and in Fig.~\ref{fig:CepCO} we have plotted their positions for clouds moving at $\leq$ --12~km~s$^{-1}$ together with the corresponding positions in our survey for two areas located outside the defined extent of the  \ion{H}{ii} region. Hence, this system of clouds have velocities in the same range as the $V_{b}$ components in the central area. This therefore opens the question of which clouds are associated with the S 171 nebula and which are not.       

None of these clouds at significant negative velocities appear as distinct dark clouds in optical images, and none is listed in the catalogue of dark clouds in {\citet{lyn62}. It appears that this system of clouds are background objects, possibly at the far side of the Orion arm, where gas can be expected to stream into the spiral arm. We suppose that the most eastern cloudlets in our survey, areas E4 and Esouth:E marked in Fig.~\ref{fig:CepCO}, may belong to the same system.

For the central clouds in Fig.~\ref{fig:171fields}, residing in areas from W10 to E3 (west to east) and S4 to N1 (south to north), there is no doubt that all belong to the dusty shell surrounding Be~59. Some of these clouds are seen in silhouette against the bright nebulosity and have bright rims and associated elephant trunks, and most individual cloudlets are connected to each other with dark nebulosity.

\begin{figure}[t]
\centering
\includegraphics[angle=00, width=9cm]{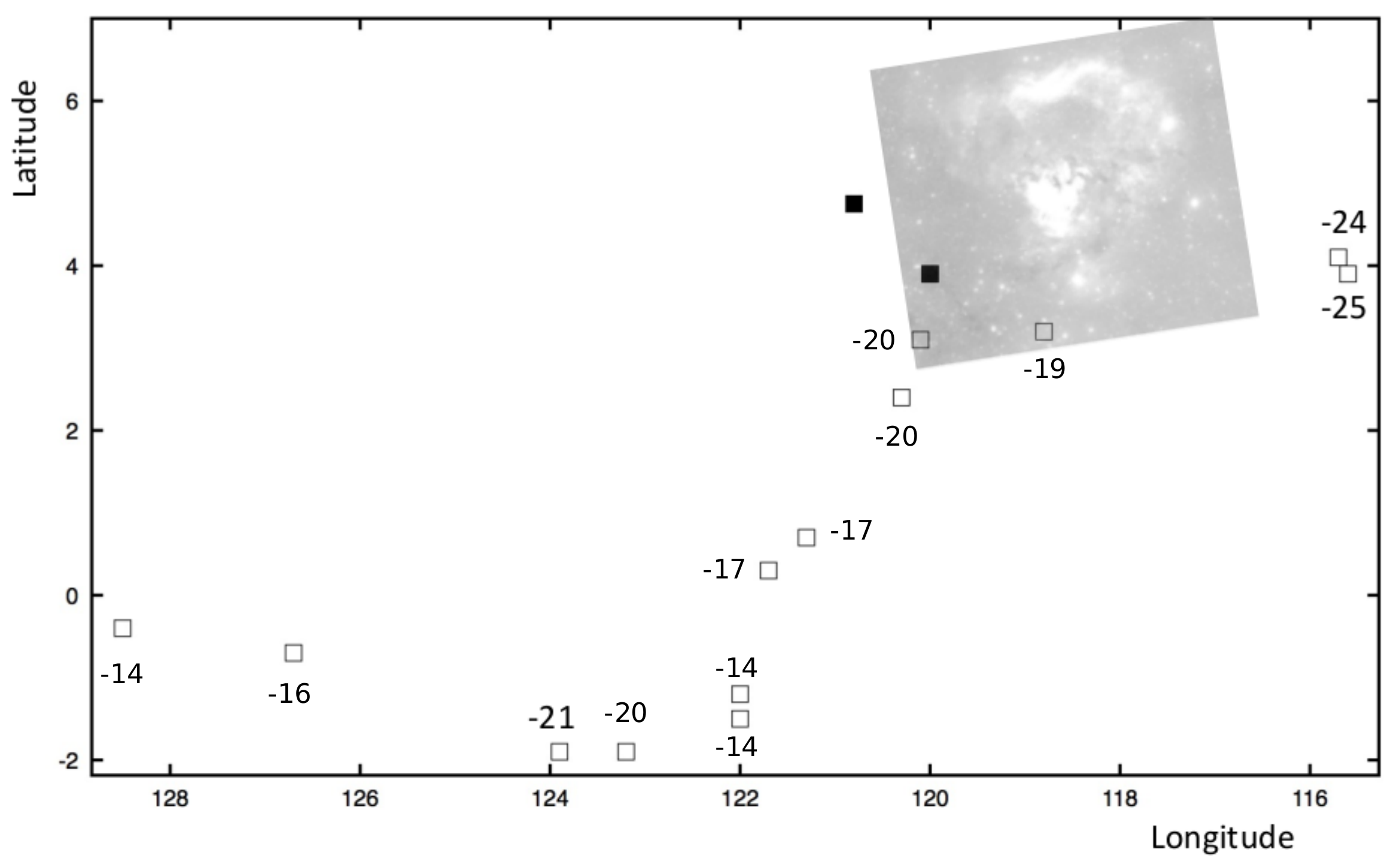}
\caption{Positions in galactic coordinates  of molecular cloudlets moving at high negative speed ($\leq$ --12~km~s$^{-1}$) in an extended area outside the central part of the S~171 complex (left image from Fig.~\ref{fig:171fields} inserted for reference). Open squares are cloudlets from \citet{yon97} and filled squares are E4 and Esouth:E from our survey. }
\label{fig:CepCO} 
\end{figure}

\section{Discussion}
\label{sec:discussion}

\subsection{Kinematics of the molecular shell}
\label{sec:patterns}

\begin{figure}[t]
\centering
\includegraphics[angle=00, width=1\linewidth]{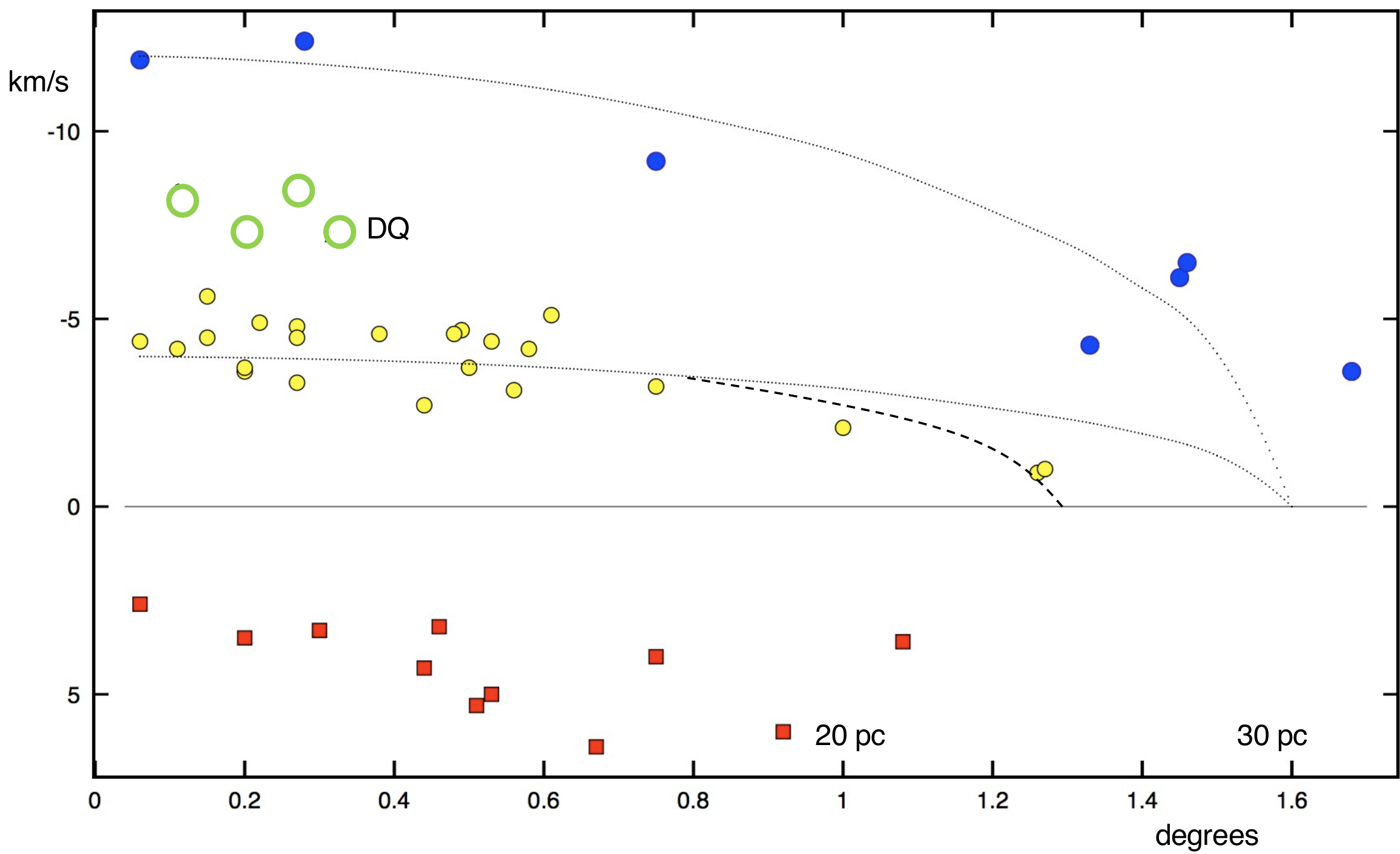}
\caption{Radial velocities relative to the cluster velocity for different sub-areas against angular distance from the centre (in all directions). Blue dots mark areas that are expanding at high velocity, $v_{exp}$ $\approx$ 12~km~s$^{-1}$, and yellow circles (smaller) areas expanding at moderate speed, $v_{exp}$ $\approx$ 4~km~s$^{-1}$, when assuming spherically expanding systems (dotted curves). Green circles mark areas with intermediate velocities (`DQ' is the Dancing Queen). Red squares show areas with components whose velocities are redshifted relative to the cluster. } 
\label{fig:RVpattern} 
\end{figure}

\begin{figure*}[t]
  \centering
\includegraphics[angle=00, width=8.5cm]{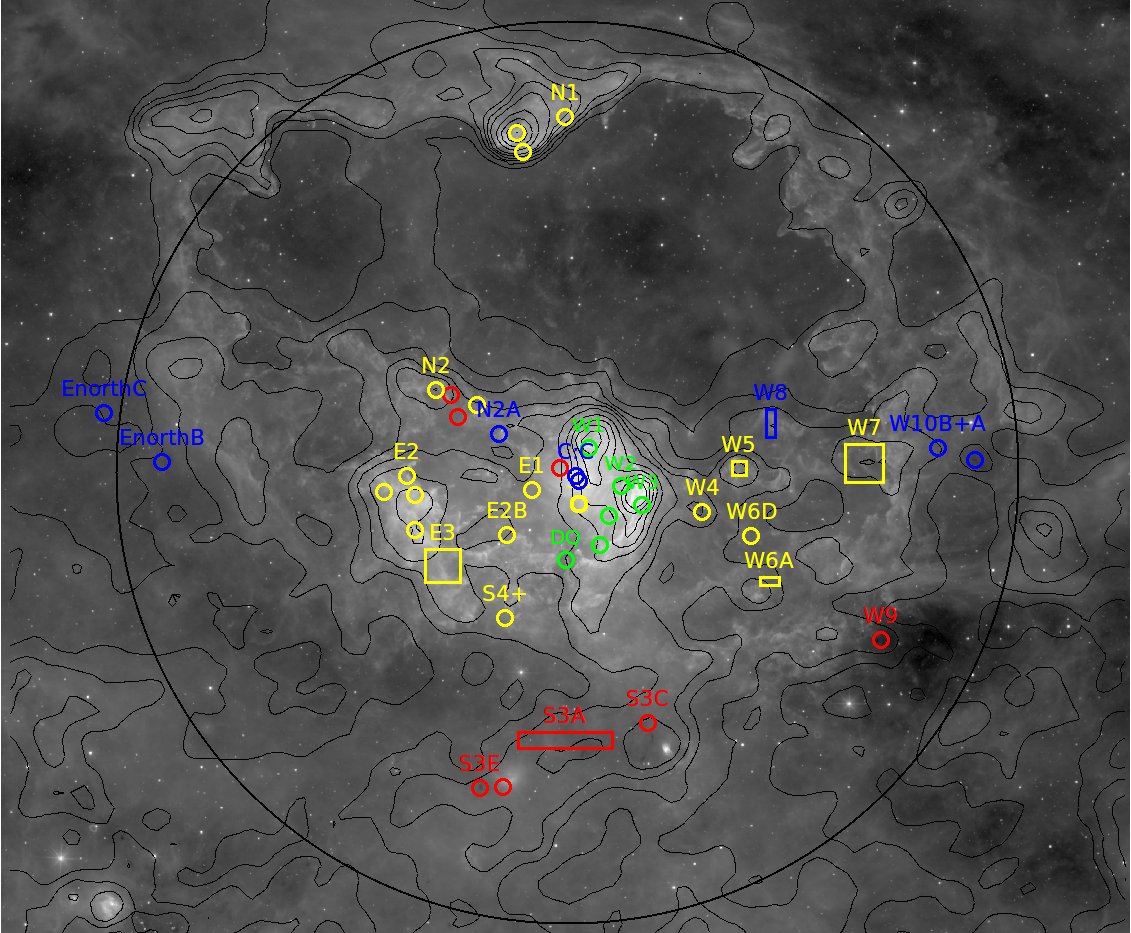}
\includegraphics[angle=00, width=9.5cm]{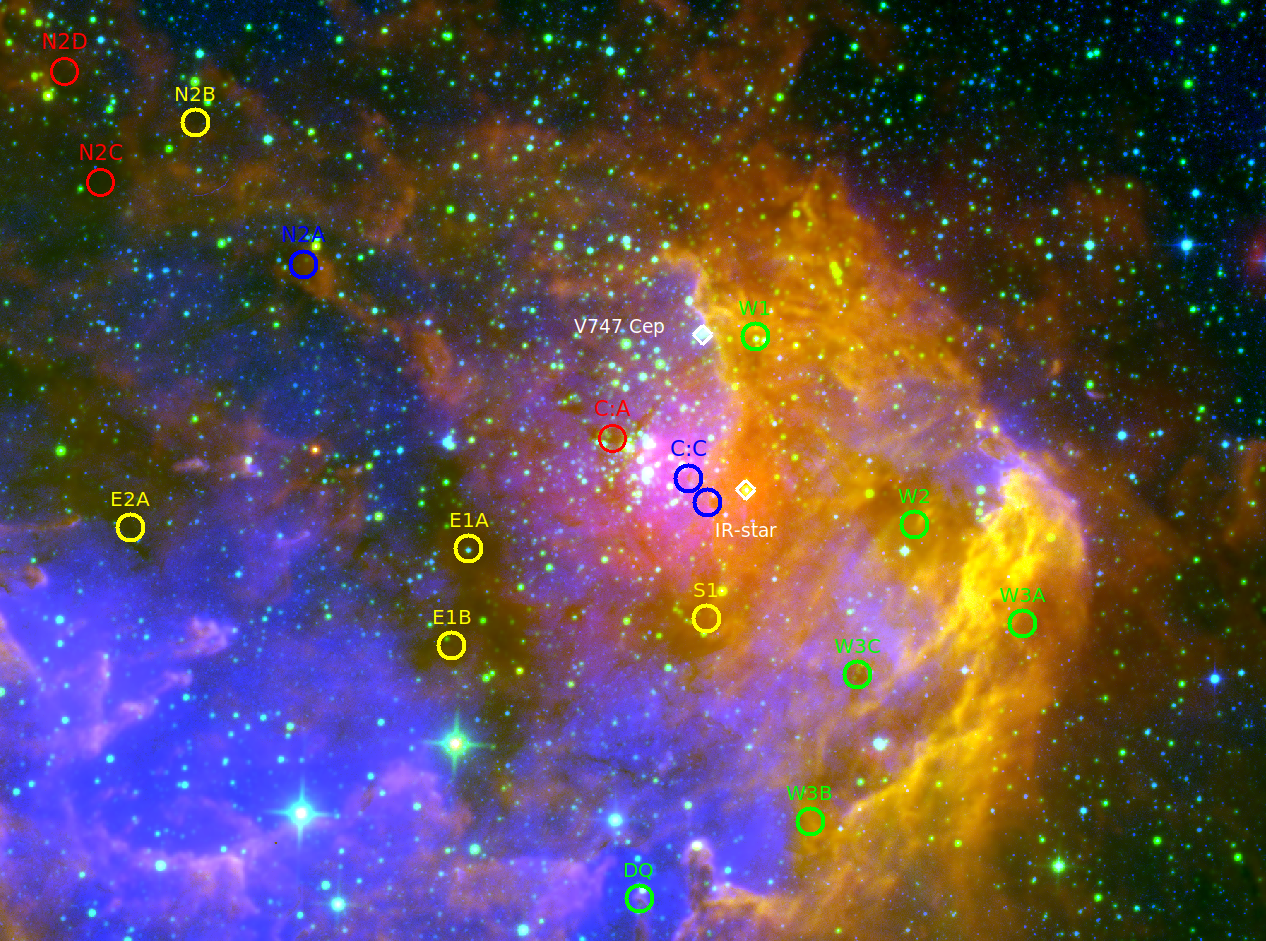}
\caption{ Map of cloudlets. Left: Overview of the region shown in Fig. \ref{fig:OBass} in 12~$\mu$m from WISE (grey scale) with the Planck 857 GHz map (black contours) overlaid to emphasise the dust emission. A black circle spanning a radius of 1.6$^{\circ}$ is centred on the Berkeley~59 cluster. The blue dots from Fig.~\ref{fig:RVpattern} are shown as blue circles and mark the locations where the high negative expansion velocities are found. Likewise, the green circles refer to the intermediate and the yellow the low negative velocity components, while the red circles mark positions where the positive component is found, possibly related to the expansion on the far side. Right: Close-up 55$\arcmin \times$ 40$\arcmin$ view of the central region where the red, green, blue coding is WISE 12 $\mu$m (red), 4.6 $\mu$m (green), and Digitized Sky Survey 2 red plates (blue). The positions of V747~Cep and the IR star are marked with white diamonds. }
\label{fig:RVpositions} 
\end{figure*}

The RV pattern over the molecular gas in the S 171 region is complex, as noted also in previous investigations referred to in Sect.~\ref{sec:S171}. In several areas listed in Table~\ref{tab:OSOvel} we find two distinct $V_{b}$ components (marked with an asterisk) at certain positions indicating that there are at least two systems of expanding clouds in the foreground of the complex. The lines are narrow but their velocities can change over small areas over our grids. When observed with a larger beam these components merge, and this appears to be the case for the broad plateau emission observed with a beam of 2$\arcmin$.7 in \citet{yan92} in W1 and W2 (their Fig. 3), which is resolved in our maps into narrow components drifting in velocity in RA and Dec. 

As described in Sect.~\ref{sec:COvel} individual spectra usually show several velocity components. Besides the two groups of components that we considered as not related to the S~171 complex we distinguish four kinematically separated groups related to the expanding shell. In Fig.~\ref{fig:RVpattern} these components are plotted as RV relative to the mean cluster velocity of --9.5~km~s$^{-1}$ against angular distance from the centre for different sub-areas (listed in Table B), where a given component was detected. Each point represents an average from all detections within a given sub-area. The spread in velocity within a given sub-area is small, $\leq$~1~km~s$^{-1}$.

The blue dots represent structures expanding at the highest speed in our direction when assuming a spherical shell expanding at $v_{exp}$ = 12~km~s$^{-1}$ and a radius of 31 pc (dotted curve). The yellow dots are consistent with a more moderate expansion velocity of $v_{exp}$ = 4~km~s$^{-1}$. The red squares in Fig.\,\ref{fig:RVpattern}, represent components that are red-shifted relative to the cluster by a few km~s$^{-1}$. Finally, we identify yet another component present in the central region (W1, W2, and W3) with velocities between the high and moderate velocity groups (marked with green circles). In these spectra the components associated with moderate velocities (yellow dots) are also present. The positions of clouds with implied high expansion velocities are shown as blue circles in both panels in Fig.~\ref{fig:RVpositions} and those at  intermediate velocities are all in the right panel as green circles.

While the blue dots in (Fig.~\ref{fig:RVpattern}) are reasonable well described by a spherically system of clouds expanding at high velocity and located at the outskirts of the \ion{H}{ii} region, the case is different for clouds moving at moderate velocities (yellow points). Not only is this system less extensive than the high-velocity system, also the outermost points fall below the curve assuming spherical expansion and a similar  radius. This cut off is indicated by the dashed curve, where RVs approach zero km~s$^{-1}$ at a radius of $\sim$ 25 pc. Hence, it appears that this system does not extend as far out from the centre as the high-velocity system. Moreover, it appears to have a counterpart in clouds moving at similar velocities but red-shifted relative to the cluster (red squares in Fig.\,\ref{fig:RVpattern}). These red-shifted clouds are also confined within a projected distances of 25 pc from the centre. A natural interpretation is that they are part of the same expanding shell and that we detect molecular emission both from its near and far sides.  It is hard to judge if there is also a remote counterpart to the high-velocity component because it would fall in the spectral region where the local components contribute. However, we note that at some positions such components are split in two.  

Clouds moving at moderate negative velocity are found in all directions over the nebula including also the dark dust cloud associated with NGC 7822. In contrast, the high-velocity clouds are found only along a band extending from west to east and crossing the centre. Sub-area Enorth:C is located just outside the adopted radius of the complex (beyond 31 pc in Fig.~\ref{fig:RVpattern}), and there is a small cloud in the foreground of the central cluster (sub-area C:C). The area over which the intermediate velocity component is detected coincides in part with the peak intensity of the continuous radio emission as mapped in \citet{har81}, outlined as yellow ovals in the Fig.~\ref{fig:RVpositions}.

To summarise, the bulk of the molecular gas in front of the bright nebulosity expands at moderate velocity. However, we also localised a system of smaller cloudlets, which appear to expand with a higher velocity than expected from simulations, although slower than could be inferred at start from some estimates of the central velocity of the complex (Table~\ref{tab:Be59}). These are the cloudlets marked with blue dots in Fig.~\ref{fig:RVpattern} and blue circles in Fig.~\ref{fig:RVpositions}.

\subsection{Masses of shell structures}
\label{sec:masses} 

Our $^{13}$CO observations do not cover the entire area of the S~171 complex. Nevertheless, we made an attempt to obtain rough estimates of the mass of individual cloudlets in the shell by the same method as used for deriving the masses of elephant trunks in \citet{gah06}. In short, the total mass of a given cloud is 

\begin{equation}
M \mathrm{[M_{\sun}]} = 1.35 \times 10^{-11}\>F(T_{ex})\>D^{2} A\>\langle I_{13}\rangle  \,,
\end{equation}
\noindent
where D [pc] is the distance, A [arcsec$^{2}$] is the cloud area, and $\langle I_{13}\rangle$ is the average integrated $^{13}$CO intensity over the area.  We have assumed an excitation temperature $T_{ex}$ = 23 K, which gives a dependence $F(T_{ex})$ = 30 (for instance \citealt{nik01}). 

We then treated clouds with high and moderate expansion velocities separately. For some clouds we included also areas outside our grids with similar prominent extinction and assuming the same value of the average integrated intensity. Most of the mass in the foreground shell is concentrated in the very dark clouds obscuring the areas of maximum continuous radio emission just to the west of the cluster, including areas W1, S1, W2, and W3 (Fig.~\ref{fig:RVpositions}). These areas are well sampled and covered also by \citet{yan92}, who derived masses for two `clumps' C1  (our positions W2 and W3) and C2 (our positions W1). In comparison with their results our estimates are similar for C2 but larger by a factor of 3 for C1, when the different adopted distance to the complex is taken into account. We regard this as a reasonable agreement, and also for area N1 in NGC 7822 our derived mass of $\sim$ 300 $M_{\sun}$ is similar (a factor of 1.5 larger) to the value quoted in \citet{elm78}. Our higher values are more consistent with column densities inferred from the large visual extinction of these clouds. For the clouds emitting the intermediate velocity components we estimate a mass of $\sim$ 200 $M_{\sun}$.

Based on our estimates the total mass of the shell in front of the \ion{H}{ii} region is of the order of 2200 $M_{\sun}$, of which $\sim$ 1500 $M_{\sun}$ is contained in the central areas described above. There is a considerable uncertainty in our mass estimates and more complete surveys of other molecular transition useful for mass determinations are warranted. Areas W9, S3, S4, and N3 do not show any components at high or moderate velocity. Either these are yet undisturbed clouds in the foreground of the nebula, or the lines are too weak for detection, which can be the case also in the relatively large areas E1 and E3, where lines at moderate expansion velocity are detected just over smaller parts of the maps.  

\begin{table}
  \centering
  \caption{Estimated properties for the clouds with high expansion velocity. The mass estimates are subject to considerable uncertainty.}
  \begin{tabular}{lcccc}
    \hline
    Position & Velocity   & Projected size  &  $\langle I_{13}\rangle$   &  Mass    \\
             & (km/s lsr) & 10$^3$ arcsec$^2$  &  (K km/s)                &  (M$_{\sun}$) \\
    \hline
    W10A     & -13.8      & 15.1           &  9.1  &       67   \\
    W10B     & -13.3      &  5.4           &  3.0  &        9   \\
    W8       & -18.2      & 10.8           &  0.9  &        5   \\
    C:C      & -21.0      &  3.6           &  4.1  &        7   \\
    N2A      & -21.5      &  9.5           & 10.1  &       47   \\
    EnorthB  & -15.6      &  3.6           &  1.8  &        3   \\
    EnorthC  & -12.9      &  3.6           &  0.9  &        2   \\ 
    \hline
    \end{tabular}
\label{tab:cloudlets}
\end{table}  

All cloudlets that we identify as moving at the highest expansion velocities have low masses. Among these, W10A and N2 are the most massive, $\sim$ 70 $M_{\sun}$ and $\sim$ 50 $M_{\sun}$, while the others have very low masses, $\leq$ 10 $M_{\sun}$. As will be discussed in the subsequent sections, this may be the reason why these cloudlets obtained a larger speed and reached farther out from the centre than the heavier clouds moving at moderate velocities.

\subsection{Dynamical evolution of the complex}
\label{sec:dynamics} 

\begin{table}
\centering 
\caption{Physical parameters for the cluster and expanding shell.} 
\begin{tabular}{lcc} 
\hline
\noalign{\smallskip}
Parameter & Value & References  \\
\noalign{\smallskip} 
\hline
\noalign{\smallskip}
\noalign{\smallskip}
{\it Cluster} & & \\
\noalign{\smallskip} 
L$_{t}$ [10$^{6}$ L$_{\sun}$] & 1.8 & 1, 2  \\ 
N$_{Lyc}$ [10$^{49}$ sec$^{-1}$] & 6.8 & 2   \\
\noalign{\smallskip}
\noalign{\smallskip}
{\it  \ion{H}{ii} region} &  & \\
\noalign{\smallskip}
radius [pc] & 31 & 1 \\
T$_{e}$ [K] & 8200 & 3  \\ 
n$_{e}$ [cm$^{-3}$] & $\sim$ 30--40 &  4  \\
\noalign{\smallskip}
\noalign{\smallskip}
{\it Foreground clouds} &  & \\
\noalign{\smallskip}
v$_{exp}$ [km s$^{-1}$] & 4 and 12 & 1  \\
total mass [M$_{\odot}$] & 2200  & 1  \\
\noalign{\smallskip}  
\hline 
  
\end{tabular} 
\label{tab:parameters} 

\tablefoot{1. This work;  2. \citet{ster03}; 3. \citet{ros80}; 4. \citet{har81}. }

\end{table}

The expansion of the S~171 region is driven by the interaction from the hot members in the Be 59 cluster. Some quantities related to the physical conditions in the complex are listed in Table~\ref{tab:parameters}, where the first entries contain estimates of  the total luminosity, L$_{t}$, and the total flux of Lyman continuum photons, N$_{Lyc}$, for all cluster members earlier than B1. The five O stars dominate the fluxes, and the contributions from members later than B0.5 are negligible. According to \citet{pan08,lat11,panw18} the ages of pre-main-sequence stars in the general area around Be~59 indicate a spread from 0.5 to 5 Myrs. However, it is also noted that the massive stars in the cluster are not significantly evolved and with inferred ages consistent to what is assumed in the present study.

The middle panel contains estimates of the electron temperature, $T_{e}$ and electron number density, $n_{e}$, related to the physical state of the plasma in the central part of the \ion{H}{ii} region followed by properties of the molecular shell in front of the nebula as estimated above. Here, the total mass of the expanding molecular shell on the remote side of the complex could not be estimated. 

As shown in both panels of Fig.~\ref{fig:RVpositions}, the areas of maximum continuous radio emission are coinciding with some highly obscured foreground clouds, as W1 and W2, which are offset to the west relative to the central cluster. This enhanced emission is most likely related to photodissociation regions (PDRs) hidden behind the dark clouds and excited by UV light from massive stars in Be~59. Possibly, a main contribution comes from the most luminous O star (V747 Cep) and the heavily obscured O star (the IR star), which are also offset to the west (by $\sim$4 pc in projected distance).
It is also over this general area our OSO spectra contain lines from the intermediate velocity component (green circles in Fig.~\ref{fig:RVpositions}).

A similar type of interaction takes place at the periphery of the \ion{H}{ii} region north of the cluster, where the bright nebula NGC 7822 borders a massive and opaques cloud, N1, only that in this case the PDR is visible as a bright nebula. The star marked with a circle just south of NGC 7822 in Fig.~\ref{fig:OBass} was noted as a member of the Cep OB4 association by \citet{mac68}. This star (MacC 9 = V399 Cep) of spectral type B2 is at the right distance to the complex according to Gaia DR 2, but we conclude that it plays a minor role for the excitation of NGC 7822 in comparison to the central O stars. 

The ionised gas is thermal \citep{ang77,ped80}, and it is unlikely that the large expansion velocities observed has been caused by an early supernova explosion, which besides would have disrupted the entire molecular shells. 

In Sect.~\ref{sec:patterns} we distinguished three systems of expanding shell structures, one containing rather massive clouds moving at 4 km s$^{-1}$ relative to the cluster and presumably with a counterpart at the remote side of the complex, another system with predominantly smaller cloudlets expanding at 12 km s$^{-1}$, and one central area where some clouds move with velocities between these two.

The 4~km~s$^{-1}$ component is in line with expectations from some numerical simulations of evolving \ion{H}{ii} regions with comparable sizes at an age of 2 Myrs  \citep[e.g.][]{fre03,hos06}. On the other hand, the 12 km s$^{-1}$ component traces a system of clouds moving outwards at a considerably higher speed compared to theoretical predictions assuming similar total luminosities and ionising fluxes as in Table~\ref{tab:parameters}. As discussed above, this system of clouds extends farther out from the centre and contains cloudlets of smaller mass than those expanding at moderate velocities. For the high-velocity clouds the crossing time to reach from the centre to the periphery of the \ion{H}{ii} region at constant speed is 2.5 10$^{6}$ yrs, which is in line with the cluster age.

Below, we consider as a working hypothesis that the high-velocity clouds were confined and detached from the more massive structures in the expanding shell early in the expansion of the shell. In our picture, the shell is very clumpy and contains small overdense clumps. The entire shell is decelerated by accumulation of material from the ISM, but dense clumps have more inertia and are more difficult to slow down. The tiny cloudlets retained a higher speed and are now located farther out than the shell, even outside the border of the \ion{H}{ii} region (Fig.~\ref{fig:RVpattern}). The clouds are confined within a relatively narrow belt extending from the western edge of the \ion{H}{ii} region over the central part to the eastern edge. The cause for this asymmetry is unclear.

\section{Numerical simulations of the complex}
\label{Sect:Simulations}

In this section we explore to what extent we can match the complex velocity pattern observed for the molecular gas in S 171 with a theoretical model simulation. As discussed in Sect.~\ref{sec:dynamics}, we have identified a system of high-velocity cloudlets of relatively small masses that have reached farther out from the centre than the more massive cloud structures that form the shell (Fig.~\ref{fig:RVpattern}). We attempt to include both these cloud systems in our simulation.

Our simulation describes how a cloud of gas is affected by the stellar winds of an embedded star cluster. This requires coupling between hydrodynamics (which describes the dynamics of the gas), gravity (which describes the dynamics of the stellar cluster), and stellar evolution (which describes the mass evolution and stellar wind output of the individual stars), plus a prescription for stellar winds.

\subsection{Multiphysics stellar wind bubble model}

\subsubsection{Model}

We simulate hydrodynamics using the {\sc fi} code \citep[][based on \citet{her89,ger04}]{pel04}, which is based on the smoothed particle hydrodynamics (SPH) formalism. Stellar winds are implemented using the prescription by \citet{helm19}, which inserts new SPH particles corresponding to the stellar wind into the hydrodynamics code. The stellar parameters required to describe the winds' mass loss rate and velocity are prescribed by the {\sc seba} code \citep[][with the adjustments by \citet{too12}]{por96}, a parameterised stellar evolution code. The stars' dynamics are simulated using the {\sc huayno} code, a second-order symplectic, direct N-body code \citep{pel12}. All of these codes were brought together in the Astrophysical MUltipurpose Software Environment \citep[AMUSE;][]{por18}. Finally, the gas and star cluster were able to affect each other through Bridge interactions \citep{fuj07}, which resolved gravitational interactions between the two components every 1 kyr, while leaving the internal evolution of each component unaffected. 

Our initial conditions consist of a number of gas and stellar components. We include a main gas cloud of 3000 M$_\odot$ (to take into account unobserved material behind the \ion{H}{ii} region) and a virial radius of 6 pc distributed following a Plummer profile \citep{plu1911}, which represents the centrally concentrated gas left over from star formation. We place two cloudlets of 10 and 50 M$_\odot$ (Plummer spheres with virial radii of 0.5 pc) at rest at $x=\pm 3$ pc, representing two cloudlets that have already detached from the main gas distribution. The velocity fields of these Plummer spheres are virialised and non-turbulent.

The stellar component consists of $10^3$ stars between 0.5 and 16 M$_\odot$ following a Salpeter initial mass function, and distributed according to a 2 pc virial radius Plummer sphere. Additionally, we include a collection of massive stars representative of the O-type (and giant B-type) stars observed. We used \citet{ster03} to associate a mass with the stellar type of each massive star. The mass assumed for every star is provided in Table \ref{tab:massivestars}. We do not initialise these stars as binaries, but we do concentrate them within the inner 0.1 pc of the cluster.

\begin{table}
\centering 
\caption{Massive stars in the simulation with their stellar types and corresponding masses from \citet{ster03}.} 
\begin{tabular}{lcc} 
\hline
\noalign{\smallskip}
Star & MK class & Mass M$_\odot$  \\
\noalign{\smallskip} 
\hline
\noalign{\smallskip}
\noalign{\smallskip}
V747 Cep A & O5.5V & 50.4 \\
V747 Cep B & O6V & 45.2 \\
BD+66$^\circ$ 1675 A & O6V & 45.2 \\
BD+66$^\circ$ 1675 B & O8V & 30.8 \\
P16 A & O7V & 37.7 \\
P16 B & B? & 30 \\
IR star A & O9V & 25.4 \\
BD+66$^\circ$ 1674 A & B0 III & 27.4 \\
\hline 
  
\end{tabular} 
\label{tab:massivestars} 
\end{table}

\subsubsection{Results}

We ran this simulation for 2 Myr at a resolution of about $3\cdot 10^{-3}$ M$_\odot$ per initial SPH particle (and $10^{-5}$ M$_\odot$ for the stellar wind particles). Figure~\ref{fig:hydro_maps} presents a number of snapshots of the simulation. The stellar winds from the centrally concentrated massive stars open a cavity in the gas cloud that expands with time and sweeps up the detached cloudlets. In the bottom panel, we can see how the core of the more massive cloudlet (50 M$_\odot$) lags behind the shell and deforms the shell immediately around it. This implies that detached cloudlets cannot be accelerated to velocities beyond that of the shell purely due to stellar winds. 

Deviations from a spherical shell have also arisen where stellar winds break through due to slight asymmetries, but these do not strongly affect the estimate of the shell radius. 

\begin{figure}[t]
\centering
\includegraphics[width=0.93\linewidth]{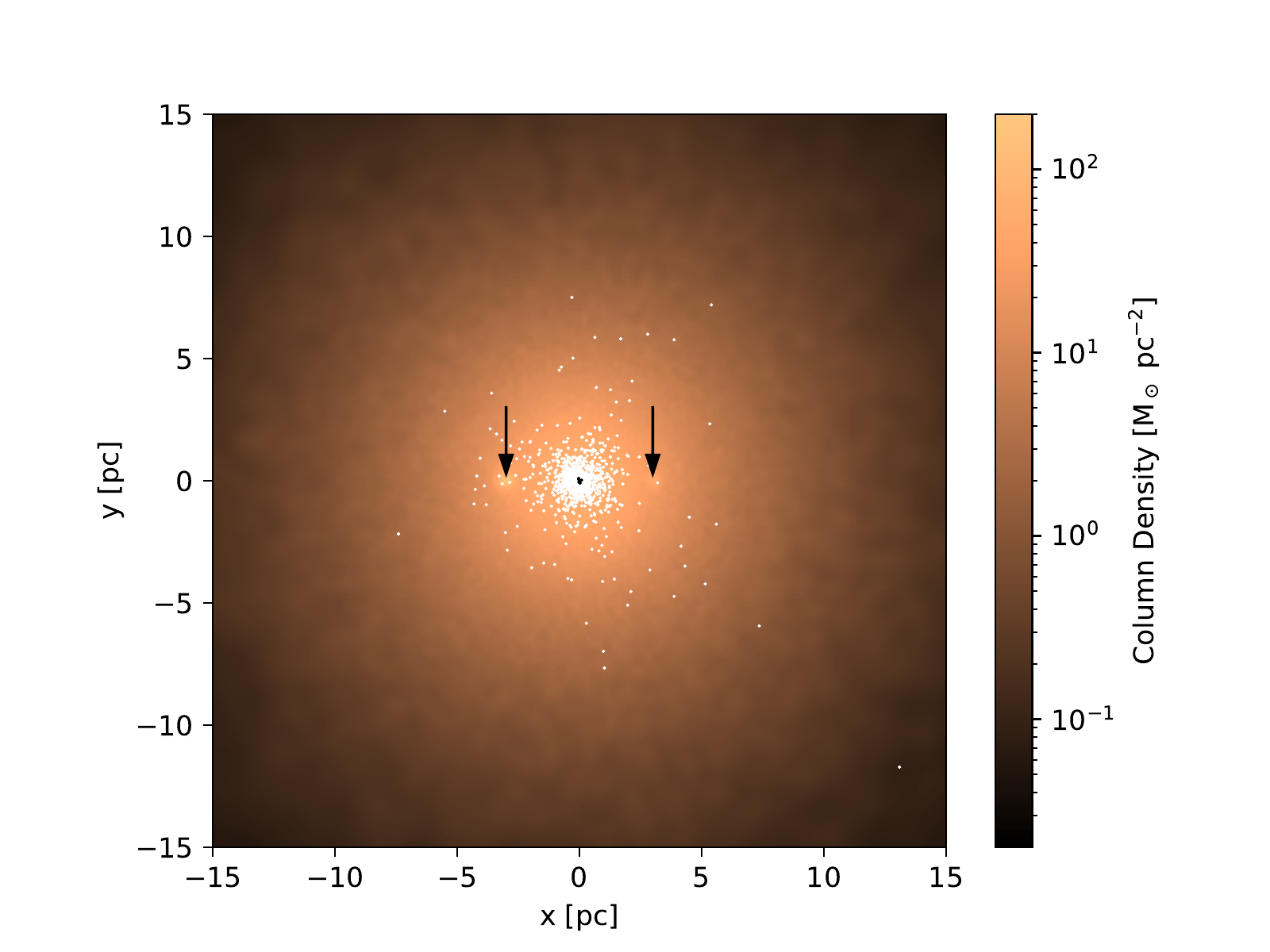}
\includegraphics[width=0.93\linewidth]{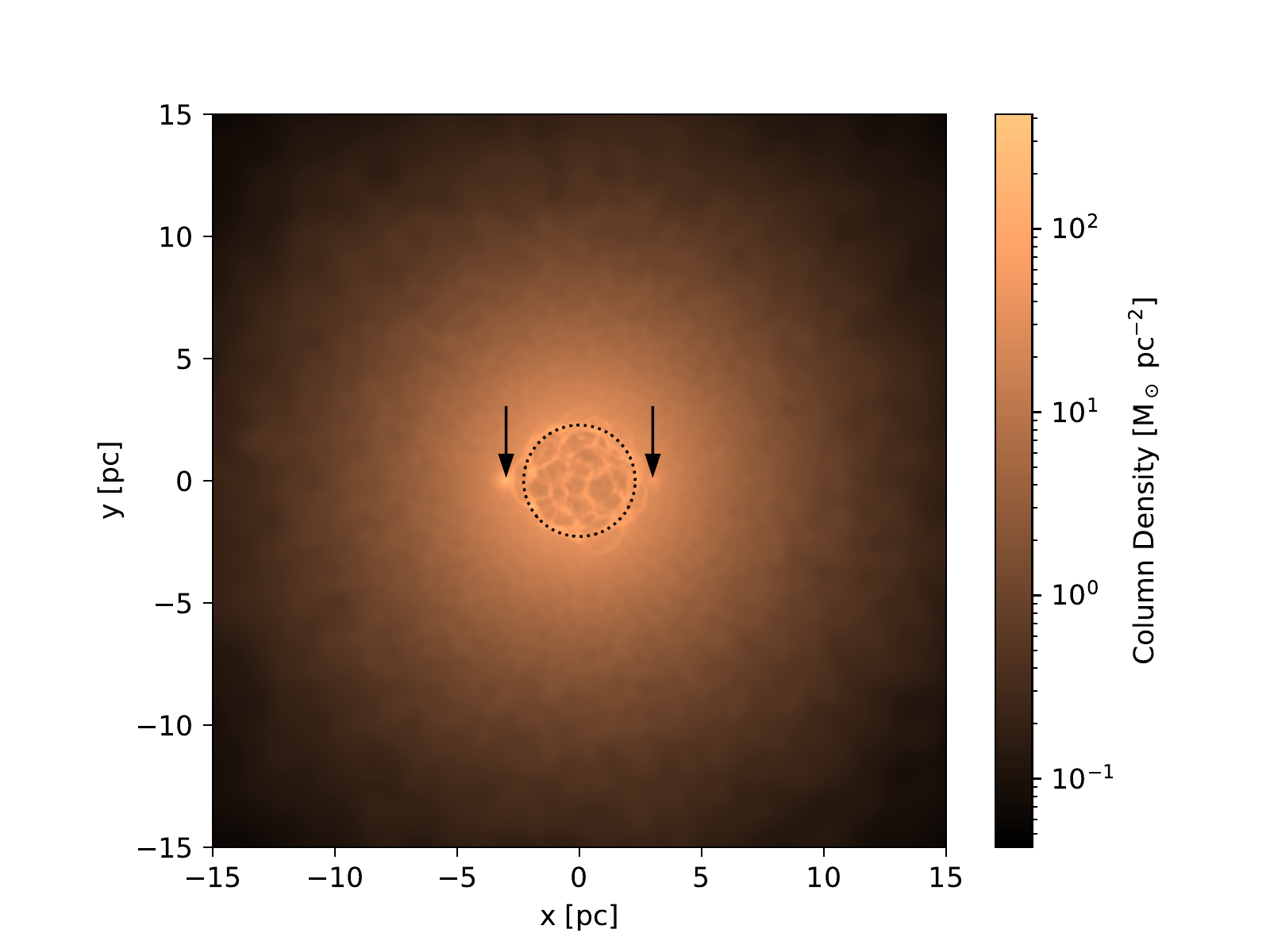}
\includegraphics[width=0.93\linewidth]{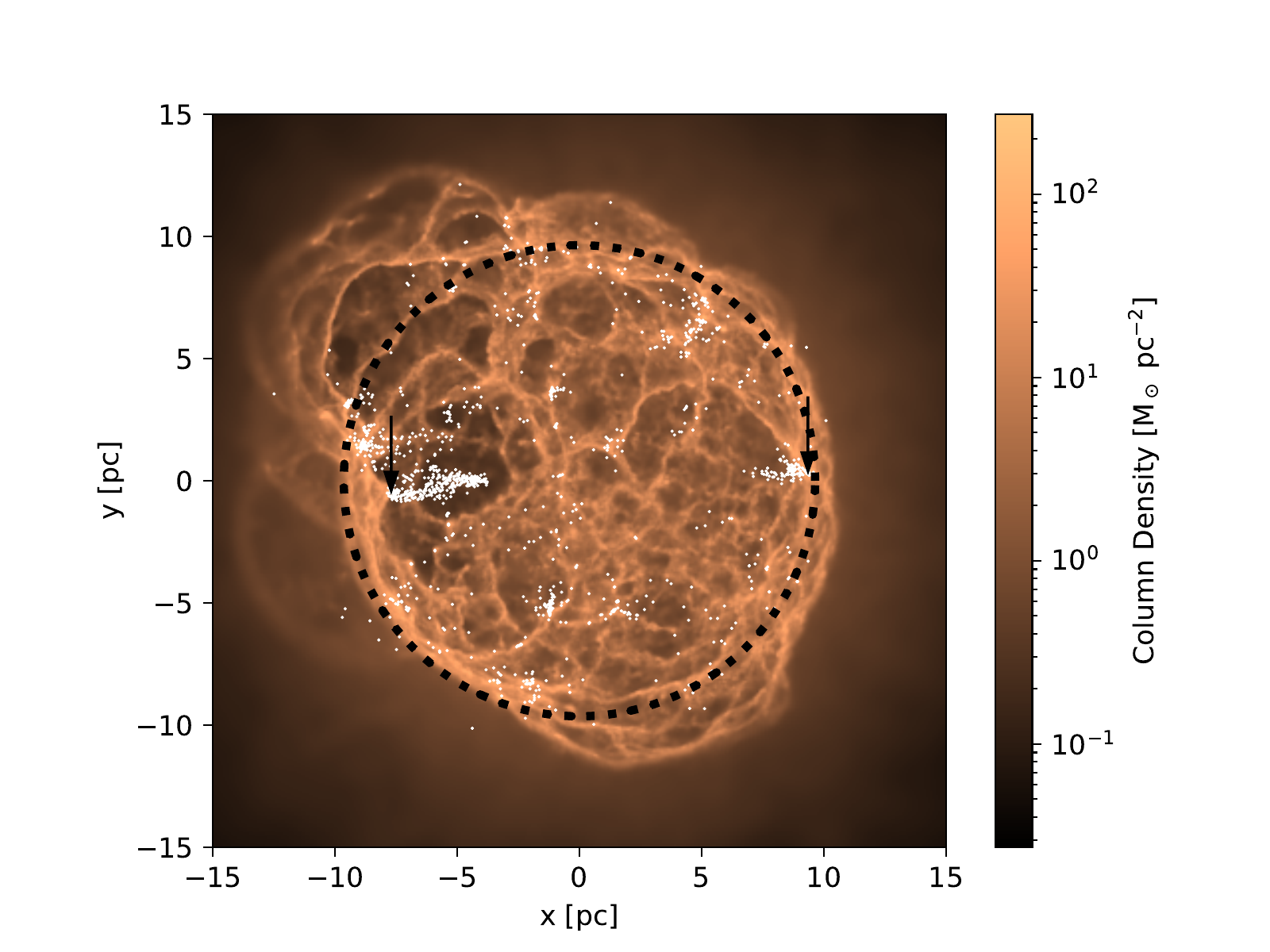}
\caption{ Gas column density maps of the numerical simulation at 0.01 Myr (top), 0.12 Myr (centre), and 0.30 Myr (bottom). The top panel also shows the projected positions of the stars more massive than 16 M$_\odot$ (black points) and those less massive than 16 M$_\odot$ (white points). The bottom panel shows the projected positions of the sink particles as white points. The arrows point out the 50 M$_\odot$ (left) and 10 M$_\odot$ (right) cloudlets. The centre and bottom panels show the position of the shell as a black dotted circle. }
\label{fig:hydro_maps} 
\end{figure}

Our simulation does not include an ambient ISM, which typically dampens the expansion of a stellar wind bubble as more ISM is swept up in the bubble's outer shell \citep{wea77}. Without this damping, the shell reached a terminal velocity of $\sim$60 km/s around 0.3 Myr. 

Our multiphysics model converts gas particles into sink particles when they reach a density of $3.3\cdot 10^6$ amu cm$^{-3}$ to prevent numerical errors. These sink particles interact with the gas component only through gravity, not through any hydrodynamical interaction. In the bottom panel of Fig.~\ref{fig:hydro_maps}, we show all sink particles as white points. A number of sink particles form radially directed lines, with the most prominent trailing the 50 M$_\odot$ globule. This is reminiscent of the pillars observed at the interface between some \ion{H}{ii} regions and molecular shells, which \citet{tre12a} propose form when the curvature of the shell becomes high and the shell collapses in on itself. The globules we inserted into the initial conditions are natural places for such a curvature to develop. As the pillar collapses our model converts it into sink particles. In contrast with a sink particle, a real pillar is able to interact hydrodynamically with its surroundings, but the fact that they lag behind the shell implies that they decouple from the processes that drive the expansion and interact with their surroundings in a similar way to sink particles.

\citet{tre12b,tre13} also discuss another structure observed near such shells, namely dense globules just within the \ion{H}{ii} region. These structures form from interactions with turbulent material outside the shell. Because our simulations do not include turbulent gas these structures do not form.

Because stellar wind bubbles expand into the ISM supersonically, the ISM beyond a certain shell has not influenced its evolution up to that point. As a result, we could use the state of the simulation at any moment as the initial conditions of a new simulation with an arbitrary ISM distribution outside the shell. Re-simulating a large number of simulation snapshots with additional ISM outside the shell would be prohibitively expensive in terms of computer resources, because the entire ISM must be represented by SPH particles. The volume required to represent S 171 and to reduce edge effects (due to the absence of SPH particles beyond the simulation box) would be so great that the number of SPH particles required would be much greater than that of the shell.

\subsection{Simplified numerical wind bubble model}

\subsubsection{Model}

Instead, we turn to a simplified (and computationally cheaper) numerical approach to a similar problem, which we can apply to the state of our simulation at different times. In our simulation, stellar winds quickly form a mostly spherical bubble in the gas cloud that is much larger than the collection of massive stars. We can approximate this as a wind bubble blown by a star with the stellar wind output of all massive stars combined (although losses in wind shocks between stars can decrease the effective input of energy and momentum into the shell). 

The classic solution to this problem is the similarity solution of \citet{wea77}, which describes the spherical expansion of a stellar wind bubble in a uniform ISM. More recently, \citet{lan21a} presented a modified solution taking into account the fractal structure of the bubble's boundary and the subsequent efficient radiative losses. We cannot use their solutions directly because our shell consists of a swept-up Plummer sphere, not a swept-up uniform ISM, but we can use the differential equations to which they are a solution. 

Both models are based on the conservation of momentum, equating the momentum of the shell to that imparted by stellar winds. Although they differ in the prescription of how momentum is transferred to the shell, they are both of the following form:

\begin{equation}
\frac{d}{dt}\left( M_s\left(R_s\right)\dot{R}_s \right) = \eta\dot{p}_w,
\end{equation}where $M_s\left(R_s\right)$ is the mass of the shell once it has reached radius $R_s$, $\dot{p}_w$ is the shell's momentum input rate due to the stellar winds, and $\eta$ is an efficiency factor. \citet{lan21a} introduce two different factors, $\alpha_p$ and $\alpha_R$, that account for enhanced momentum transfer and non-spherical expansion, respectively. Both of these factors are variable but of order unity \citep[as demonstrated by][]{lan21b}. For simplicity, we combine these into one value, which can additionally account for losses in momentum transfer due to stellar wind shocks between stars. 

In a uniform ISM of density $\rho_0$, the mass of the shell is simply $M_s=\frac{4\pi}{3}R_s^3\rho_0$, resulting in a neat power law solution. In the case of our model, the density distribution follows a Plummer sphere interior to the initial shell radius, $R_{s,0}$. The density distribution outside this radius is undefined as of yet. For the sake of simplicity we assume a uniform ambient ISM density distribution. The shell mass is then

\begin{equation}
M_s\left(R_s\right) = \begin{cases}
M_P\left(<R_s\right) \quad\quad\quad\quad\quad\quad\quad\quad\quad \textrm{ if } R_s < R_{s,0} \\
M_P\left(<R_{s,0}\right) + \frac{4\pi}{3}\left( R_s^3 - R_{s,0}^3 \right)\rho_0 \textrm{ if } R_s > R_{s,0},
\end{cases}
\end{equation}where $M_P\left(<R\right)$ is the mass of a Plummer sphere within a radius $R$ and the ambient ISM density is $\rho_0$. 

The resulting differential equations lack neat power law solutions, and require numerical integration. We use fourth-order Runge-Kutta integration for this, with time steps equal to $0.01\cdot\textrm{min}\left(x_i/\dot{x}_i\right)$, where $x_i$ denotes a vector of all integrated variables (e.g. shell radius and velocity). 

Given a simulation snapshot, we can now continue the evolution of the wind bubble using this method. As initial conditions, we need the shell's radius and velocity, and in the case of the Weaver model, the internal energy contained within the shell radius. We define the shell's radius as the location of the peak of the histogram of all gas particle's distance from the origin (and the shell's velocity in a similar way). The mechanical luminosity of the central wind source is fixed by the stellar population. Finally, we set the ISM density exterior to the shell to be equal to the density of the main Plummer sphere at the shell radius. This ensures continuity in the density distribution, and uniquely relates each shell radius to an ISM density.

If the evolution of the bubble using the simplified numerical model reaches the observed shell radius and velocity at the same moment in time, the combined model is consistent with the observations. This moment in time is then our estimate of the system age, and the density of the main Plummer sphere at the initial shell radius is our estimate of the ambient ISM density. Because we have a limited set of initial conditions, the shell radius and velocity will never match the observed values at the exact same moments. However, for every set of initial conditions (each corresponding to an ISM density), we can record the times at which the observed shell radius and velocity are reached, and find the intercept of these two curves.

\subsubsection{Results}

Figure~\ref{fig:numerical_evolution} shows the evolution through time of the shell radius (orange) and velocity (blue) following the \citet{wea77} (solid) and \citet{lan21a} (dashed) models. We note that all times quoted in this section are with respect to the start of the multiphysics simulation. For comparison, we have also plotted the analytic solutions of the Weaver (dotted) and Lancaster (dash-dotted) models using the same stellar wind output and ISM density. The initial conditions of the numerical solutions are offset from these solutions, but the numerical solutions do converge to the analytic solutions. This is more obvious for the Weaver solution than the Lancaster solution; in the latter case, the shell radius is initially too large, and the velocity must be below the analytic shell velocity so the radius can eventually converge to the analytic shell radius. Further evolution of the system does show that the numerical solution converges to the analytic solution.

The shaded regions denote the radius and velocity of the massive, moderate velocity shell component (25 pc and 4 km/s, respectively) with 10\% errors. For the initial conditions used for this plot, the combined multiphysics and Lancaster models agree with the observations because both shell radius and velocity cross the observed values at the same moment, at about 2 Myr. Notably, the velocity of the Weaver solution is much greater than observed, and does not reach the observed value even at 10 Myr. For this instance of the simplified model, we used the state of the multiphysics model at 0.23 Myr, when the shell radius was 6.5 pc, the shell velocity 41 km/s, and the ISM density at the shell edge was 19.4 amu cm$^{-3}$.

\begin{figure}[t]
\centering
\includegraphics[width=\linewidth]{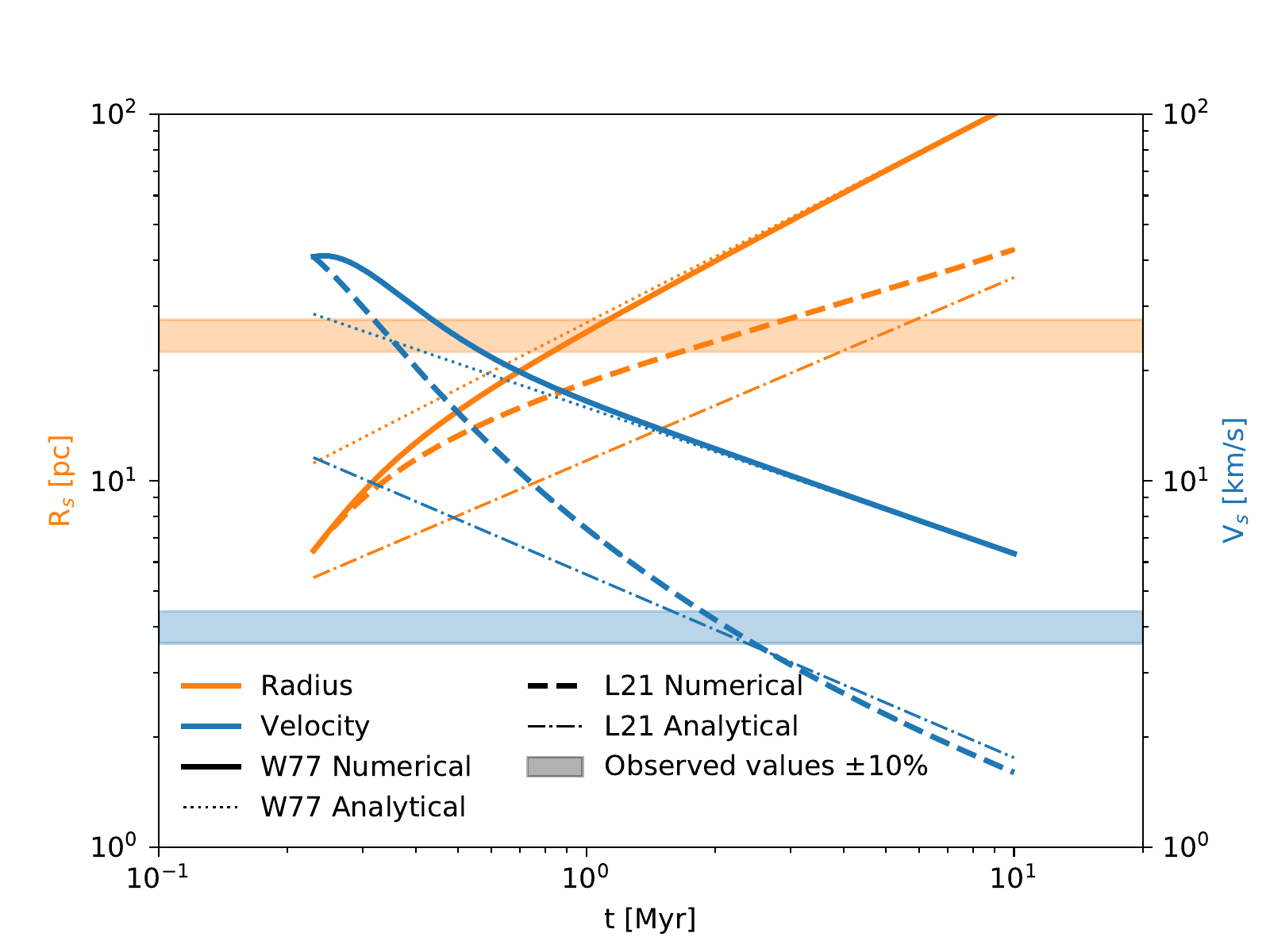}
\caption{ Evolution through time of the stellar wind bubble radius (orange) and velocity (blue) using the \citet{wea77} (solid) and \citet{lan21a} (dashed) models, starting from the bubble's radius and velocity at 0.23 Myr (with a corresponding density at the shell of 16.0 amu cm$^{-3}$). The efficiency $\alpha$ parameters in the Lancaster model have both been set to 1. The dotted and dash-dotted lines denote the analytical Weaver and Lancaster solutions, respectively. The shaded regions show 25 pc and 4 km/s, with 10\% errors. }
\label{fig:numerical_evolution} 
\end{figure}

Fig.~\ref{fig:density_age} shows, for a number of simulation snapshots, the density of a Plummer sphere at the shell radius and the moments that the numerically integrated (using the Lancaster model for the top panel, and the Weaver model for the bottom panel) shell radius and velocity reach values of 25 pc and 4 km/s (corresponding to the massive, moderate velocity shell). The shaded regions denote the range of values allowed by the efficiencies shown and by letting the observed radius and velocity vary by 10\%. We point out that the two panels cover different ranges of efficiency $\eta$. This is expected to be of order unity, but in the case of the Weaver model there were no solutions with $\eta=1$ or $\eta=3$ across the available range of densities. The radius and velocity curves did approach each other at high densities, but did not cross at the greatest density possible, which corresponds to the density of the Plummer sphere's core (661 atomic mass units (amu) cm$^{-3}$). 

\begin{figure}[t]
\centering
\includegraphics[width=\linewidth]{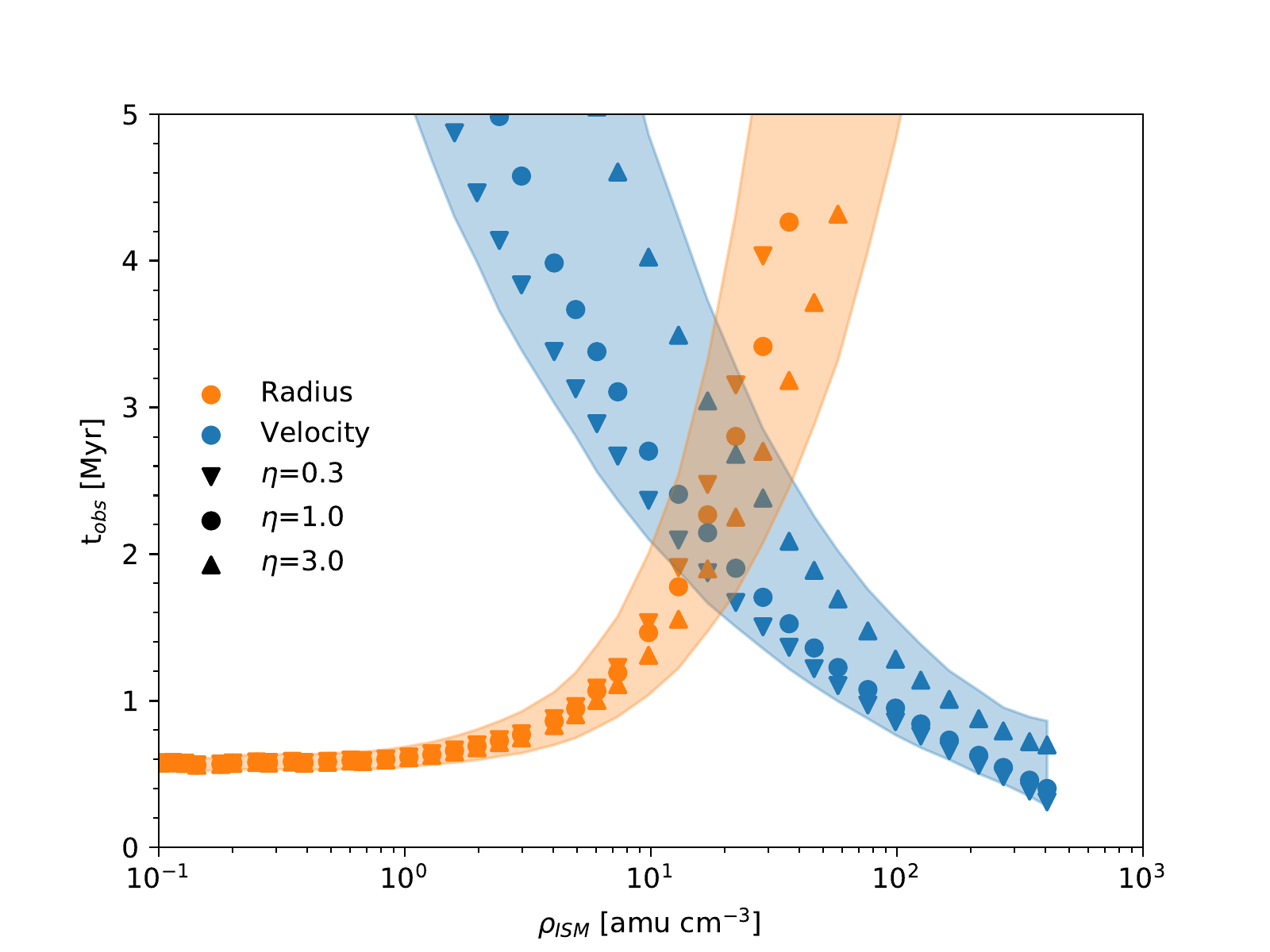}
\includegraphics[width=\linewidth]{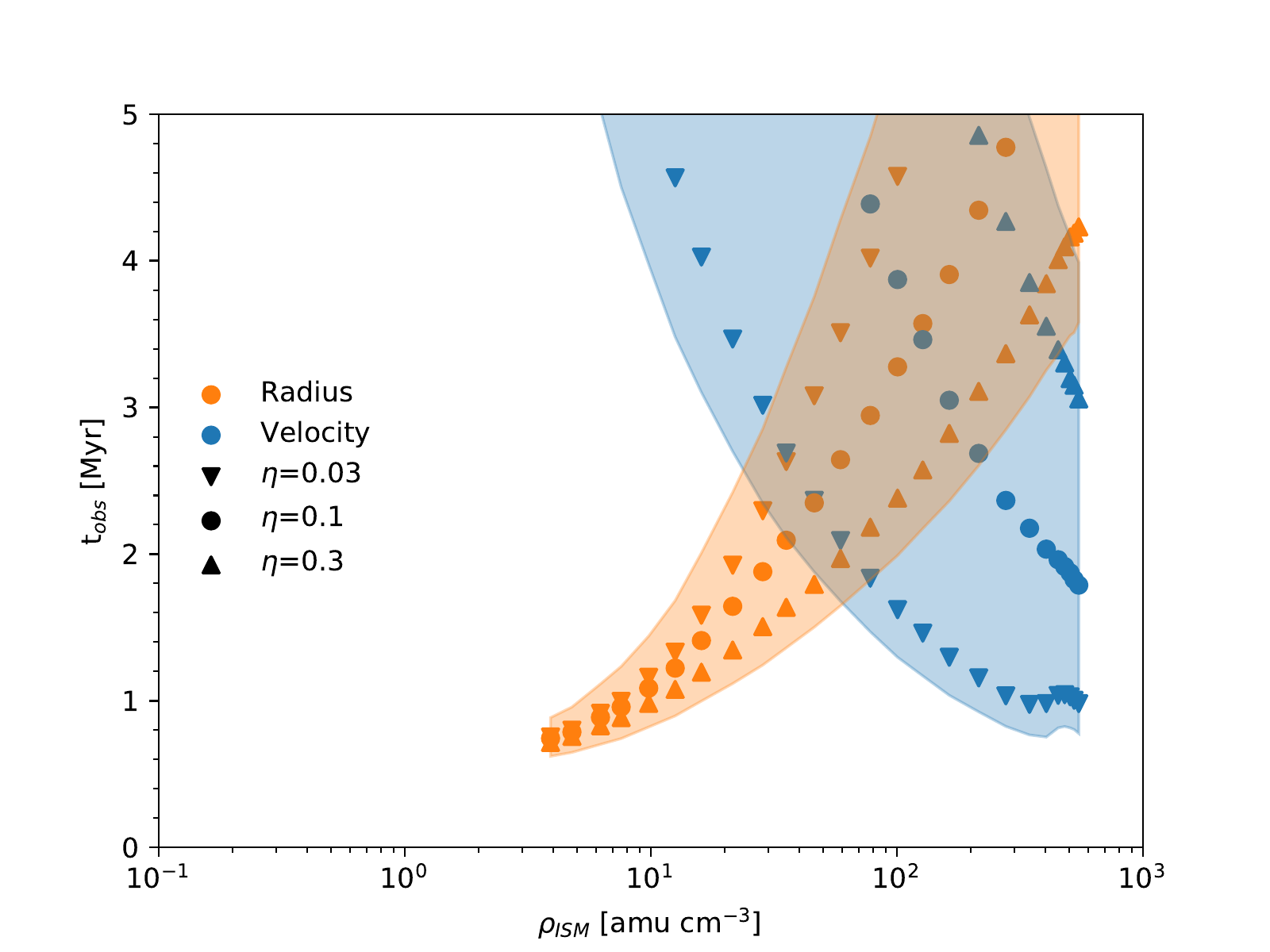}
\caption{ For a number of snapshots of the simulation, the density of a Plummer sphere at the shell position (assumed to be the ISM density) versus the moment at which in the numerical integration a shell velocity of 4 km/s and a radius of 25 pc is reached. Different symbols denote different stellar wind efficiencies $\eta$. Shaded regions denote the envelope allowed for all efficiencies plotted and a 10\% error in the observed velocity and radius. The top panel shows the results obtained using the Lancaster model, and the bottom panel shows the results from the Weaver model. }
\label{fig:density_age} 
\end{figure}

From the Lancaster model, the ISM density inferred is in the range of $10-30$ amu cm$^{-3}$, slightly below the typical density of molecular clouds. The age is in the range of $2-2.5$ Myr, similar the age assumed for the system.

From the Weaver model, the ISM density inferred is in the range $30-300$ amu cm$^{-3}$, of the order of a typical molecular cloud. The age is in the range $2.5-4$ Myr, slightly older than estimated earlier. It should be noted that the lower ISM density and age correspond to a very low stellar wind efficiency, one that we have not explored with the Lancaster model. 

We do not expect such a low efficiency to be plausible; \citet{lan21b} only finds an efficiency of similarly low order of magnitude for a very dense, very low star formation efficiency cloud. In their work, they use the cloud mass and star formation efficiency to compute an effective stellar wind mechanical luminosity. Inverting this, with our actual cloud mass and mechanical luminosity, we get an effective star formation efficiency of $\sim$50\%, which implies a stellar wind efficiency close to unity. 

For comparison, we also present the results of matching the observed radii and velocities to the analytic solutions of Weaver and Lancaster. These results are in Table~\ref{tab:analyticsolutions}. We match both the massive, moderate velocity shell and the high-velocity cloudlets. This disregards the presence of leftover cluster material, but is independent of potential numerical errors in our simulation.

Fitting the Weaver model to the moderate velocity shell results in a relatively old age and a very dense ISM (about twice the density of the main Plummer sphere in our simulation), unless $\eta$ is very low. Fitting the Lancaster model, on the other hand, leads to a somewhat younger system age (though still much older than estimated for the system), and an ISM more diffuse than a GMC. For completeness, we also included the ages and densities resulting from fitting to the high-velocity cloudlets, although it is difficult to explain them as a shell when the moderate velocity shell is much more massive and closer in. 

\begin{table}
\centering 
\caption{System age ($t_{\textrm{obs}}$) and ISM density ($\rho_{\textrm{ISM}}$) estimates by solving the Weaver and Lancaster models for the observed radii and velocities of the shell (25 pc and 4 km/s, respectively) and cloudlets (31 pc and 12 km/s, respectively). These values were computed assuming $\eta=1$; different values for $\eta$ do not modify the age but do modify the density linearly. }
\begin{tabular}{lcc} 
\hline
\noalign{\smallskip}
Model & $t_{\textrm{obs}}$ (Myr) & $\rho_{\textrm{ISM}}$ (amu cm$^{-3}$) \\
\hline
Weaver & 3.67 & $1.23\cdot 10^3$ \\
Shell & & \\
\hline
Lancaster & 3.06 & 6.74 \\
Shell & & \\
\hline
Weaver & 1.52 & $2.96\cdot 10^1$ \\
Cloudlets & & \\
\hline
Lancaster & 1.26 & $4.87\cdot 10^{-1}$ \\
Cloudlets & & \\
\hline 
  
\end{tabular} 
\label{tab:analyticsolutions} 
\end{table}

We did not observe a clear velocity bimodality in the simulation. The amount of material with velocities greater than 2.5 times that of the shell itself (which would be an analogue to the observed high-velocity structures) is only $\sim$0.01\% of the total shell material, or $\sim$0.3 M$_\odot$, less massive than the individual high-velocity cloudlets. Additionally, as can be seen in Fig.~\ref{fig:hydro_maps}, the cloudlets that were detached from the main cloud at initialisation typically lagged behind the shell and have similar velocities to the shell itself, and so did not end up as a high-velocity component.

\subsection{Simplified numerical cloudlet model}

\subsubsection{Model}

\citet{lan21b} show that stellar wind bubbles blown in a uniform but turbulent medium are fractal shaped rather than uniform spherical shells. Even in our smooth\footnote{Up to the inherent discretisation in SPH particles.} and virialised Plummer sphere asymmetries arise, from secondary bubbles where the winds break through the shell into low-density material (shown in the top left and bottom right of the bottom frame of Fig.~\ref{fig:hydro_maps}) to substructures in the shell itself. 

In the models above, the fundamental mechanism slowing down the expanding bubble is the accumulation of ISM material in the shell. The magnitude of this deceleration is related to the ratio of the masses of the shell and of the swept-up ISM. In a non-uniform shell, different sections will have different ratios, and as a result, different rates of deceleration. Viewed differently, a denser section of the shell will have more inertia and be able to maintain its velocity for longer than a less dense section. In this section we investigate whether this varying deceleration can result in the multiple components observed in S 171.

We derive the equation of motion of a section of a shell, or cloudlet (which we use from now on), through a uniform medium, using a similar method to \citet{lan21a}, that is, exploiting conservation of momentum.

This momentum equation is of the following form:

\begin{equation}
\begin{split}
& \frac{d}{dt} \left( \left( M_{c,0} + \rho_0 \int_{t_0}^{t} \sigma_c\left(t'\right) \dot{R}_c\left(t'\right) dt' \right) \dot{R}_c\left(t\right) \right) \\
& = \eta\dot{p}_w \frac{\sigma_c\left(t\right)}{4\pi R_c\left(t\right)^2},
\end{split}
\end{equation}where $M_c$ and $R_c$ are the cloudlet mass and distance to the cloud centre, respectively, and $\sigma_c$ is the cloudlet's cross-section as seen from the cloud centre. This is both the surface being struck by stellar winds, and the surface accumulating ISM material. Also, in this expression, and all hereafter, a subscript 0 denotes the initial value of a quantity.

The evolution of $\sigma_c$ is dictated by hydrodynamical processes, but for simplicity we consider two interesting edge cases: constant cross-section (corresponding to a bound, non-expanding cloudlet), and quadratically growing with distance (corresponding to a shell structure growing with the shell). These have the respective forms

\begin{equation}
\sigma_c\left(t\right) = \sigma_{c,0},
\end{equation}

\begin{equation}
\sigma_c\left(t\right) = \sigma_{c,0} \left(\frac{R_c\left(t\right)}{R_{c,0}}\right)^2.
\end{equation}

They reduce to the following differential equations in $R_c$, for the non-expanding and expanding cases, respectively

\begin{equation}
\ddot{R}_c = \left(\frac{\eta\dot{p}_w}{4\pi R_c^2} - \rho_0\dot{R}_c^2\right)/\left(\frac{M_{c,0}}{\sigma_{c,0}} + \rho_0\left(R_c-R_{c,0}\right)\right)
\end{equation}

\begin{equation}
\begin{split}
\ddot{R}_c = & \left(\frac{\eta\dot{p}_w}{4\pi R_{c,0}^2} - \rho_0\dot{R}_c^2\left(\frac{R_c}{R_{c,0}}\right)^2\right)/ \\
& \left(\frac{M_{c,0}}{\sigma_{c,0}} + \frac{\rho_0}{3R_{c,0}^2}\left(R_c^3-R_{c,0}^3\right)\right).
\end{split}
\end{equation}

We note that the cloudlet's initial mass and cross-section only appear as the ratio $\frac{M_{c,0}}{\sigma_{c,0}}$. This implies that a cloudlet's evolution is determined by its initial average (mass) surface density, and its actual dimensions are not important. 

Another point to note is that the expanding case is equivalent to the Lancaster model when the initial average surface density of the cloudlet is the same as that of the shell.

\subsubsection{Results}

Figure~\ref{fig:globule_propagation} shows the distance and velocity of cloudlets of varying surface density, using the differential equations described above. The initial conditions for each efficiency were taken from the snapshots best coinciding with the observations of the shell after continued integration, and evolved until the moment the shell coincided with the observations. 

\begin{figure}[t]
\centering
\includegraphics[width=\linewidth]{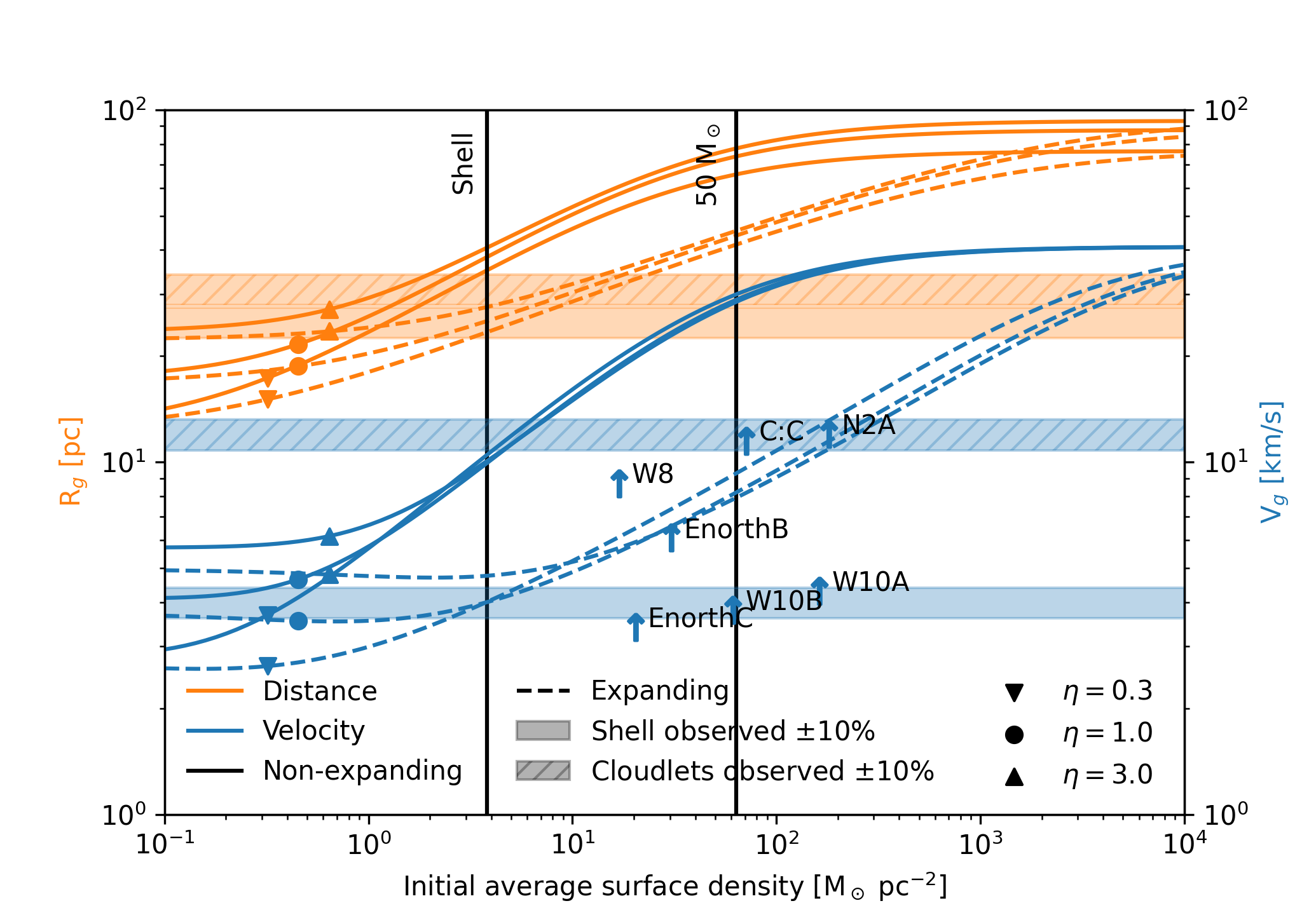}
\caption{ Distance to the centre of the complex (orange) and velocity (blue) of cloudlets as a function of initial average surface density, after evolution to 1.9, 2.17, and 2.3 Myr (for $\eta=0.3$, 1, and 3, respectively). Solid lines represent cloudlets that do not expand as they travel, i.e. bound structures. Dashed lines represent cloudlets that expand with the shell, with their cross-sections growing with the distance squared, i.e. features on the shell that grow with it. Symbols overplotted on these lines indicate the stellar wind efficiency $\eta$. The hatched shaded regions denote 31 pc and 12 km/s, the radius and velocity of the high velocity component of S 171, and the clear shaded regions denote 25 pc and 4 km/s, the radius and velocity of the moderate velocity component. All observed values are shown with 10\% errors. The surface densities and line-of-sight velocities (with respect to the cluster centre) of the cloudlets in Table \ref{tab:cloudlets} are shown as upward blue arrows. This velocity is a lower limit of their expansion velocity. The initial conditions were taken from the simulation snapshot at 0.22, 0.23, and 0.25 Myr (again for $\eta=0.3$, 1, and 3, respectively), which resulted in a bubble radius of 25 pc and a velocity of 4 km/s at the respective end time using the Lancaster model. The two vertical black lines denote the average surface density of the shell at 0.23 Myr (left) and the surface density at the core of a 50 M$_\odot$ (and 0.5 pc) Plummer sphere cloudlet.}
\label{fig:globule_propagation} 
\end{figure}

Non-expanding cloudlets always reach farther distances and retain higher velocities than expanding cloudlets because they sweep up less of the ISM. This effect is most pronounced near the centre of the range of densities explored, where non-expanding cloudlets can reach distances about twice as far as expanding cloudlets and velocities about four times as fast. Compared to the average shell, non-expanding cloudlets with high surface densities can reach distances almost four times the shell radius and velocities about ten times as great. 

We note that there is no density for which the distance and velocity match the observed high-velocity cloudlets. The configuration closest to matching the observations is a non-expanding cloudlet of about the same initial average surface density as the shell. However, this cloudlet's distance is more than 10\% greater than the observed distance, and its velocity is more than 10\% smaller than observed. Allowing a cloudlet to expand more slowly than quadratic would put it in between the expanding and non-expanding cases; this would alleviate disagreement with the distance, but increase disagreement with the velocity. Allowing the stellar wind efficiency to vary cannot alleviate this disagreement either, for similar reasons. However, assuming a slightly younger system age (at which the radius was smaller, and the velocity greater) could put the cloudlets in the observed range.

We also show the high-velocity cloudlets. We derive their average surface densities from Table \ref{tab:cloudlets} using a distance of 1.1 kpc, and correct the line-of-sight velocity for that of the cluster itself (-9.5 km/s). This is a lower limit on their expansion velocity, which can also have a component in the plane of the sky. In our spherical expansion scenario, the line-of-sight velocity is closer to the expansion velocity the closer a cloudlet is to the cluster centre. The cloudlets C:C and N2A are very close to the centre, while EnorthB, EnorthC, W10A, and W10B are nearly at the 30 pc outer circle of the projected shell. In terms of the surface density, the cloudlets are in the range of 20-200 M$_\odot$ pc$^{-2}$. This combination of velocity and surface density puts the cloudlets closer to the expanding limit than the non-expanding limit.

\subsection{Summary}

To summarise, through a hybrid numerical modelling approach we demonstrate that stellar winds from the massive stars of Be 59 can give rise to the complex gas structure of S 171. We used a coupled hydrodynamical, gravitational, and stellar evolutionary model to simulate the initial formation of a stellar wind bubble in the concentrated gas distribution left over from the formation of a star cluster. We then used simplified numerical models to continue the evolution of the wind bubble through uniform ISM of different densities in an effort to match the observed radius and velocity of the massive shell of S 171, and so provide an estimate of the system age and ambient ISM density. Finally, we introduced a new simplified numerical model to describe the dynamic evolution of shell substructures through a uniform ISM, and used it to show that dense and detached shell fragments are able to reach farther distances and higher velocities than the main shell. 

We show that using the recent stellar wind bubble model by \citet{lan21a} to continue the bubble evolution results in more reasonable values for the system age and ambient ISM density compared to the classical model of \citet{wea77}, assuming reasonable values for the efficiency of momentum transfer. As derived from the Lancaster model, the system age is consistent with earlier estimates, and the ambient ISM density is reasonable for the larger surroundings of a star forming regions. The Weaver model requires very inefficient momentum transfer from stellar winds to the shell to reproduce the moderate velocity shell, and then results in a system age older than estimated earlier and a very dense ambient ISM. 

We also show that variations in density within the shell can lead to components propagating through the ISM at different velocities. We are unable to reproduce the observed ratios between the distances and velocities of the observed structures, but the case of non-expanding cloudlets with initial surface density similar to the shell comes closest. These variations can arise naturally from fragmentation of the shell, as is demonstrated in our multiphysics simulation. If the shell is only able to partially fragment into small cloudlets we would naturally obtain high-velocity, low-mass cloudlets and a moderate-velocity, high-mass shell.

Our results qualitatively show that stellar winds can reproduce the gas structure of S 171, but we refrain from drawing strong quantitative conclusions. The SPH method of modelling hydrodynamics is infamously bad at modelling shocks, being unable to properly resolve thin shock layers. This makes our estimates of the shell radius and velocity that serve as the initial conditions of the simplified numerical models uncertain. Notably, the shell velocity is greater than expected from the Weaver and Lancaster models, even though the shell mass is greater due to the swept-up Plummer sphere. Although our model slightly favours the Lancaster model over the Weaver model, we also refrain from drawing conclusions about this.

A fully consistent reproduction of the system would require a grid based hydrodynamical model \citep[such as the one used by][]{lan21b} or a hybrid method (such as moving mesh, or meshless finite mass). Models that start from the collapse of a giant molecular cloud \citep[such as by][]{Wall20,gru21} can improve the authenticity of the gas distribution at the moment the bubble begins being blown. A full hydrodynamical simulation, including an ambient ISM, can also self-consistently demonstrate the origin of the dense cloudlets that end up at high velocity.

\section{Conclusions}
\label{sec:conclusions}

We have investigated the kinematics of an expanding \ion{H}{ii} region powered by a young stellar cluster for which the inferred expansion velocity appears to be higher than predicted in theoretical model simulations. However, published estimates of the central velocity of this complex can be highly divergent, and more detailed information on the motion of the surrounding shell is warranted. In this paper we draw attention to the nebula Sharpless 171 (with NGC 7822), which surrounds the cluster Berkeley 59. Optical spectra of 27 stars were obtained in the cluster area, providing a new estimate of the RV of the cluster. The velocity pattern over the molecular shell was mapped in $^{13}$CO(1-0) at OSO. From these data we deduce the expansion velocities of different cloud fragments in the shell relative to the cluster. As described below, we could define three systems of such clouds expanding at different rates. One system expands at a high rate, although not as high as one could expect from some of the published estimates of the central velocity. 
 
 We assigned a spectral class to each star, including the separate components in double-lined spectroscopic binaries. The majority of the stars had not been classified before, including one central O star hidden behind foreground dust.
   
 From spectral classifications and RVs, plus existing photometric data and Gaia parallaxes, we selected 19 of the 27 stars as likely members of Be~59.  
 
Repeated observations were done, especially for members, which revealed that five of the massive stars are spectroscopic binaries, of which one is double-lined and one triple-lined. We followed these stars over longer times and derived periods, semi-amplitudes, and systemic velocities. The most luminous O star is the eclipsing binary V747 Cep, and our RV curve agrees with the photometric period. 

From the RVs of members we obtained a mean heliocentric velocity of -- 18.75 km~s$^{-1}$ (-- 9.5 km~s$^{-1}$ in lsr), which we define as the central RV of the complex. None of the members deviate significantly from the mean value.

The S 171 complex extends over 3.2$\degr$ in the sky, and for our $^{13}$CO mapping we primarily covered areas with distinct dark clouds. The velocity pattern over the nebula is complex, and in most directions several velocity components are present in the spectra. In addition to components that originate in gas not related to the complex, we identified two systems of shell structures expanding at different velocities. These components can be described as located at the peripheries of spherically expanding shells. One system, composed primarily of rather massive shell fragments, expands at a moderate velocity, 4~km~s$^{-1}$, and includes two velocity components that are related to the front and back sides of the complex. This system is confined within a radius of 25 pc. The second system expands at a much higher velocity, 12~km~s$^{-1}$, and is concentrated in a belt that crosses the entire \ion{H}{ii} region, extending to a radius $\geq$ 30 pc. In addition, we find some cloudlets in the central area with expansion velocities in between those of these systems. 

The luminosity and UV flux is dominated by the central O stars. Our interpretation of the observed complex kinematic structure is based on the assumption that the expansion started about 2 million yrs ago, driven by radiation and winds from the central massive stars. Our model simulations show that cloudlets of smaller masses obtained very high initial velocities, while more massive shell fragments dragged behind.

The velocity and radius of the moderate velocity shell are generally consistent with theoretical predictions for the expansion of a stellar wind bubble driven by the cluster's population of massive stars.

The cloudlets observed with higher velocity and at greater distance from the massive shell can arise through interaction with the ambient ISM, which decelerates dense and bound substructures of the shell less efficiently than low-density substructures that co-expand with the shell.

Although our models are able to qualitatively reproduce the structure of S 171 for plausible parameters, we refrain from drawing quantitative conclusions about the value of these parameters. This requires further work using more accurate methods.

\begin{acknowledgements}

We thank the anonymous referee for helpful comments that led to an improvement of the paper.
This work was supported by the Magnus Bergvall Foundation and the L\"angmanska Kulturfonden in Sweden. We made use of the SIMBAD database operated at Centre de Donnees Astronomique de Strasbourg and data from the European Space Agency (ESA) mission {\it Gaia} (https://www.cosmos.esa.int/gaia), processed by the {\it Gaia} Data Processing and Analysis Consortium (DPAC, https://www.cosmos.esa.int/web/gaia/dpac/consortium). Funding for the DPAC has been provided by national institutions, in particular the institutions participating in the {\it Gaia} Multilateral Agreement. This publication made use of data products from the Wide-field Infrared Survey Explorer, which is a joint project of the University of California, Los Angeles, and the Jet Propulsion Laboratory/California Institute of Technology, funded by the National Aeronautics and Space Administration. We also used a map obtained with Planck (http://www.esa.int/Planck), an ESA science mission with instruments and contributions directly funded by ESA Member States, NASA, and Canada.
We made use of the Binary Star Solver v1.2.0 python program \citep{mil20}.

In this work we use the matplotlib \citep{hun07}, NumPy \citep{oli06}, AMUSE \citep{por18}, SeBa \citep{por12}, Huayno \citep{pel12}, and Fi \citep{pel04} packages.

We thank John Telting for very useful help, comments and suggestions related to FIES, and in addition we thank the following observers at the NOT for contributing to obtaining the FIES data set: Pere Blay Serrano, Rosa Clavero, Abel de Burgos, Sune Dyrbye, Grigori Fedorets, Emanuel Gafton, Francisco Jose Galindo-Guil, Nicholas Emborg Jahnsen, Raine Karjalainen, Anni Kasikov, Marcelo Aron Keniger, Jyri Lehtinen, Teet Kuutma, Illa Losada, Julia Martikainen, Shane Moran, Christina Konstantopoulou, Viktoria Pinter, Tapio Pursimo, Rene Tronsgaard Rasmussen, Lauri Siltala, Auni Somero, and Joonas Viuho.

Based on observations made with the Nordic Optical Telescope, owned in collaboration by the University of Turku and Aarhus University, and operated jointly by Aarhus University, the University of Turku and the University of Oslo, representing Denmark, Finland and Norway, the University of Iceland and Stockholm University at the Observatorio del Roque de los Muchachos, La Palma, Spain, of the Instituto de Astrofisica de Canarias.

The authors would like to state that this manuscript was submitted before Christmas.

\end{acknowledgements}

\bibliographystyle{aa}
\bibliography{biblio}

\begin{appendix}

\section{Individual stellar radial velocities}
\label{AppendixA1}

\vspace{1cm}

\begin{table}
  \caption{Individual RV measurements  at multiple epochs for each star, and where possible for each stellar component.
    The second column gives the heliocentric Julian date of the measurement.}
  \label{tab:RVstars1}
  %Added by TeX Support 
  % Table rv_table.tab  Fri 13:47:59 19-Nov-2021
% 1st part
%\documentstyle{article}
%\begin{document}
\newcommand\cola {\null}
\newcommand\colb {&}
\newcommand\colc {&}
\newcommand\cold {&}
\newcommand\eol{\\}
\newcommand\extline{&&&\eol}

\begin{tabular}{lrrr}
\cola ID\colb HJD\colc V$_{\rm hel}$\cold $\sigma_v$\eol
\cola \colb days\colc km/s\cold km/s\eol
\extline
\cola V747Cep\colb   2458830.473741\colc   34.1\cold  9.8\eol
\cola V747Cep\colb   2458835.519666\colc   42.5\cold  7.8\eol
\cola V747Cep\colb   2458852.312770\colc  -10.2\cold  4.6\eol
\cola V747Cep\colb   2458864.371491\colc -131.4\cold  3.1\eol
\cola V747Cep\colb   2458869.355782\colc -117.9\cold  4.4\eol
\cola V747Cep\colb   2458876.422214\colc  -16.3\cold  9.0\eol
\cola V747Cep\colb   2458885.395874\colc -122.2\cold  4.6\eol
\cola V747Cep\colb   2459037.695436\colc   50.4\cold  5.9\eol
\cola V747Cep\colb   2459049.582579\colc    3.9\cold  4.8\eol
\cola V747Cep\colb   2459233.458153\colc  -21.6\cold  8.2\eol
\cola V747Cep\colb   2459465.671689\colc  -20.0\cold  4.7\eol
\cola V747Cep\colb   2459502.622502\colc    0.3\cold  4.2\eol
\cola BD+66/1675A\colb   2457321.394955\colc  -10.4\cold  1.3\eol
\cola BD+66/1675A\colb   2457333.634011\colc  -16.1\cold  1.1\eol
\cola BD+66/1675A\colb   2457630.476572\colc  -10.5\cold  1.7\eol
\cola BD+66/1675A\colb   2457631.455812\colc  -12.5\cold  2.4\eol
\cola BD+66/1675A\colb   2457635.464969\colc  -17.0\cold  1.8\eol
\cola BD+66/1675A\colb   2457764.406728\colc   -7.5\cold  2.8\eol
\cola BD+66/1675A\colb   2457771.349602\colc  -10.7\cold  1.3\eol
\cola BD+66/1675A\colb   2457778.363977\colc  -11.8\cold  2.1\eol
\cola BD+66/1675A\colb   2457779.369691\colc  -15.2\cold  1.6\eol
\cola BD+66/1675A\colb   2457973.579881\colc   -3.8\cold  2.4\eol
\cola BD+66/1675A\colb   2457978.617800\colc   -7.9\cold  1.7\eol
\cola BD+66/1675A\colb   2458015.679685\colc  -15.7\cold  1.9\eol
\cola BD+66/1675A\colb   2458026.636645\colc  -11.8\cold  1.4\eol
\cola BD+66/1675B\colb   2457321.394955\colc  122.0\cold 12.0\eol
\cola BD+66/1675B\colb   2457333.634011\colc -170.0\cold  6.0\eol
\cola BD+66/1675B\colb   2457630.476572\colc   29.0\cold  3.0\eol
\cola BD+66/1675B\colb   2457631.455812\colc   51.0\cold 12.0\eol
\cola BD+66/1675B\colb   2457635.464969\colc -144.0\cold  9.0\eol
\cola BD+66/1675B\colb   2457764.406728\colc   89.0\cold 10.0\eol
\cola BD+66/1675B\colb   2457771.349602\colc -102.0\cold 10.0\eol
\cola BD+66/1675B\colb   2457778.363977\colc  -13.0\cold  6.0\eol
\cola BD+66/1675B\colb   2457779.369691\colc -136.0\cold  9.0\eol
\cola BD+66/1675B\colb   2457973.579881\colc   64.0\cold 14.0\eol
\cola BD+66/1675B\colb   2457978.617800\colc  124.0\cold 14.0\eol
\cola BD+66/1675B\colb   2458015.679685\colc -140.0\cold  7.0\eol
\cola BD+66/1675B\colb   2458026.636645\colc  -92.0\cold 17.0\eol
\cola BD+66/1675C\colb   2457321.394955\colc -287.0\cold 11.0\eol
\cola BD+66/1675C\colb   2457333.634011\colc  281.0\cold 14.0\eol
\cola BD+66/1675C\colb   2457631.455812\colc -267.0\cold 16.0\eol
\cola BD+66/1675C\colb   2457635.464969\colc  253.0\cold 15.0\eol
\cola BD+66/1675C\colb   2457764.406728\colc -245.0\cold  5.0\eol
\cola BD+66/1675C\colb   2457771.349602\colc  193.0\cold  5.0\eol
\cola BD+66/1675C\colb   2457779.369691\colc  274.0\cold 15.0\eol
\cola BD+66/1675C\colb   2457973.579881\colc -230.0\cold 41.0\eol
\cola BD+66/1675C\colb   2457978.617800\colc -295.0\cold 15.0\eol
\cola BD+66/1675C\colb   2458015.679685\colc  266.0\cold 14.0\eol
\cola BD+66/1675C\colb   2458026.636645\colc  196.0\cold 16.0\eol
\cola P16\colb   2457315.499880\colc  -28.9\cold  5.6\eol
\cola P16\colb   2457321.435010\colc  -20.4\cold  2.2\eol
\cola P16\colb   2457899.678400\colc  -28.3\cold  3.0\eol
\cola P16\colb   2457977.614250\colc  -25.9\cold  2.1\eol
\cola P16\colb   2457978.643920\colc  -25.7\cold  3.2\eol
\cola P16\colb   2457987.580688\colc  -12.4\cold  3.5\eol
\cola P16\colb   2457992.442982\colc  -31.1\cold  4.0\eol

\end{tabular}

\end{table}

\begin{table}
  %\caption{Table~\ref{tab:RVstars1} continued.}
  {\bf Table~\ref{tab:RVstars1}} continued. \\
  \label{tab:RVstars2}
  %Added by TeX Support 
  % Table rv_table.tab  Fri 13:47:59 19-Nov-2021
% 2nd part
%\documentstyle{article}
%\begin{document}
\newcommand\cola {\null}
\newcommand\colb {&}
\newcommand\colc {&}
\newcommand\cold {&}
\newcommand\eol{\\}
\newcommand\extline{&&&\eol}

\begin{tabular}{lrrr}
\cola ID\colb HJD\colc V$_{\rm hel}$\cold $\sigma_v$\eol
\cola \colb days\colc km/s\cold km/s\eol
\extline
\cola IR-star\colb   2458407.696605\colc  -14.0\cold  1.9\eol
\cola IR-star\colb   2458521.393966\colc  -30.4\cold  2.2\eol
\cola IR-star\colb   2458525.342204\colc   17.7\cold  2.0\eol
\cola IR-star\colb   2458643.710819\colc  -52.5\cold  3.1\eol
\cola IR-star\colb   2458661.625497\colc  -35.9\cold  3.3\eol
\cola IR-star\colb   2458662.618199\colc  -52.6\cold  1.3\eol
\cola IR-star\colb   2458664.604441\colc   -7.7\cold  1.5\eol
\cola IR-star\colb   2458665.608394\colc   26.4\cold  1.1\eol
\cola IR-star\colb   2458698.560491\colc  -12.2\cold  1.5\eol
\cola IR-star\colb   2458707.729192\colc   -9.6\cold  1.8\eol
\cola IR-star\colb   2458709.521672\colc  -45.8\cold  1.6\eol
\cola IR-star\colb   2458766.710392\colc  -39.8\cold  2.4\eol
\cola IR-star\colb   2458779.477079\colc   19.0\cold  1.6\eol
\cola IR-star\colb   2458805.489840\colc  -53.3\cold  2.2\eol
\cola IR-star\colb   2458806.537612\colc  -44.2\cold  1.3\eol
\cola IR-star\colb   2458864.347646\colc   -9.8\cold  1.7\eol
\cola IR-star\colb   2459037.671826\colc   39.3\cold  5.9\eol
\cola IR-star\colb   2459049.606064\colc   12.6\cold  3.3\eol
\cola BD+66/1674A\colb   2457321.404873\colc   83.0\cold  3.3\eol
\cola BD+66/1674A\colb   2457333.644406\colc  -24.8\cold  2.0\eol
\cola BD+66/1674A\colb   2457630.486733\colc -140.4\cold  6.7\eol
\cola BD+66/1674A\colb   2457631.466193\colc -177.6\cold  3.6\eol
\cola BD+66/1674A\colb   2457635.474552\colc  -31.8\cold  3.5\eol
\cola BD+66/1674A\colb   2457764.417211\colc -192.5\cold  2.3\eol
\cola BD+66/1674A\colb   2457778.373338\colc  -17.0\cold  2.5\eol
\cola BD+66/1674a\colb   2457779.379109\colc  -66.9\cold  4.1\eol
\cola BD+66/1674A\colb   2457973.512879\colc  -25.5\cold  3.9\eol
\cola BD+66/1674A\colb   2457978.719895\colc  125.7\cold  5.6\eol
\cola BD+66/1674A\colb   2458015.689182\colc  113.8\cold 15.0\eol
\cola BD+66/1674A\colb   2458026.626092\colc -119.0\cold  4.3\eol
\cola BD+66/1674B\colb   2457321.404873\colc   12.9\cold  7.7\eol
\cola BD+66/1674B\colb   2457333.644406\colc  -11.0\cold 12.0\eol
\cola BD+66/1674B\colb   2457630.486733\colc   75.0\cold  9.0\eol
\cola BD+66/1674B\colb   2457631.466193\colc   76.0\cold  7.0\eol
\cola BD+66/1674B\colb   2457764.417211\colc  106.0\cold  8.0\eol
\cola BD+66/1674B\colb   2457778.373338\colc    1.0\cold 10.0\eol
\cola BD+66/1674B\colb   2457978.719895\colc  -70.0\cold 12.0\eol
\cola BD+66/1674B\colb   2458015.689182\colc -102.0\cold 20.0\eol
\cola BD+66/1674B\colb   2458026.626092\colc  112.0\cold 18.0\eol
\cola P66\colb   2457321.488730\colc  -22.8\cold  3.2\eol
\cola P66\colb   2457630.530613\colc  -18.0\cold  3.7\eol
\cola P3\colb   2457315.539495\colc  -13.7\cold  4.0\eol
\cola P3\colb   2457321.418428\colc  -11.8\cold  2.8\eol
\cola P3\colb   2457630.403301\colc  -14.3\cold  4.2\eol
\cola P3\colb   2457979.698944\colc  -14.5\cold  4.8\eol
\cola P3\colb   2458007.590076\colc  -15.3\cold  3.9\eol
\cola P3\colb   2458015.663140\colc  -17.2\cold  9.2\eol
\cola P3\colb   2458093.393343\colc  -15.1\cold  4.8\eol
\cola P3\colb   2458117.513020\colc  -12.4\cold  3.2\eol
\cola P3\colb   2458119.410332\colc  -12.0\cold  4.2\eol
\cola P3\colb   2458131.513260\colc  -13.6\cold  3.8\eol
\cola P3\colb   2458146.345105\colc  -11.3\cold  1.8\eol
\cola P3\colb   2458174.356811\colc  -21.6\cold  2.5\eol
\cola P310\colb   2457635.488352\colc  -21.1\cold  2.4\eol
\cola P130\colb   2457321.587786\colc  -19.0\cold  4.3\eol
\cola P130\colb   2457630.556358\colc  -24.9\cold  3.3\eol
\cola P23\colb   2457630.646009\colc   17.3\cold  9.1\eol
\cola P62\colb   2457630.607987\colc  -16.2\cold  2.2\eol
\cola P114\colb   2457321.515300\colc  -19.6\cold  4.3\eol
\cola P114\colb   2457630.581211\colc  -17.7\cold  8.4\eol

\end{tabular}
 
\end{table}

\begin{table}
  %\caption{Table~\ref{tab:RVstars1} continued.}
  {\bf Table~\ref{tab:RVstars1}} continued. \\
  \label{tab:RVstars3}
  %Added by TeX Support 
  % Table rv_table.tab  Fri 13:47:59 19-Nov-2021
% 3rd part
%\documentstyle{article}
%\begin{document}
\newcommand\cola {\null}
\newcommand\colb {&}
\newcommand\colc {&}
\newcommand\cold {&}
\newcommand\eol{\\}
\newcommand\extline{&&&\eol}

\begin{tabular}{lrrr}
\cola ID\colb HJD\colc V$_{\rm hel}$\cold $\sigma_v$\eol
\cola \colb days\colc km/s\cold km/s\eol
\extline
\cola P15\colb   2457315.610013\colc  -15.1\cold  3.6\eol
\cola P15\colb   2457630.423668\colc  -19.9\cold  2.8\eol
\cola P13\colb   2457321.661803\colc  -25.9\cold  4.7\eol
\cola P13\colb   2457635.584728\colc  -25.4\cold  2.7\eol
\cola P105\colb   2457321.699161\colc  -21.2\cold  4.6\eol
\cola P20\colb   2457321.462303\colc  -15.3\cold  4.4\eol
\cola P20\colb   2457635.636119\colc  -17.8\cold  3.6\eol
\cola P59\colb   2457635.681985\colc  -21.8\cold  5.8\eol
\cola P195\colb   2457631.492884\colc  -14.2\cold  9.5\eol
\cola P7\colb   2457400.378166\colc  -16.1\cold  3.9\eol
\cola P7\colb   2457630.705540\colc  -14.4\cold  8.5\eol
\cola P316\colb   2457635.508443\colc  -20.4\cold  1.9\eol
\cola P233\colb   2457631.584917\colc  -12.0\cold  0.9\eol
\cola P164\colb   2457321.563540\colc   -6.0\cold  0.2\eol
\cola P239\colb   2457631.630876\colc  -30.0\cold  1.0\eol
\cola P121\colb   2457321.612064\colc   -7.2\cold  0.3\eol
\cola P201A\colb   2457635.534809\colc  -26.6\cold  1.2\eol
\cola P201B\colb   2457635.534809\colc  -61.7\cold  0.5\eol
\cola P109\colb   2457631.680620\colc  -30.5\cold  0.7\eol
\cola P106\colb   2457631.539249\colc  -12.9\cold  1.5\eol
\end{tabular}
 
\end{table}

\begin{figure}[h]
\centering
\includegraphics[angle=00, width=\linewidth]{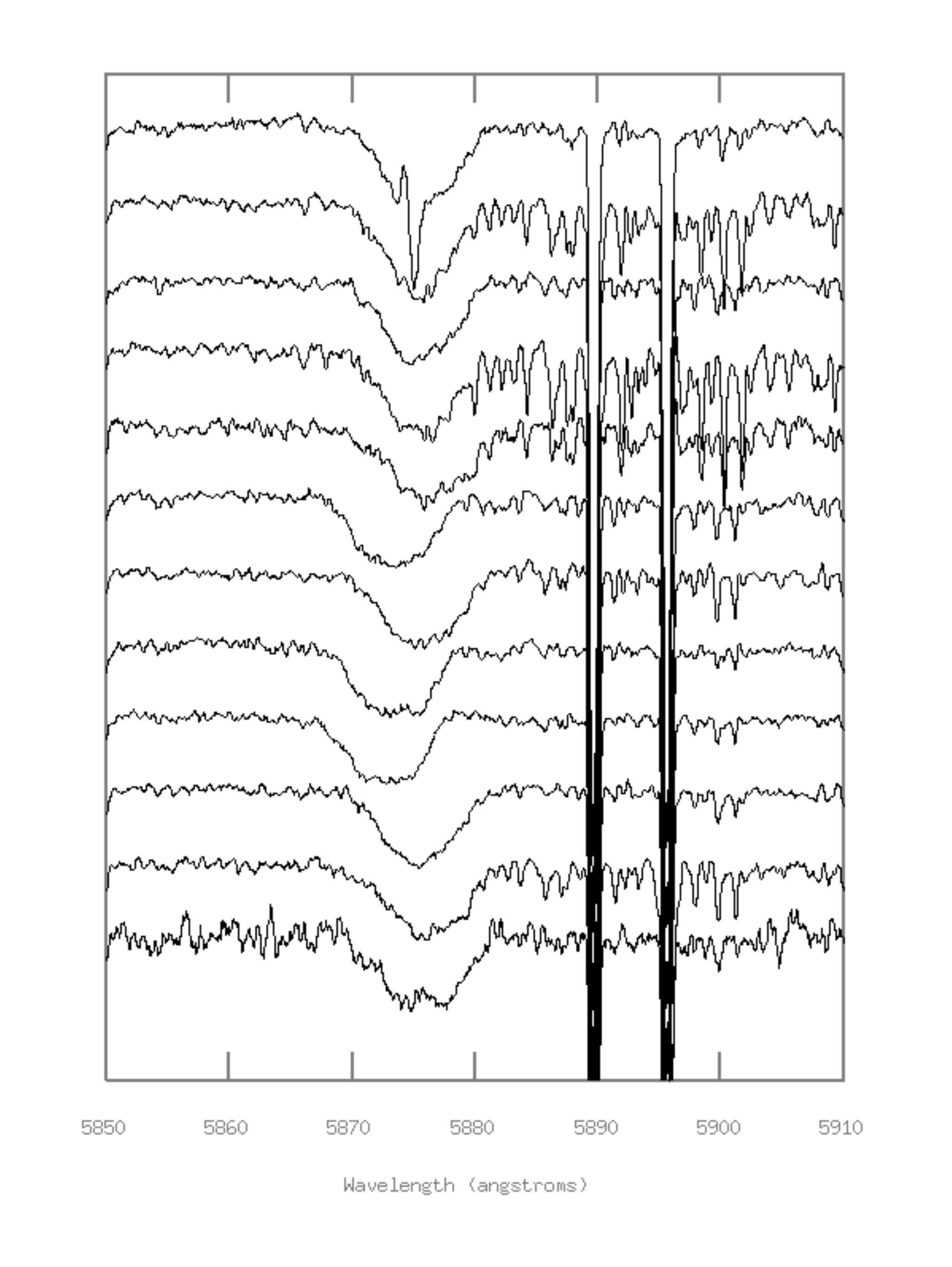}
\caption{He~I 5875 \AA~line at 12 epochs for V747 Cep (left). Note the inverse P Cygni profile in the top spectrum.}
\label{fig:multiepoch1}
\end{figure}

\begin{figure}
\centering
\includegraphics[angle=00, width=\linewidth]{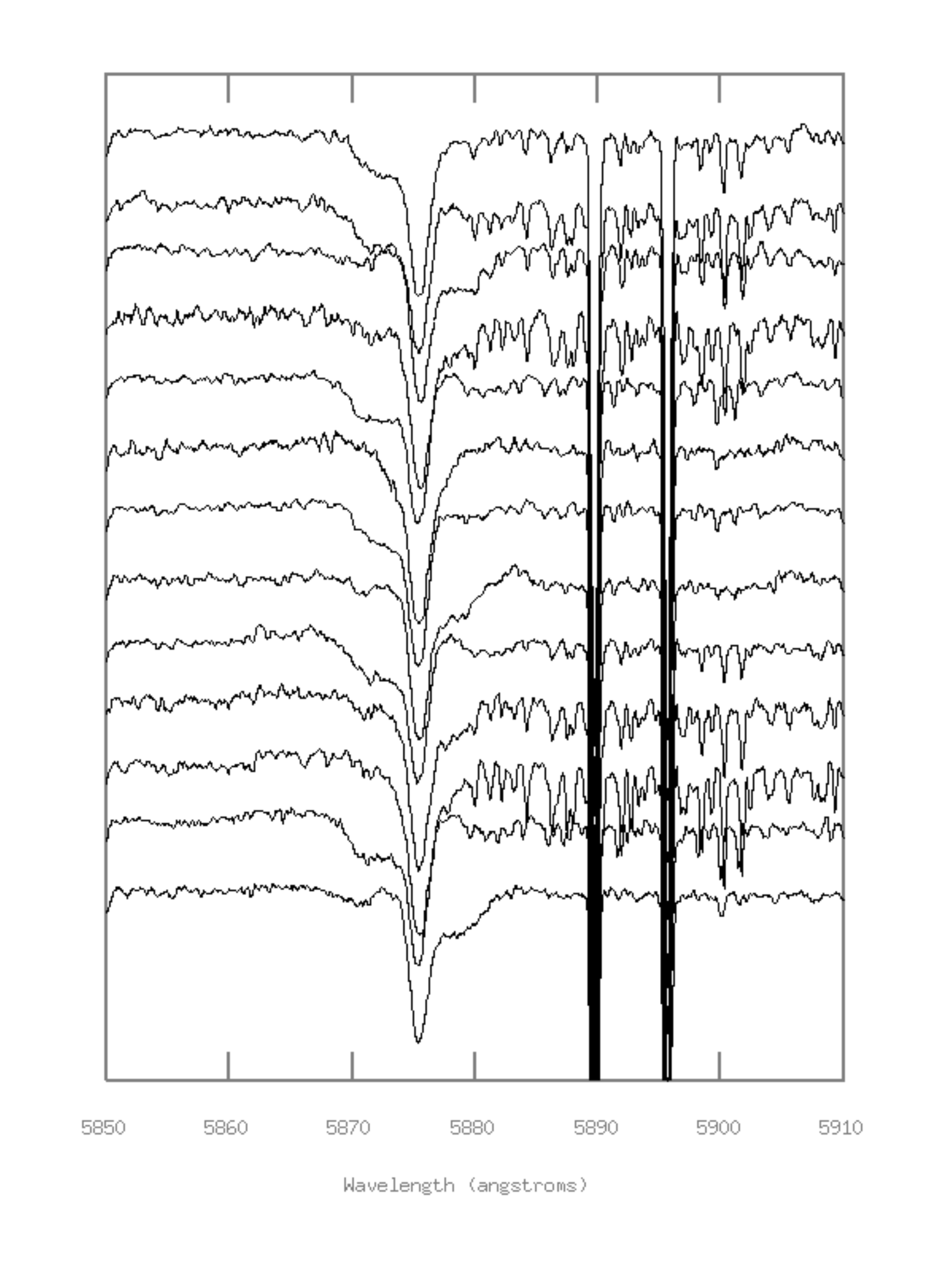}
\caption{He~I 5875 \AA~line at 13 epochs for the spectroscopic triple BD+66$\degr$1675 (right).}
\label{fig:multiepoch2}
\end{figure}

\clearpage

\section{Positions observed for $^{13}$CO}
\label{AppendixB}

\begin{table}[h]
\centering
\caption{Central positions and offsets for northern and southern areas observed for $^{13}$CO.}
\begin{tabular} {lccllll}
 \hline\hline
\noalign{\smallskip}
Area &  Central position & Total & Offsets RA & Offsets Dec. & Offsets Dec. & Offsets Dec.   \\
&  RA, Dec. (2000.0)  & & arcmin. & arcmin. & $V_{b}$ & $V_{r}$  \\ 
\noalign{\smallskip}     
\hline
\noalign{\smallskip}

N3:A & 23 50 14.0 +68 10 19  & 6 &  0 & -20,0,+13 & \ldots & \ldots  \\
 &   & &  +6,-6 & +13 & \ldots & \ldots  \\ 
N3:B & 23 52 12.0 +67 55 19  & 1 & 0 & 0 & \ldots & \ldots  \\ 
N3:C & 23 53 00.0 +67 36 37 & 1 & 0 & 0 & \ldots & \ldots  \\
N3:D & 23 53 56.0 +68 17 43 & 1 & 0 & 0 & \ldots & \ldots  \\ 
N2:A & 00 04 43.0 +67 33 16 & 1 & 0 & 0 & 0$^{(a)}$ & \ldots  \\
N2:B & 00 05 32.0 +67 39 14 & 1 & 0 & 0 & \ldots & \ldots  \\
N2:C & 00 06 14.0 +67 36 37 & 1 & 0 & 0 & 0 & 0  \\ 
N2:D & 00 06 31.0 +67 41 20 & 1 & 0 & 0 & \ldots & 0   \\
N2:E & 00 07 04.0 +67 42 25 & 1 & 0 & 0 & \ldots & \ldots  \\ 
N1:A & 00 04 07.5 +68 37 08 & 1 & 0 & 0 &  0 & \ldots   \\
N1:B & 00 02 15.3 +68 40 41 & 1 & 0 & 0 & 0 & \ldots  \\
N1:C & 00 04 06.0 +68 27 53 & 1 & 0 & 0 & \ldots & \ldots  \\ 
S1 & 00 01 44.3 +67 18 17 & 5 & 0 & -1,0,+1 & all & \ldots   \\  
 &  & & +1,-1 & 0 & all & \ldots   \\  
S2:A & 00 01 18.3 +66 56 24 & 25 & 0 & -6,-4,-2,0,+2,+4,+6 &  +2,+6 & \ldots  \\ 
&  &  & +3 & -8,-6,0,+2,+6 & 0 & \ldots  \\
&  &  & +11 & -6,-4,0,+2,+6 & -4 & \ldots  \\
&  &  & +17 & -4,0 & -4,0 & \ldots  \\
&  &  & -3 & -2,0,+2,+4,+6 & 0 & \ldots  \\
&  &  & -8 & 0,+2,+6 &\ldots & \ldots  \\
S2:B & 00 03 53.0 +66 47 05 & 1 & 0 & 0 & \ldots & \ldots   \\ 
S2:C & 00 04 53.0 +67 02 18 & 1 & 0 & 0 & 0 & \ldots  \\ 
S3:A & 00 01 36.0 +66 26 36 & 31 & +3,0,-3 & 0,-2 & \ldots & \ldots  \\ 
 &  &  & +1.5,+3,+4.5 & -1.5, 0,+1.5,+3,+4.5 & \ldots & all  \\ 
 &  &  & +12 & -2, 0 & \ldots & \ldots  \\
 &  &  & +15 & -2 & \ldots & \ldots  \\
 &  &  & -9 & -2,+2.5 & \ldots & \ldots  \\
 &  &  & -12 & 0,+2.5 & \ldots & \ldots  \\
 &  &  & -15 & -2,+2.5 & \ldots & \ldots  \\
 &  &  & -18 & 0,+2.5 & \ldots & \ldots  \\
S3:B & 23 58 26.0 +66 39 04 & 1 & 0 & 0 & \ldots & \ldots  \\
S3:C & 23 59 19.3 +66 31 43 & 1 & 0 & 0 & \ldots & \ldots  \\
S3:D & 00 04 23.4 +66 19 20 & 1 & 0 & 0 & \ldots & \ldots  \\
S3:E & 00 05 15.1 +66 17 53 & 1 & 0 & 0 &  \ldots & \ldots  \\
S4 & 00 04 26.1 +66 36 08 & 3 & 0 & -18,0,+18 & +18 &  -18 \\
                  
\noalign{\smallskip} 

\hline

\end{tabular}
\label{tab:OSOpos} 

\tablefoot{
\tablefoottext{a}{Two $V_{b}$ components present at one or several positions.}  
}

\end{table}

\begin{table*}
\centering
\caption{Central positions and offsets for western and central areas observed for $^{13}$CO.}
\begin{tabular} {lccllll}
 \hline\hline
\noalign{\smallskip}
Area &  Central position & Total & Offsets RA & Offsets Dec. &  Offsets Dec.  & Offsets Dec.   \\
&  RA, Dec. (2000.0)  & & arcmin. & arcmin. & $V_{b}$ & $V_{r}$ \\ 
\noalign{\smallskip}     
\hline
\noalign{\smallskip}

W10:A &  23 47 08.0 +67 25 03  & 1 & 0 & 0 &  0 & \ldots   \\
W10:B &  23 48 28.0 +67 28 01 & 1 & 0 & 0 &  0 & \ldots   \\
W10:C &  23 49 11.0 +67 47 51  & 1 & 0 & 0 &  \ldots & \ldots    \\        
W9 &  23 50 54.0 +66 47 51 & 13 & 0 & -4,-2,0,+2,+4 & \ldots &  -4,0,+2,+4$^{(a)}$  \\ 
 &  &  & +2,-2 & 0 & \ldots & all$^{(a)}$ \\
 &  &  & +4,-4 & -4,0,+4 & \ldots & -4,0  \\ 
W8 &  23 54 36.0 +67 34 56 & 19 & 0 & -2,0,+1,+2,+3,+4 & +1,+2,+3 & 0,+1  \\
 &  &  & +1 & +1,+2,+3 & +2,+3 & \ldots  \\ 
 &  &  & +2 & 0  & \ldots & \ldots   \\ 
 &  &  & +4 & -4,0,+4 & \ldots & \ldots   \\
  &  &  & -1 & +1,+2,+3 & +1,+2 & \ldots   \\
 &  &  & -2 & 0  & 0 & \ldots   \\ 
 &  &  & -4 & 0,+4 & \ldots & \ldots   \\
W7 & 23 51 12.0 +67 25 35 & 13 & 0 & -4,-2,0,+2,+4 & -4,0,+2 & \ldots  \\
&  & & +1,-1 & 0  & all & \ldots   \\ 
 & & & +4 & -4,0,+4 & 0,+4 & \ldots  \\
 &  & &  -4  & -4,0,+4 & -4 & \ldots  \\ 
W6:A &  23 54 50.2 +67 01 22 & 16 & 0 & -6,-4,-2,0 & \ldots & \ldots   \\  
 &  & &  +2,+4 & -4, 0,+2,+4 & 0 & \ldots  \\ 
 &  & &  +6 & -4,-2,0,+2 & \ldots & \ldots  \\ 
W6:B &  23 52 41.2 +67 04 25 & 3 & +2,0,-4 & 0  & \ldots & \ldots  \\
W6:C &  23 55 26.4 +67 10 58 & 1 & 0 & 0 & 0 & \ldots  \\
W6:D &  23 57 59.5 +66 54 42 & 1 & 0 & 0 & \ldots & \ldots  \\
W6:E &  23 58 02.1 +67 02 08 & 1 & 0 & 0 & \ldots & \ldots  \\
W5 &  23 55 50.0 +67 25 33 & 7 & 0 & -2,-1 ,0,+1 & all & \ldots  \\
&  & 7 & +1,-1,-2 & 0 & all & \ldots  \\ 
W4 &  23 57 12.8 +67 16 16 & 9 & 0 & -2,-1,0,+1 & +1 & \ldots  \\
&   &  & +1 & -2,-1,0,+1 & +1 & \ldots  \\ 
&   &  & -1 &  0 & 0 & \ldots  \\
W3:A &  23 59 25.1 +67 18 00 & 66 & 0 & -8,-6,-4,-2 & \ldots & \ldots  \\  
 &   &  & 0 & 0,+1,+2,+3,+4,+5 &  all$^{(b)}$ & \ldots  \\ 
 &   &  & 0 & +6,+7,+8,+9 &  all$^{(b)}$ & \ldots  \\ 
 &   &  & 0 & +10,+11,+12,+13,+14,+15 &  \ldots & \ldots  \\ 
 &   &  & 0 & +16,+17,+18,+19 & all  & \ldots  \\ 
 &   &  & 0 & +20,+21,+22 & all  & \ldots  \\ 
 &   &  & +2 & -9,-7,-5,-3,+3,+4,+5,+6 & -7,-3,+3,+4,+6$^{(b)}$ & \ldots  \\ 
 &   &  & +2 & +9, +12,+15,+17,+19,+22 & +9,+12 & \ldots  \\ 
 &   &  & +5 & -9,-8,-7 & -8 & \ldots  \\
 &   &  & -4 & -4,0,+2,+3,+4,+5 & 0,+2,+3,+4,+5 & +2,+3,+4,+5  \\ 
 &   &  & -4 & +6,+7,+8,+9,+10 & +6,+10 & all  \\
&   &  & -4 & +11,+12,+13,+14,+15,+16 & all & \ldots  \\ 
&   &  & -4 & +17,+18,+19,+20,+21,+22 & all & \ldots  \\ 
W3:B &  00 00 59.0 +67 09 39 & 5  & 0 & -4,-3,-2,-1,0 & -2,-1,0 & \ldots  \\
W3:C$^{(c)}$ &  00 00 37.7 +67 15 53 & 2 & 0 & 0,+2 & 0 & \ldots \\ 
W2 &  00 00 12.3 +67 22 15 & 9 & -1,0,1 & -1,0,1 & all$^{(b)}$ & \ldots  \\ 
W1 &  00 01 22.3 +67 30 15 & 47  & 0 & -7,-6,-5,-4,-3,-2,-1,0 & all$^{(b)}$ & \ldots   \\
 &   &   & 0 & +1,+2,+3,+4,+5,+6,+7 &  all & \ldots   \\
 &   &   & +1 &  -5,-1 & all & \ldots  \\
  &   &   & +2 &  +2,+3,+6 & all & \ldots  \\  
 &   &   & -1 & -4,-3,-1,0,+1,+2,+3 &all$^{(b)}$ & \ldots  \\
 &   &   & -1 & +4,+5,+6,+7 &all$^{(b)}$ & \ldots  \\
  &   &   & -2 & +6 & +6 & \ldots  \\
 &   &   & -2.5 &  -1,0,+1,+2,+3,+4,+5,+6,+7 & all$^{(b)}$ &  0  \\
 &   &   & -4 &  -1,0,+1,+2,+3,+4 & all$^{(a)}$ & 0,+1  \\
C:A & 00 02 25.4 +67 25 57 & 1 &  0 & 0  & \ldots & 0   \\
C:B & 00 02 25.8 +67 23 47 & 1 & 0 & 0   & \ldots & 0   \\
C:C & 00 02 07.1 +67 25 09 & 4 & 0 & 0 & 0 & \ldots   \\ 
  &  & & -1 & 0 & \ldots & \ldots   \\ 
 &  & & -1.5 & -1 & -1 & \ldots   \\  
 &  & & -2.3 & -2 & -2$^{(b)}$ & \ldots   \\
                   
\noalign{\smallskip} 

\hline

\end{tabular}
\label{tab:OSOpos} 

\tablefoot{
\tablefoottext{a}{ $V_{r}$ component blended with the local component.}
\tablefoottext{b}{Two $V_{b}$ components present at one or several positions.}  
\tablefoottext{c}{Centred at a globule.} 
}

\end{table*}

\begin{table*}
\centering
\caption{Central positions and offsets for eastern areas observed for $^{13}$CO.}
\begin{tabular} {lccllll}
 \hline\hline
\noalign{\smallskip}
Area &  Central position & Total & Offsets RA & Offsets Dec. & Offsets Dec.  & Offsets Dec.    \\
&  RA, Dec. (2000.0)  & & arcmin. & arcmin. & $V_{b}$ & $V_{r}$ \\ 
\noalign{\smallskip}     
\hline
\noalign{\smallskip}

E1:A        & 00 03 29.0 +67 21 15  &  9 &  0   & -3,-2,-1 & \ldots & \ldots  \\
            &                       &    & +1   & +2,+3    & \ldots & \ldots  \\
            &                       &    & -1   & -3,-4    & \ldots & \ldots  \\
E1:B        & 00 03 36.2 +67 17 07  & 45 &  0   & -7,-6,-5,-4,-3,-2,-1 & \ldots & \ldots  \\ 
            &                       &    & +1   & -5,-1 & \ldots & \ldots  \\
            &                       &    & +2   & +2,+3,+6 & \ldots & \ldots  \\
            &                       &    & -1   & -4,-3,-1,0,+1,+2,+3,+4,+5,+6,+7 & \ldots & \ldots  \\ 
            &                       &    & -2.5 & -1,0,+1,+2,+3,+4,+5,+6,+7 & \ldots & \ldots  \\ 
            &                       &    & -4   & -1,0,+1,+2,+3,+4 & +1,+4  & \ldots  \\ 
E2:A        & 00 05 58.5  +67 22 00 &  6 &  0   & -4,0,+1,+2.5 &  +2.5 & \ldots \\ 
            &                       &    & 2.8  & +1,+2.+5 & \ldots & \ldots   \\
E2:B        & 00 06 46.6  +67 27 50 &  4 & 0    & 0 & \ldots & \ldots  \\ 
            &                       &    & +1   & +1 & \ldots & \ldots  \\
            &                       &    & +5   & +1 & +1 & +1  \\ 
            &                       &    & +6   & +1.5 & +1.5 & +1.5  \\
E2:C        & 00 07 46.9 +67 19 50  &  1 &  0   & 0 & 0 & \ldots  \\ 
E2:D        & 00 08 05.8 +67 23 50  &  1 &  0   & 0 & \ldots & \ldots \\
E2:E        & 00 08 55.9 +67 20 26  &  9 &  0   & -2,+2,+12 & +2$^{(a)}$ & \ldots \\
            &                       &    & +2   & +2.3 & +2.3 & \ldots  \\
            &                       &    & +3   & 0,+1 & 0,+1 & 0  \\
            &                       &    & +5   & -1,0,+1 & -1,0,+1 & 0,+1  \\ 
E3:A        & 00 07 44.0 +67 10 00  & 76 & 0,-3 & -14.-12,-10,-8,-6,-4,-2 & \ldots & \ldots  \\
            &                       &    & 0,-3 & 0,+2,+4,+6,+8 & \ldots & \ldots   \\
            &                       &    & +3,+6,+9 & -12,-8,-4,0,+4,+8 & \ldots & \ldots  \\
            &                       &    & -6   & -14.-12,-10,-8,-6,-4,-2& \ldots  \\
            &                       &    & -6   & 0,+2,+4,+6,+8  & -14,-10,-6  & \ldots  \\  
            &                       &    & -9   & -16,-14,-12,-10,-8,-6,-4,-2,0,+2 &  -10,-8$^{(a)}$ & \ldots   \\ 
            &                       &    & -12  & -14,-12,-10,-8,-6 & -14,-12,-10 & \ldots  \\
            &                       &    & -15  & -16,-14,-12,-10,-8,-6,-4 & -4  & \ldots   \\
E3:B$^{(b)}$ & 00 04 12.5 +67 10 47  &  1 & 0,+3 & 0 &  \ldots & \ldots  \\
Esouth:A    & 00 09 11.0 +66 13 53  &  1 &  0   & 0 &  \ldots & \ldots\\
Esouth:B    & 00 12 47.0 +66 06 31  &  1 &  0   & 0 &  \ldots & \ldots  \\
Esouth:C    & 00 18 36.0 +66 07 52  &  1 &  0   & 0 &  \ldots  & \ldots \\
Esouth:D    & 00 19 53.0 +66 18 44  &  1 &  0   & 0 &  \ldots & \ldots  \\ 
Esouth:E    & 00 21 55.0 +66 34 39  &  1 &  0   & 0 & 0 & \ldots \\
Enorth:A    & 00 14 19.0 +67 20 13  &  1 &  0   & 0 & 0 & \ldots   \\
Enorth:B    & 00 17 08.0 +67 24 39  &  1 &  0   & 0 & 0 & \ldots \\
Enorth:C    & 00 19 23.0 +67 34 19  &  1 &  0   & 0 & 0 & \ldots   \\
Enorth:D    & 00 23 11.0 +67 10 54  &  1 &  0   & 0 &  \ldots & \ldots  \\ 
E4:A        & 00 27 45.0  +67 37 25 &  3 &  0   & 0,+2 & all & \ldots   \\
            &                       &    &  2   & -0.5 &  -0.5 & \ldots   \\
E4:B        & 00 28 50.0  +67 33 13 & 21 &  0   & 0 & 0$^{(a)}$ & 0   \\
            &                       &    &  -6  & -14,-12,-10.-8,-6,-4,-2 & all & \ldots   \\
            &                       &    &  -9  & -14,-12,-10.-8,-6,-4,-2,0 & -10.-8,-6,-4,-2,0 & \ldots   \\
            &                       &    & -12  & -14,-12,-10.-8,-6 & \ldots & \ldots   \\
E4:C        & 00 30 00.0  +67 30 00 & 5  &   0  & -1.5,0,+1.5 & all & all  \\
            &                       &    & +1.5 & 0 & 0 & \ldots     \\ 
            &                       &    & -1.5 & 0 & 0 & 0   \\ 
                   
\noalign{\smallskip} 

\hline

\end{tabular}
\label{tab:OSOpos} 

\tablefoot{
\tablefoottext{a}{Two $V_{b}$ components present at this position.}   
\tablefoottext{b}{Centred at a globule.}
}

\end{table*}

\clearpage

\section{Representative $^{13}$CO spectra for each region}
\label{AppendixC}

\begin{figure*}[t]
\includegraphics[angle=00, width=6cm]{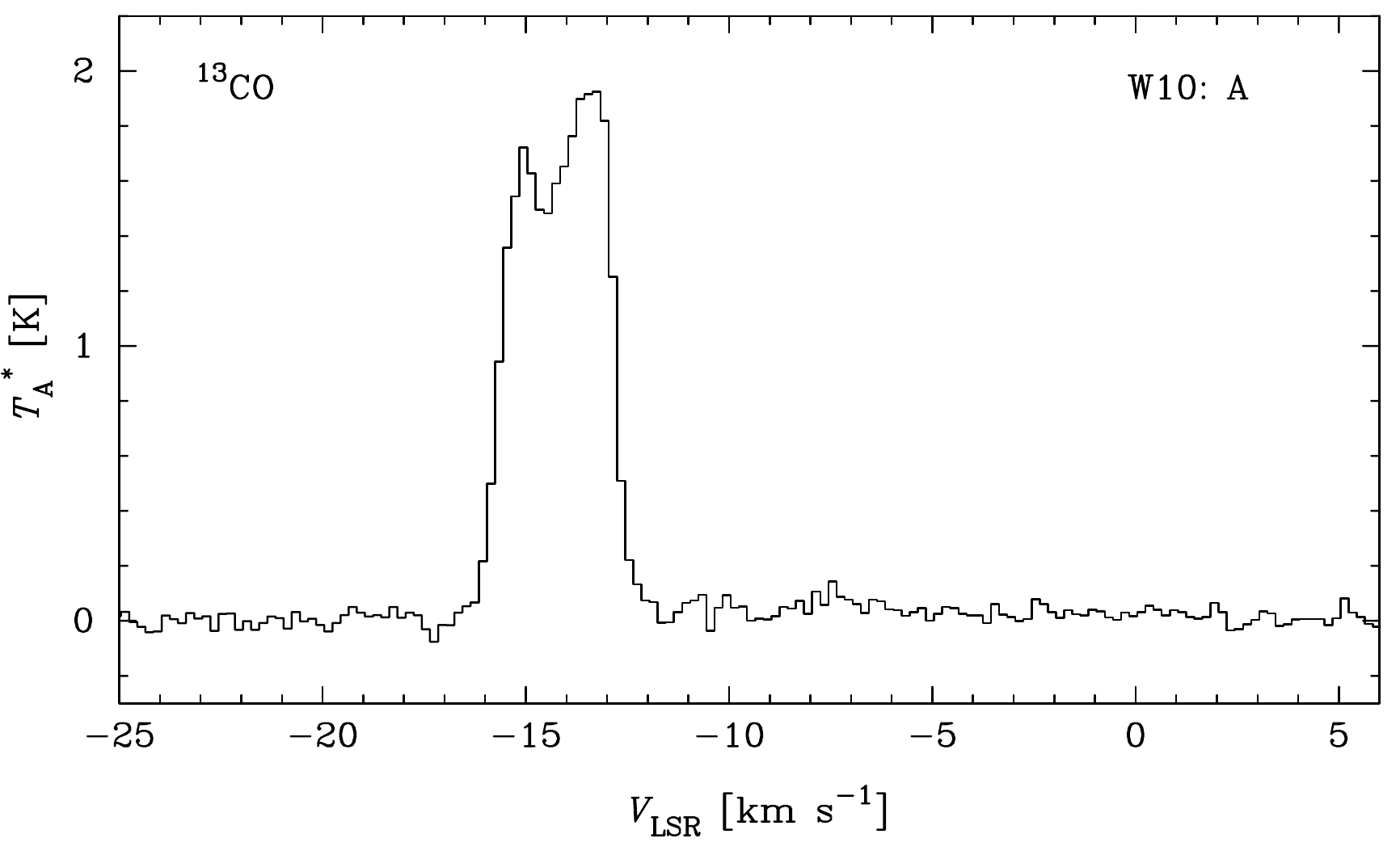}
\includegraphics[angle=00, width=6cm]{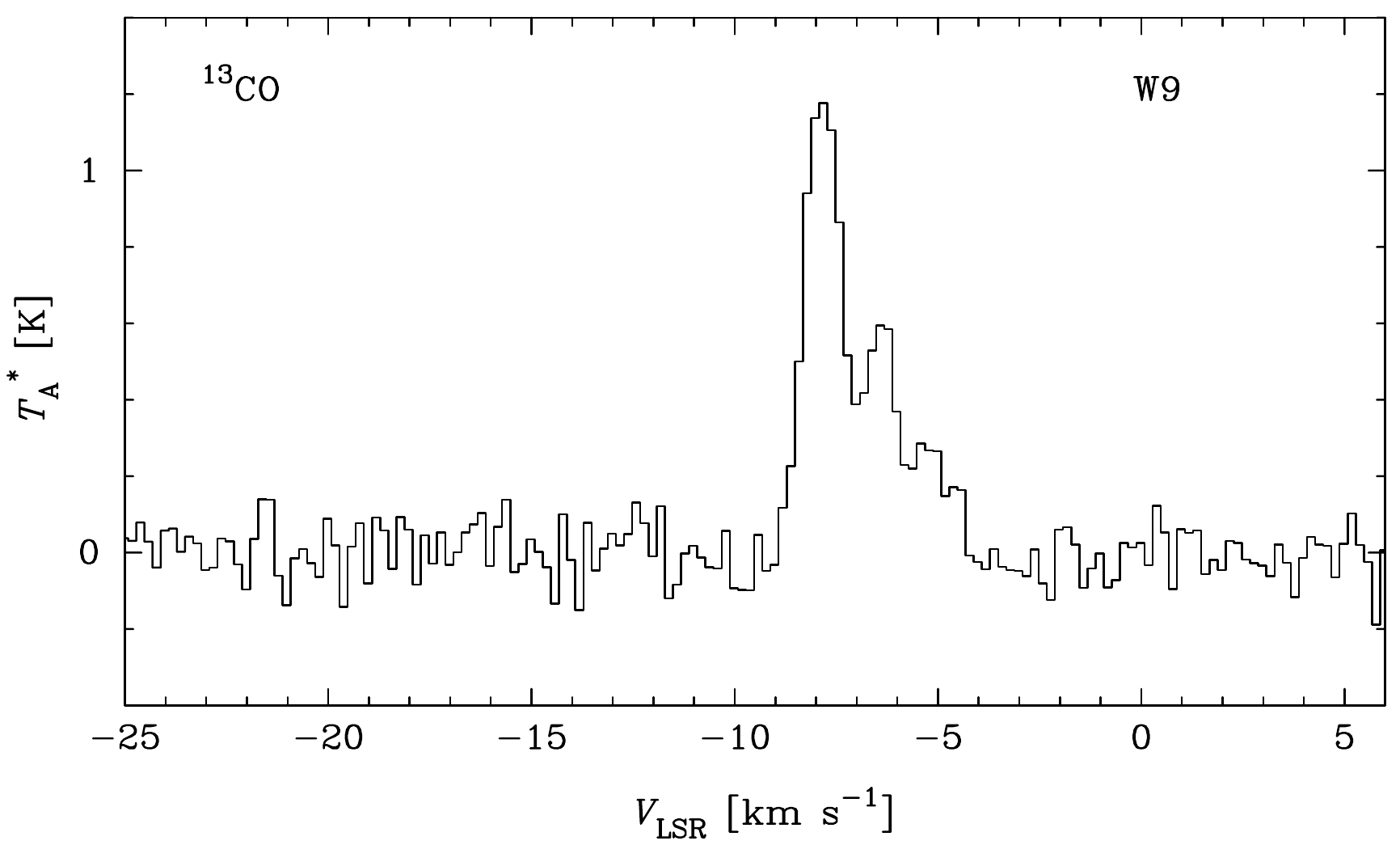}
\includegraphics[angle=00, width=6cm]{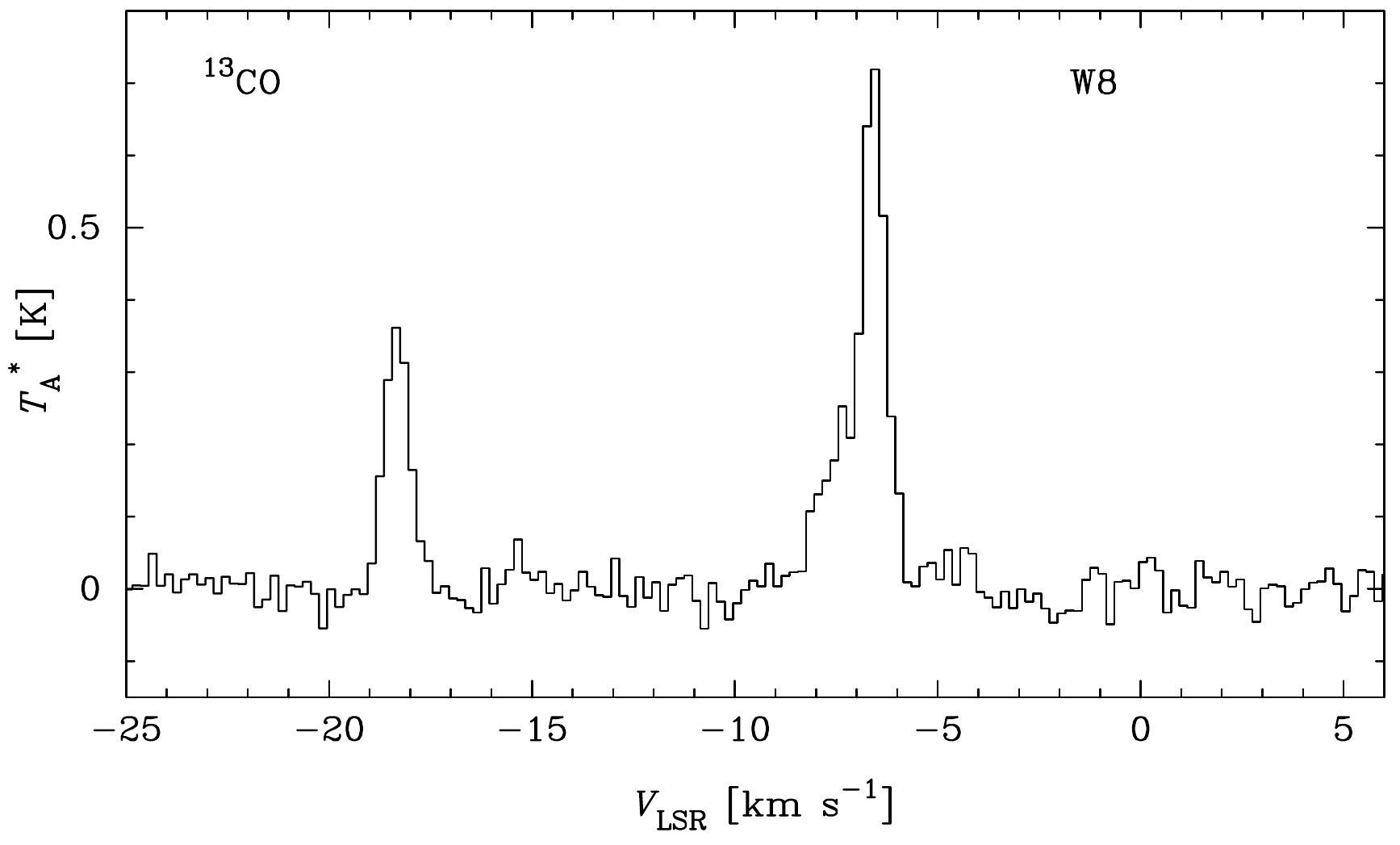}
\includegraphics[angle=00, width=6cm]{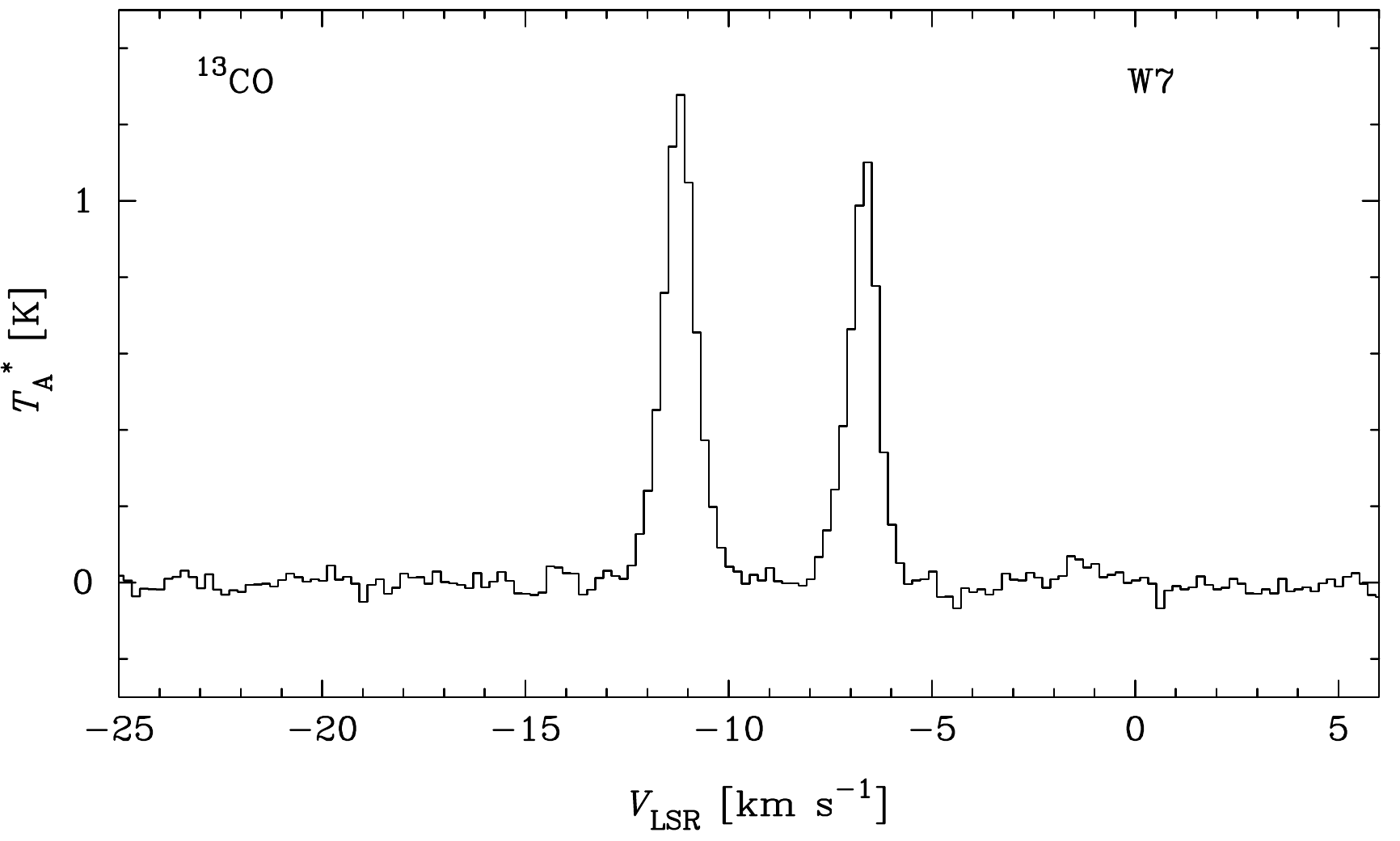}
\includegraphics[angle=00, width=6cm]{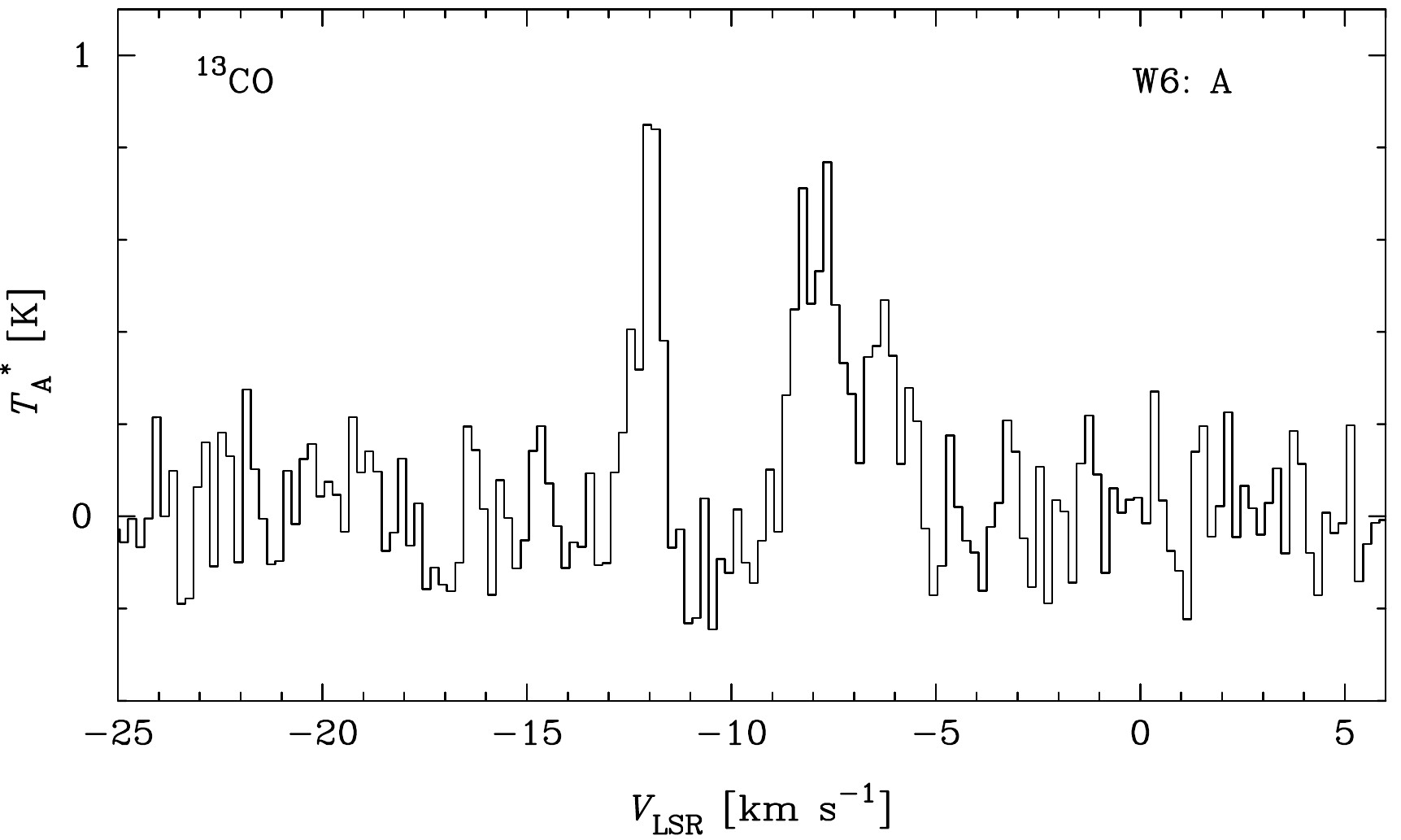}
\includegraphics[angle=00, width=6cm]{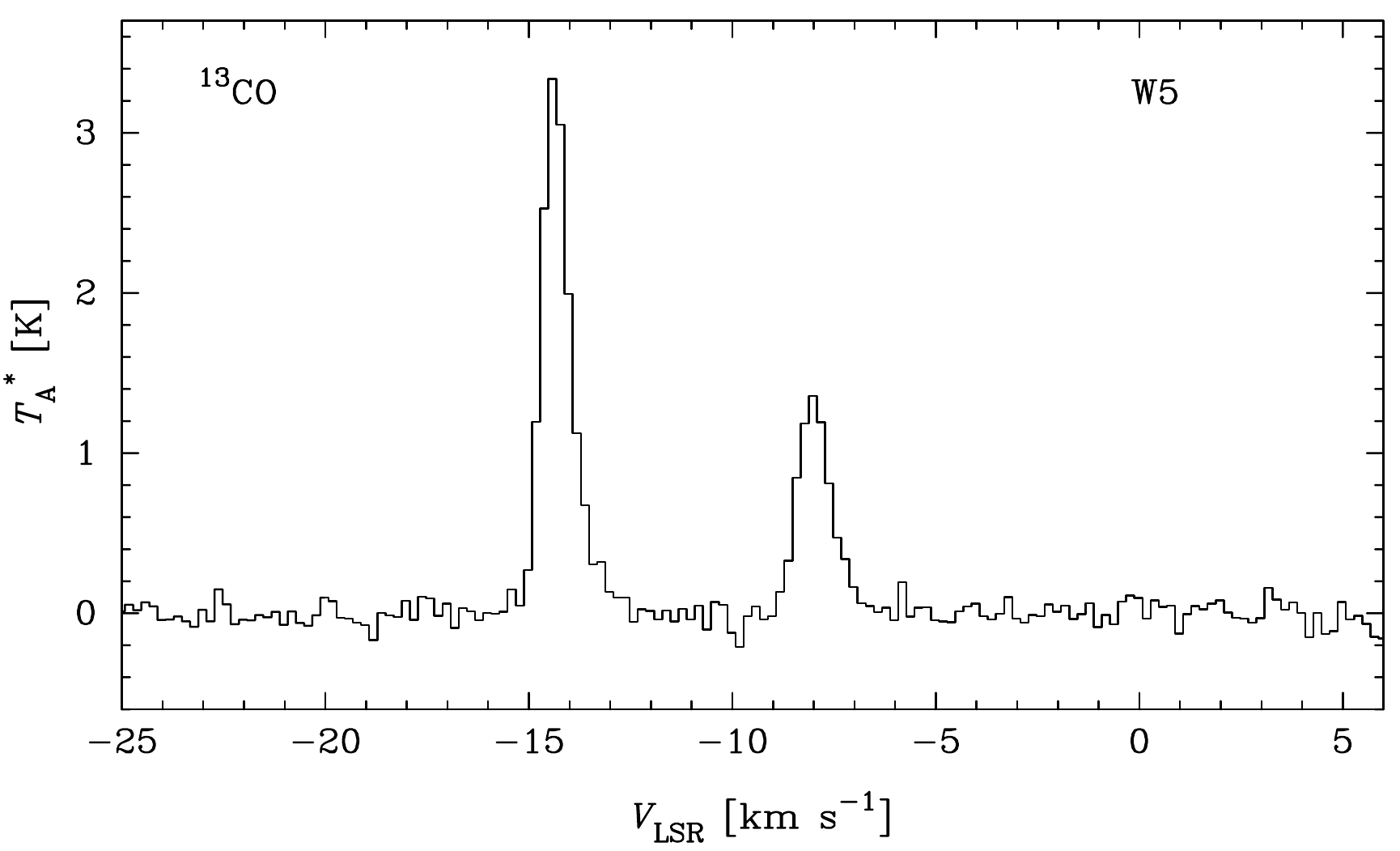}
\includegraphics[angle=00, width=6cm]{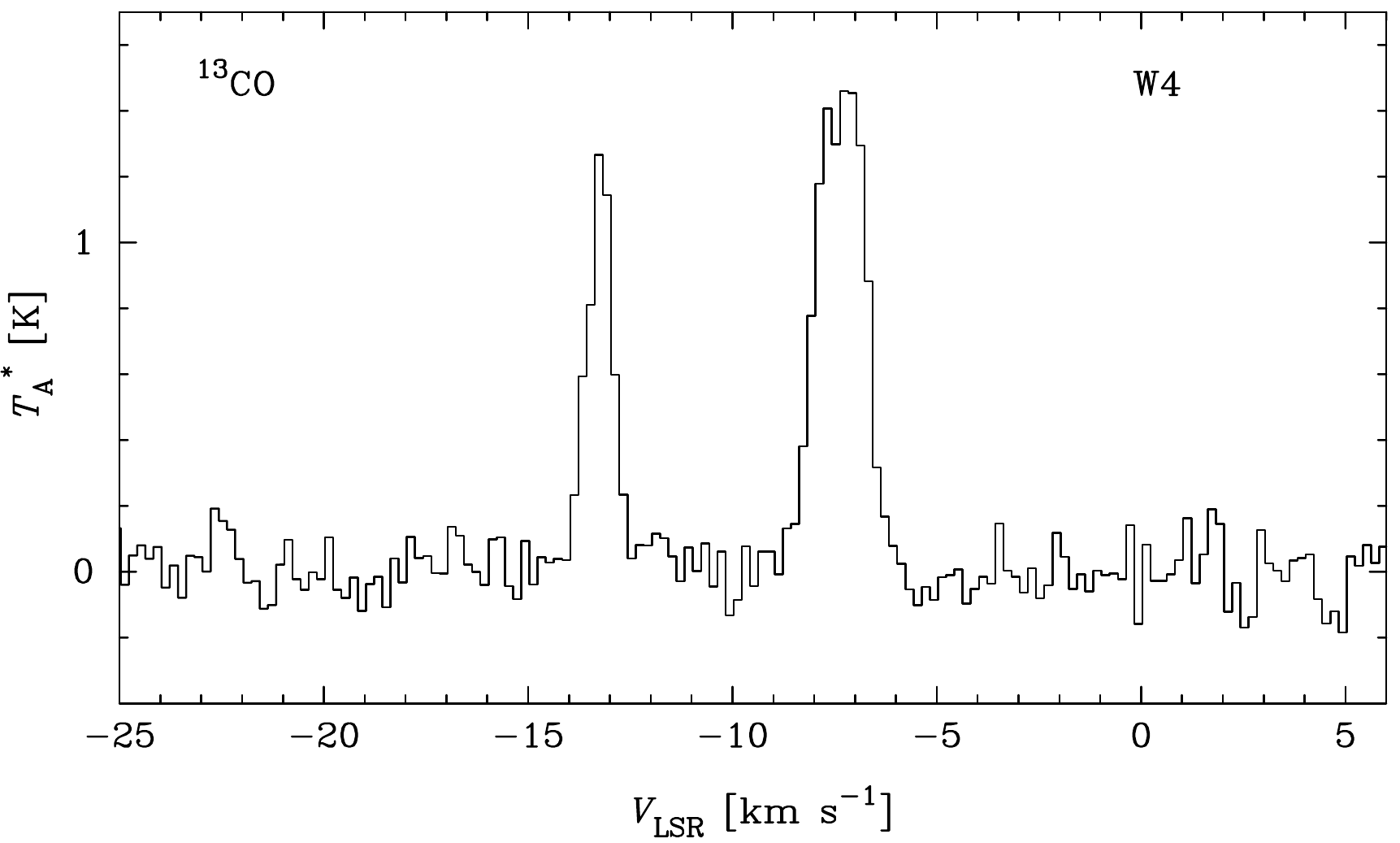}
\includegraphics[angle=00, width=6cm]{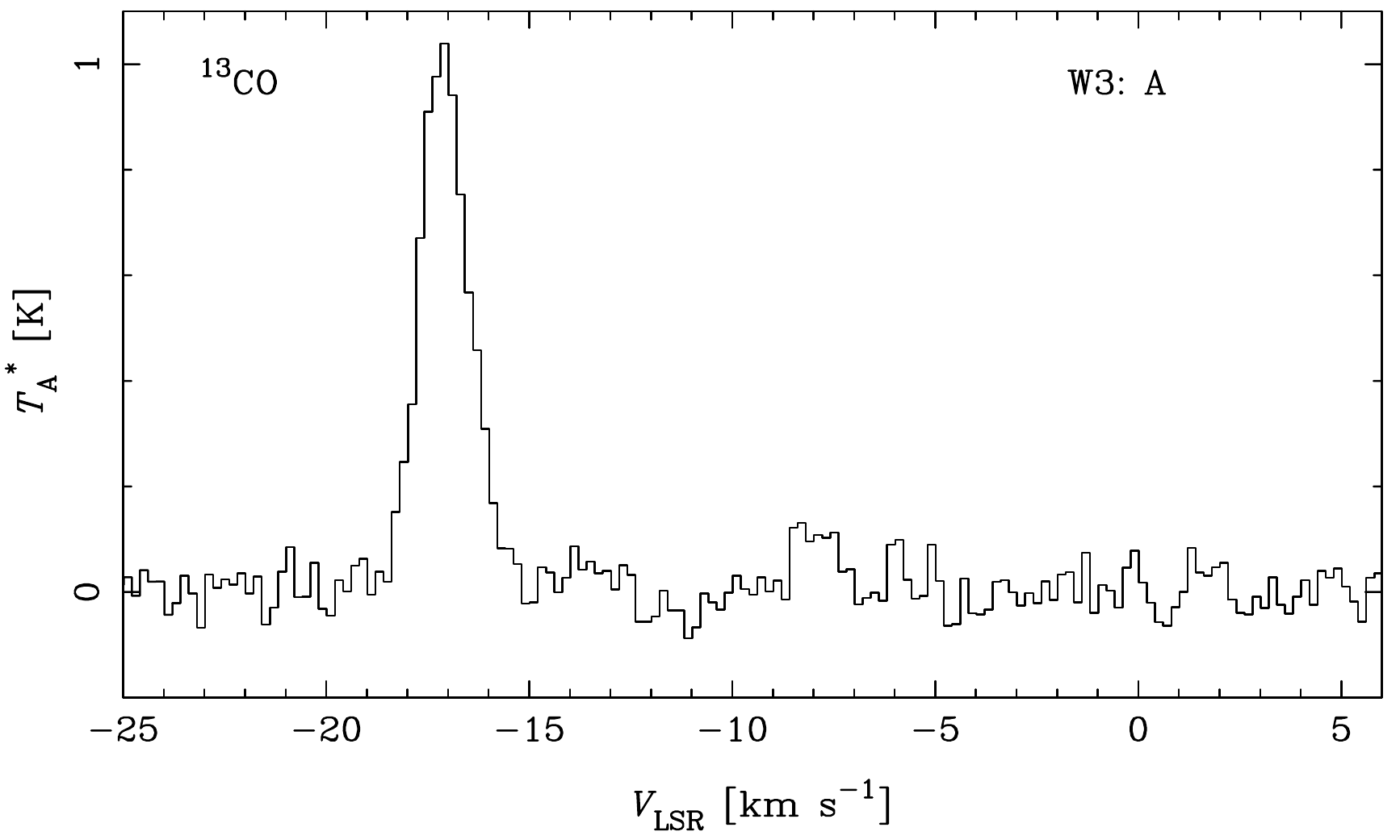}
\includegraphics[angle=00, width=6cm]{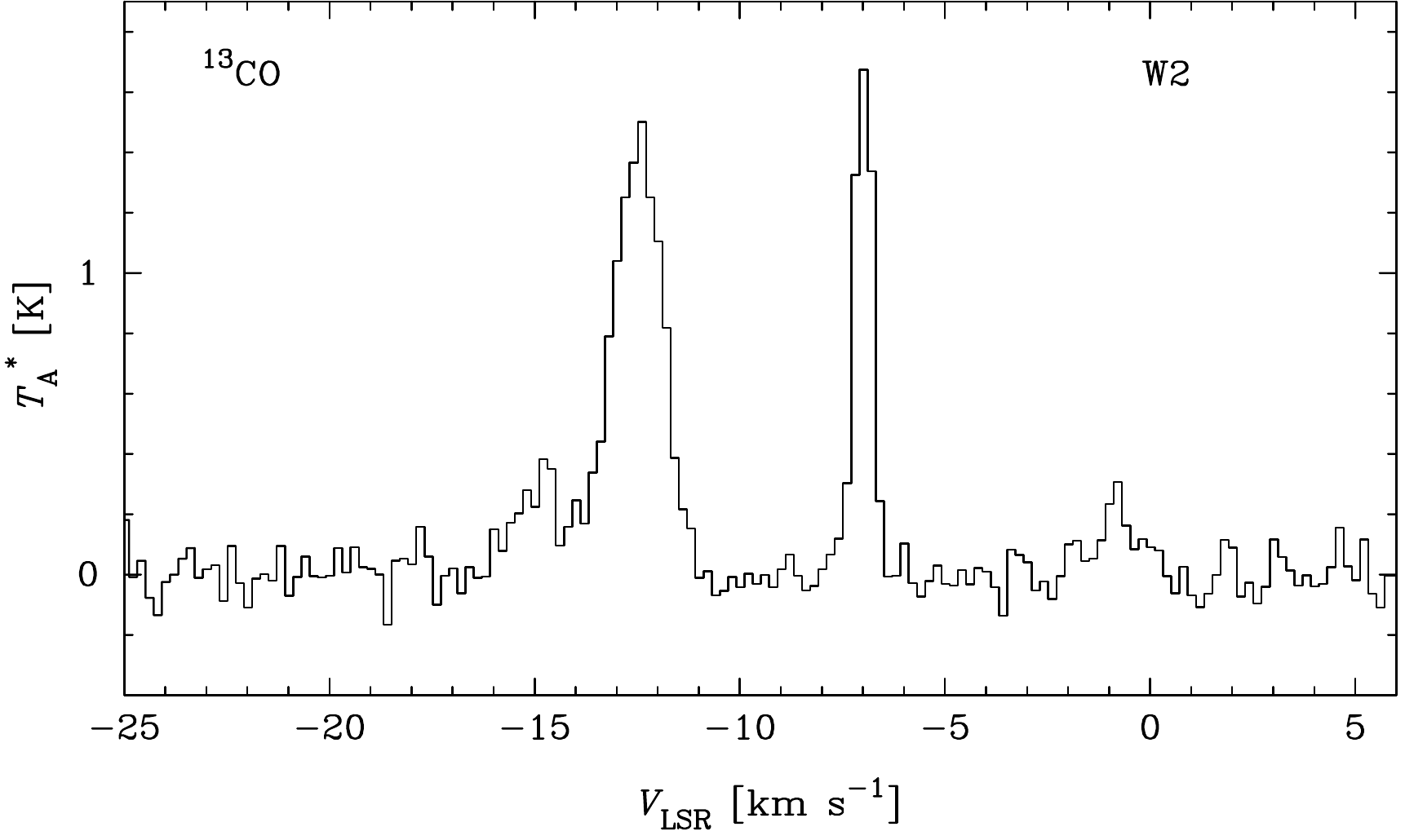}
\includegraphics[angle=00, width=6cm]{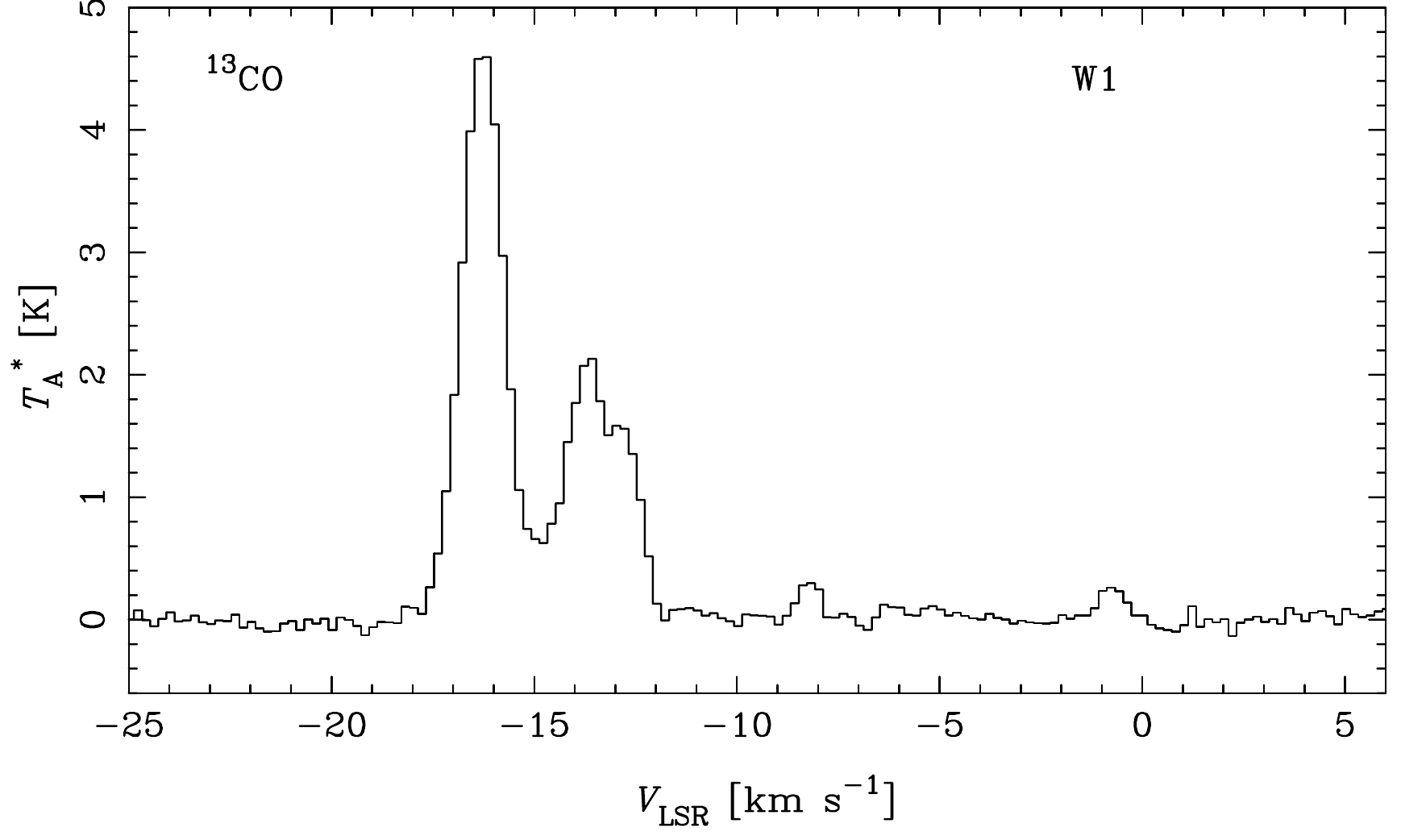}
\includegraphics[angle=00, width=6cm]{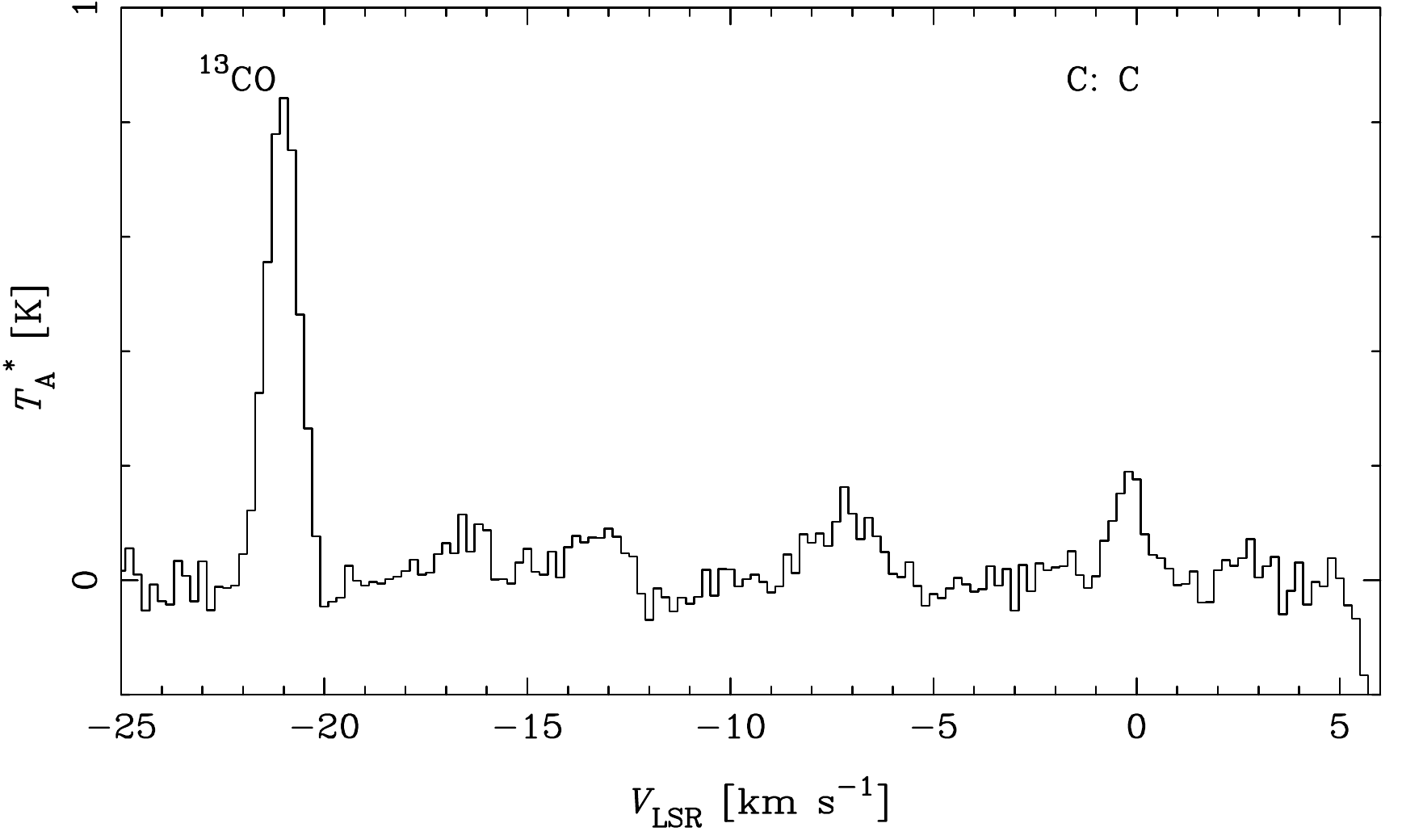}
\includegraphics[angle=00, width=6cm]{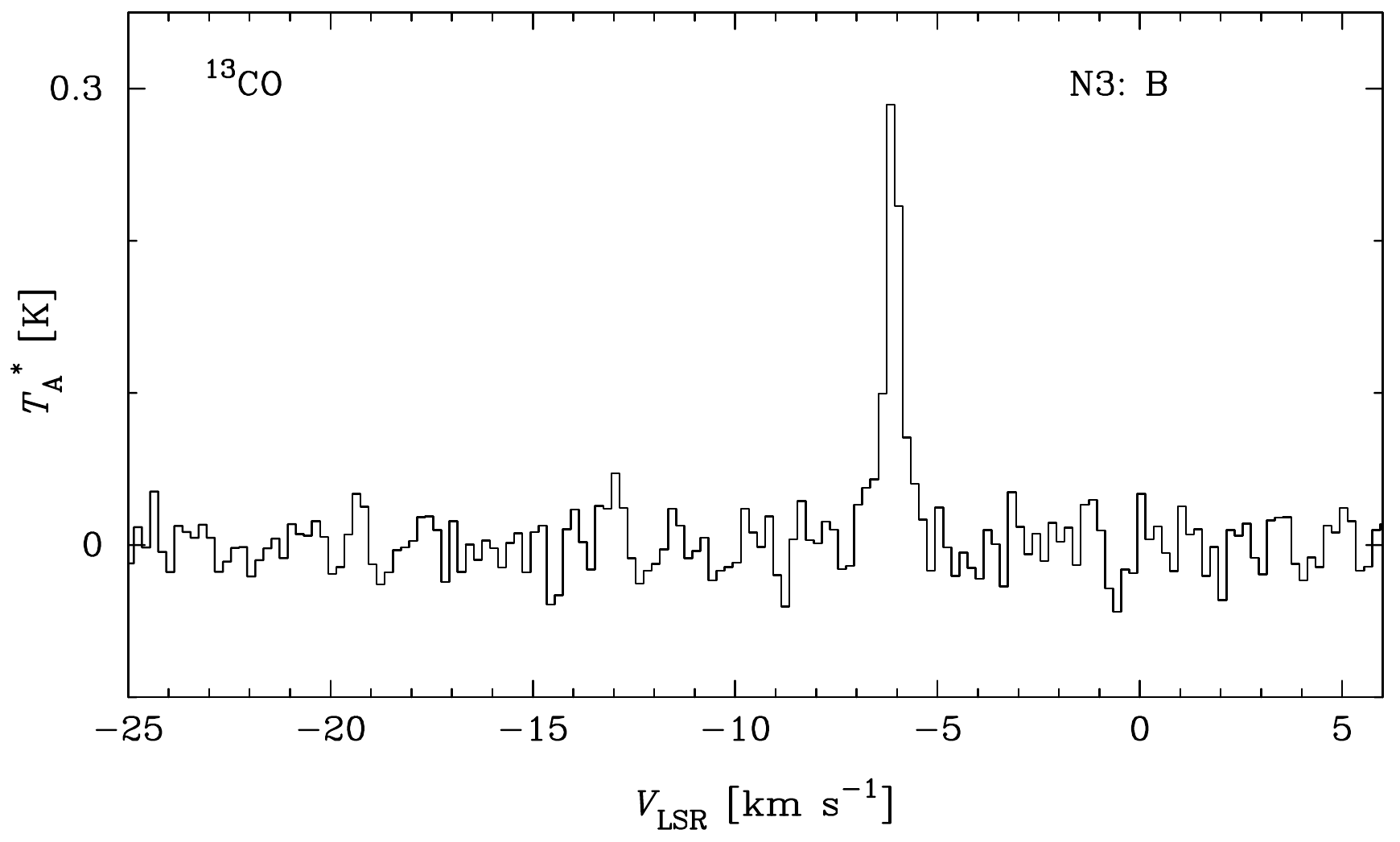}
\includegraphics[angle=00, width=6cm]{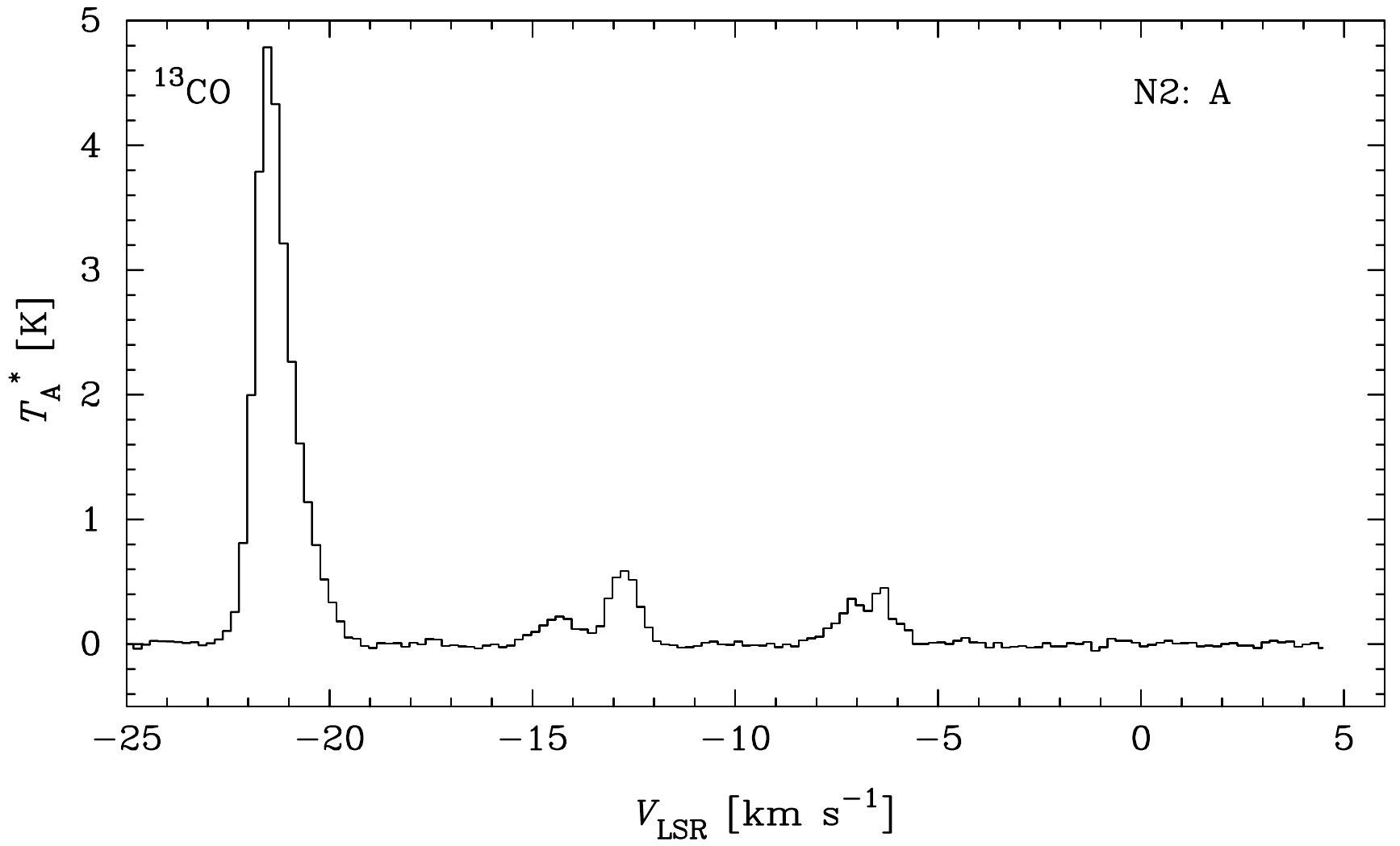}
\includegraphics[angle=00, width=6cm]{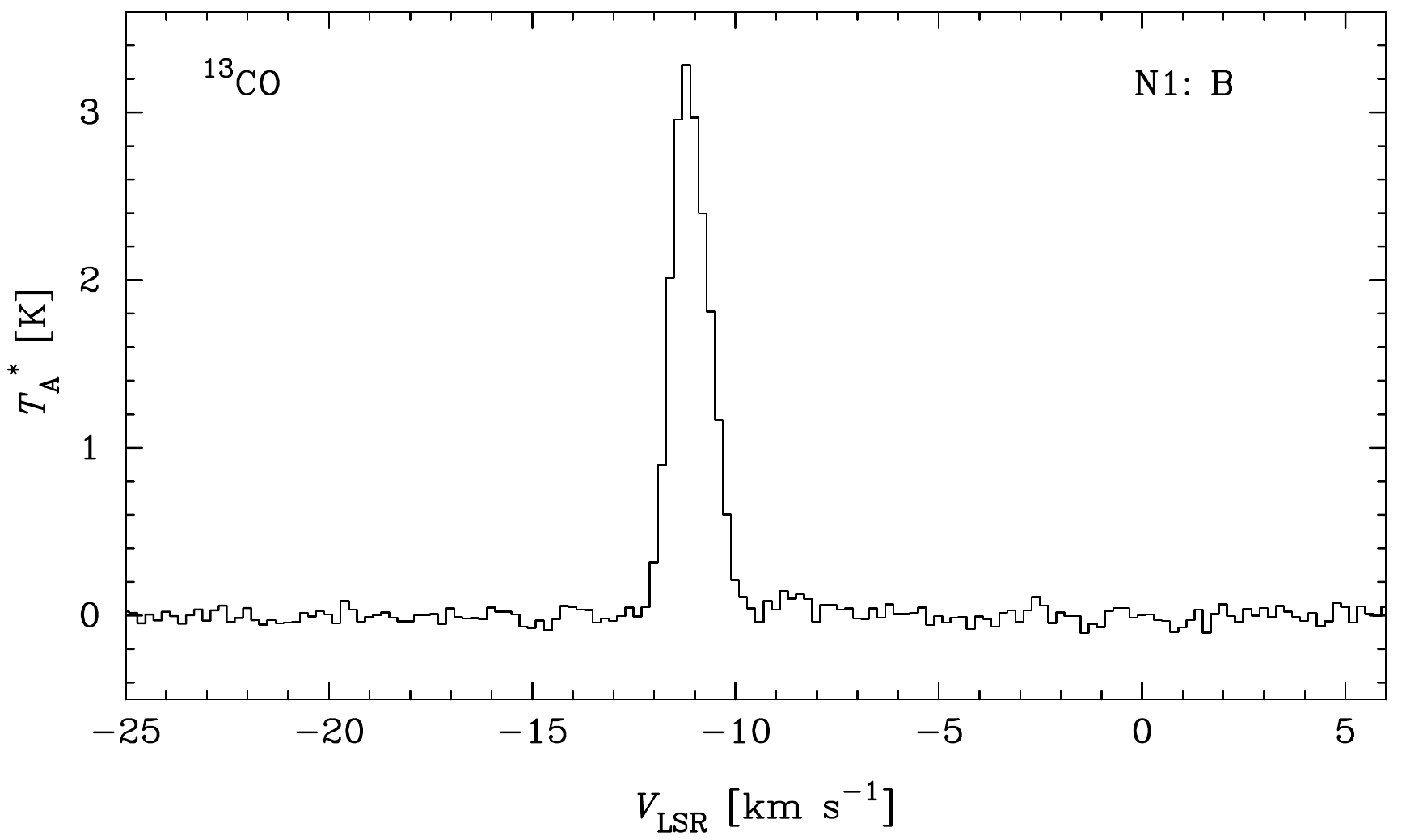}
\includegraphics[angle=00, width=6cm]{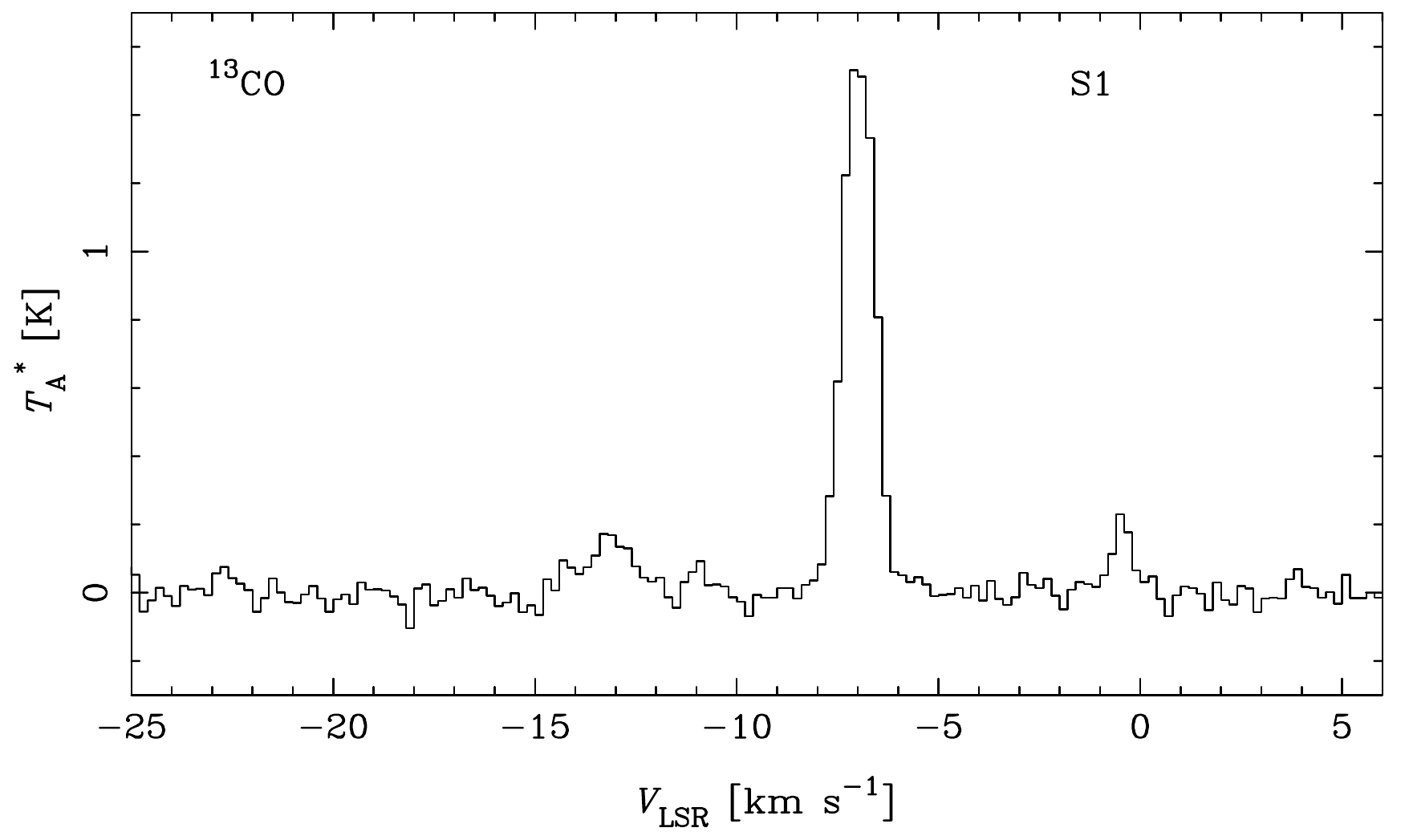}
\includegraphics[angle=00, width=6cm]{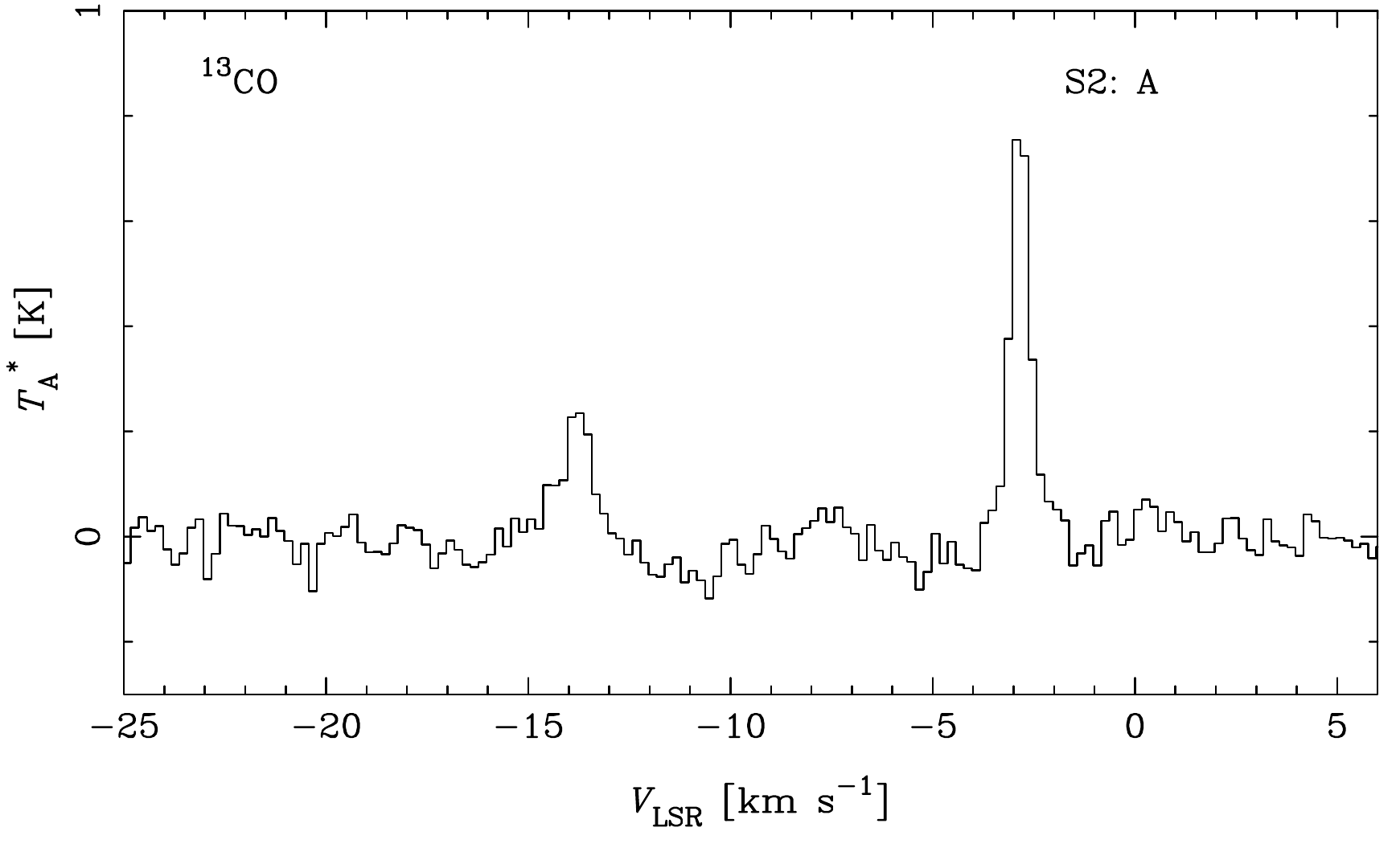}
\includegraphics[angle=00, width=6cm]{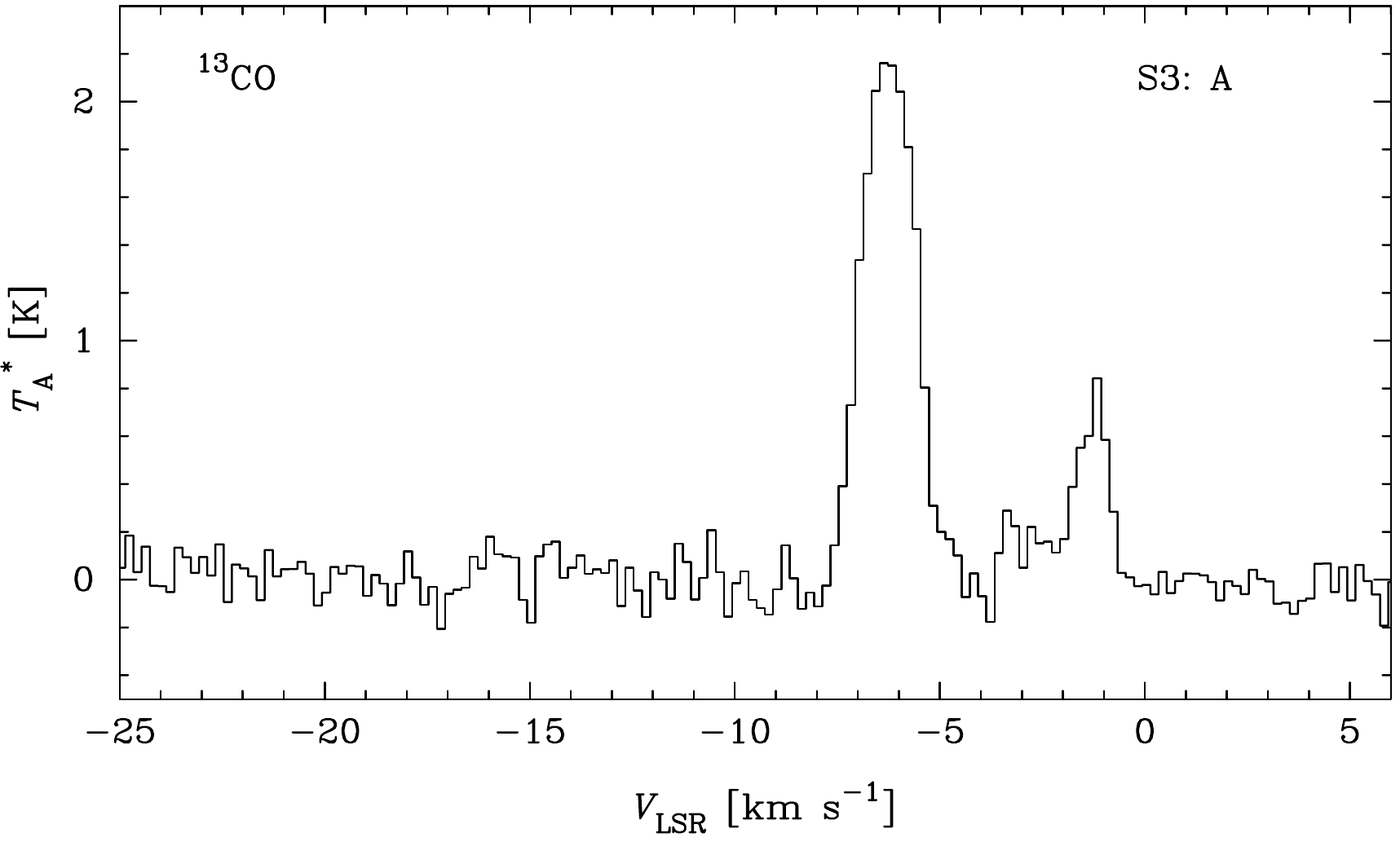}
\includegraphics[angle=00, width=6cm]{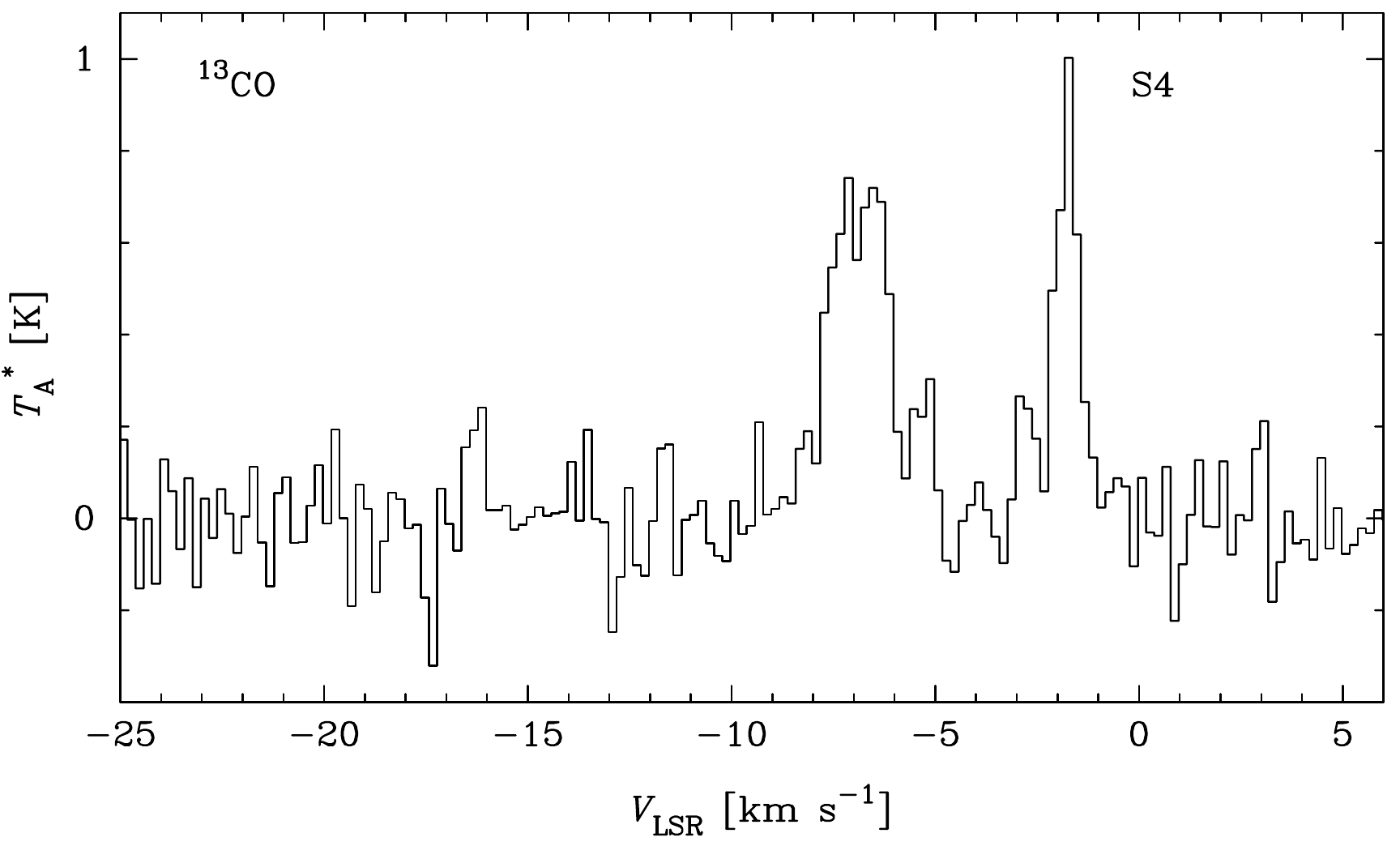}
\caption{Selected $^{13}$CO spectra from each area. Components associated with the expanding shell are typically at $<$ --10~km~s$^{-1}$, components related to the local gas (Gould's Belt) are found around 0~km~s$^{-1}$, and those related to the Orion arm at around --6~km~s$^{-1}$. Offsets are W10:A (0,0), W9 (+240,0), W8 (0, +120), W7 (0,0), W6:A (+240,0), W5 (0,0), W4 (0,+60), W3:A  (+120,+540), W2 (0,-60), W1 (-150,+180), C:C (-90,-60), N3:B (0,0), N2:A (0,0), N1:B (0,0), S1 (0,-60), S2:A (+180,0), S3:A (0,0), and S4 (0,0).}
\label{fig:COsp1}
\end{figure*}

\begin{figure*}[t]
\includegraphics[angle=00, width=5.8cm]{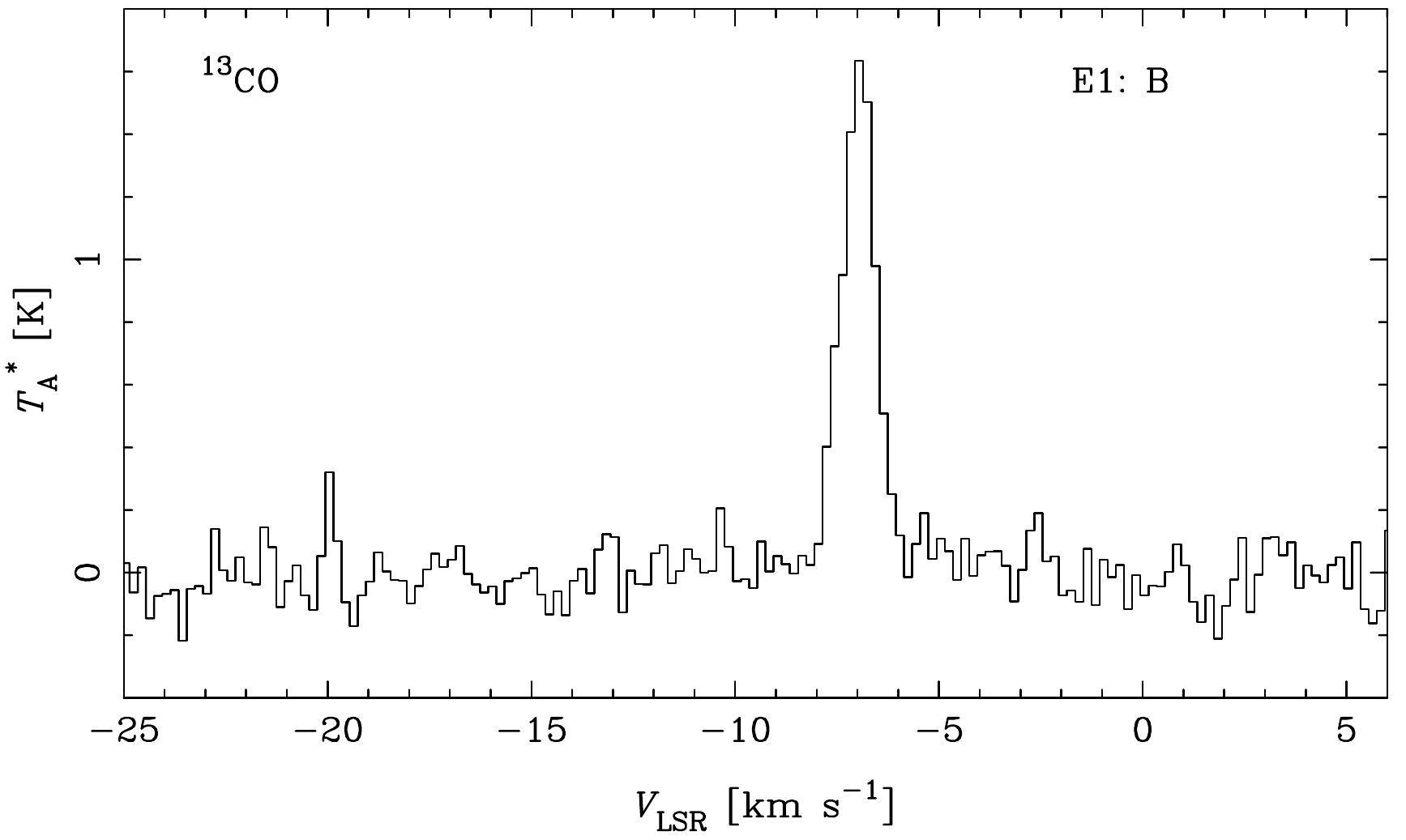}
\includegraphics[angle=00, width=5.75cm]{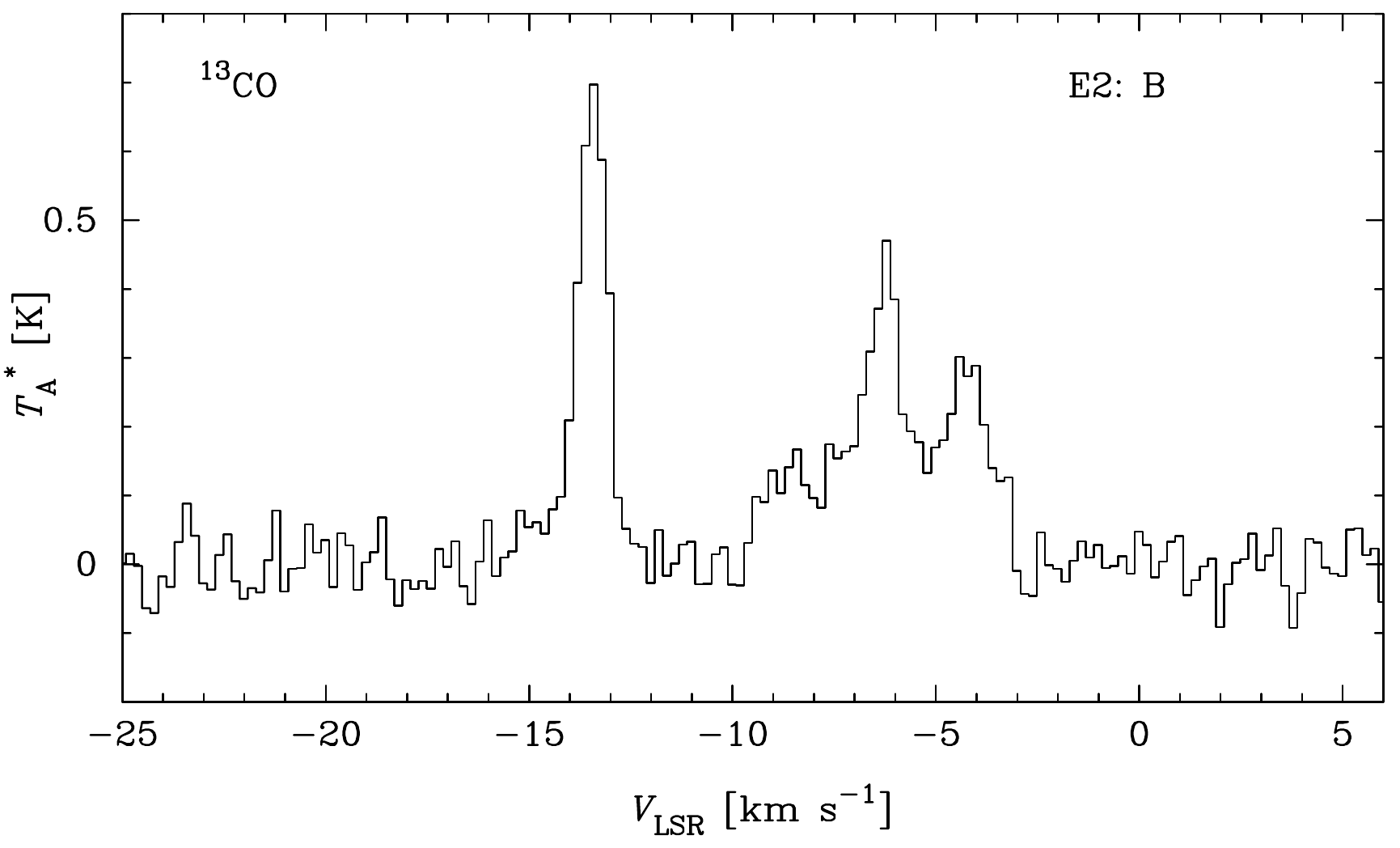}
\includegraphics[angle=00, width=5.65cm]{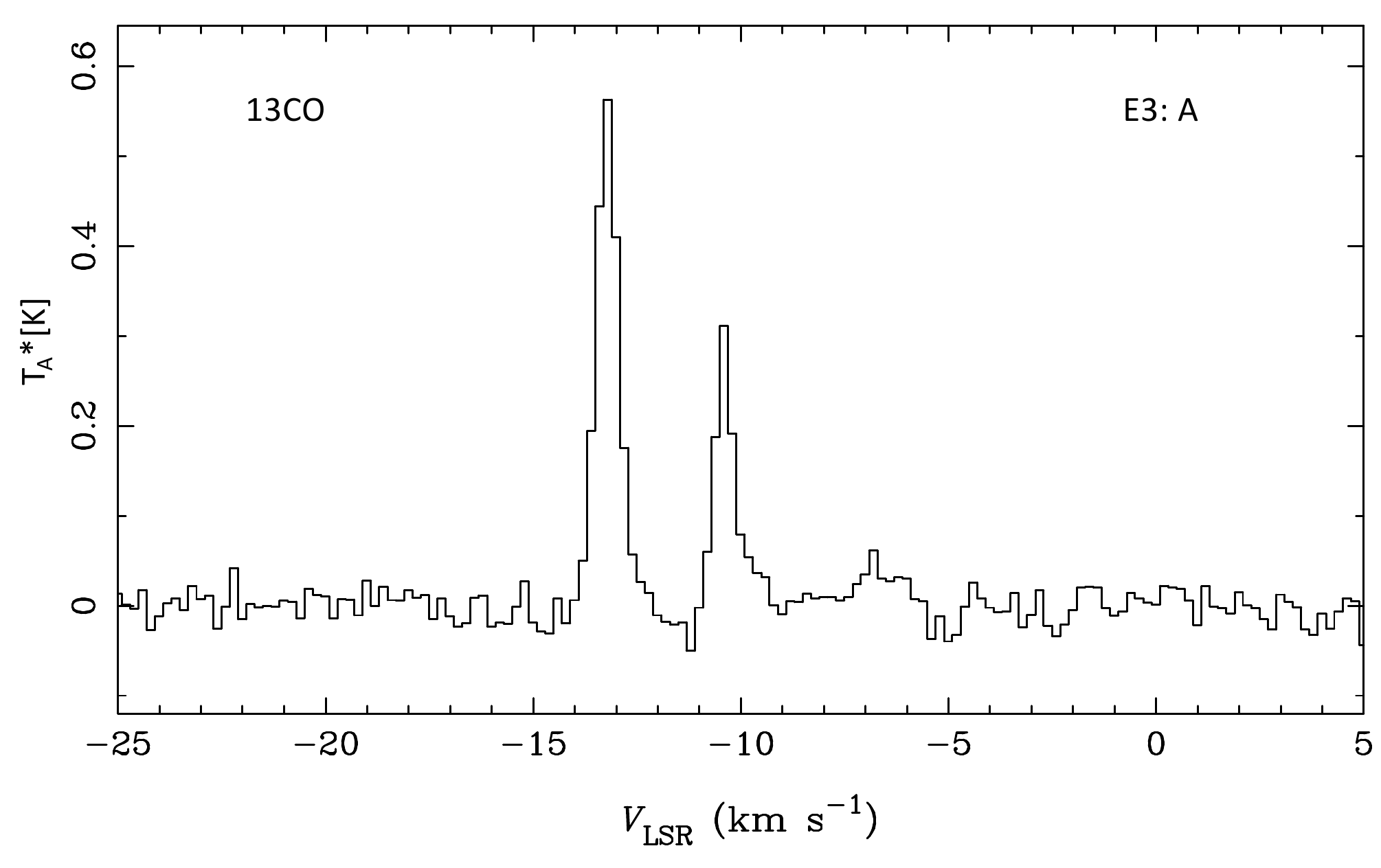}
\includegraphics[angle=00, width=6cm]{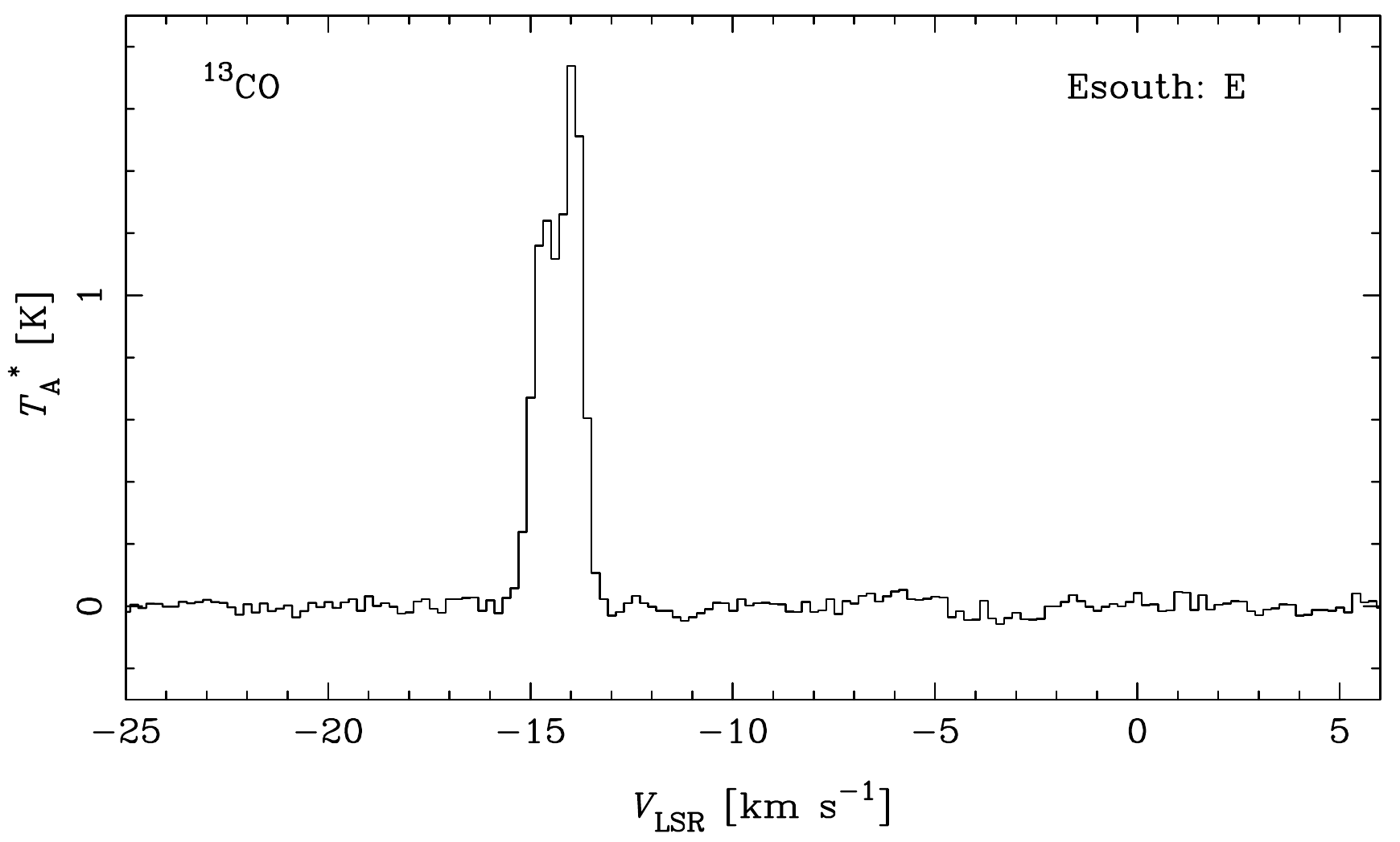}
\includegraphics[angle=00, width=6cm]{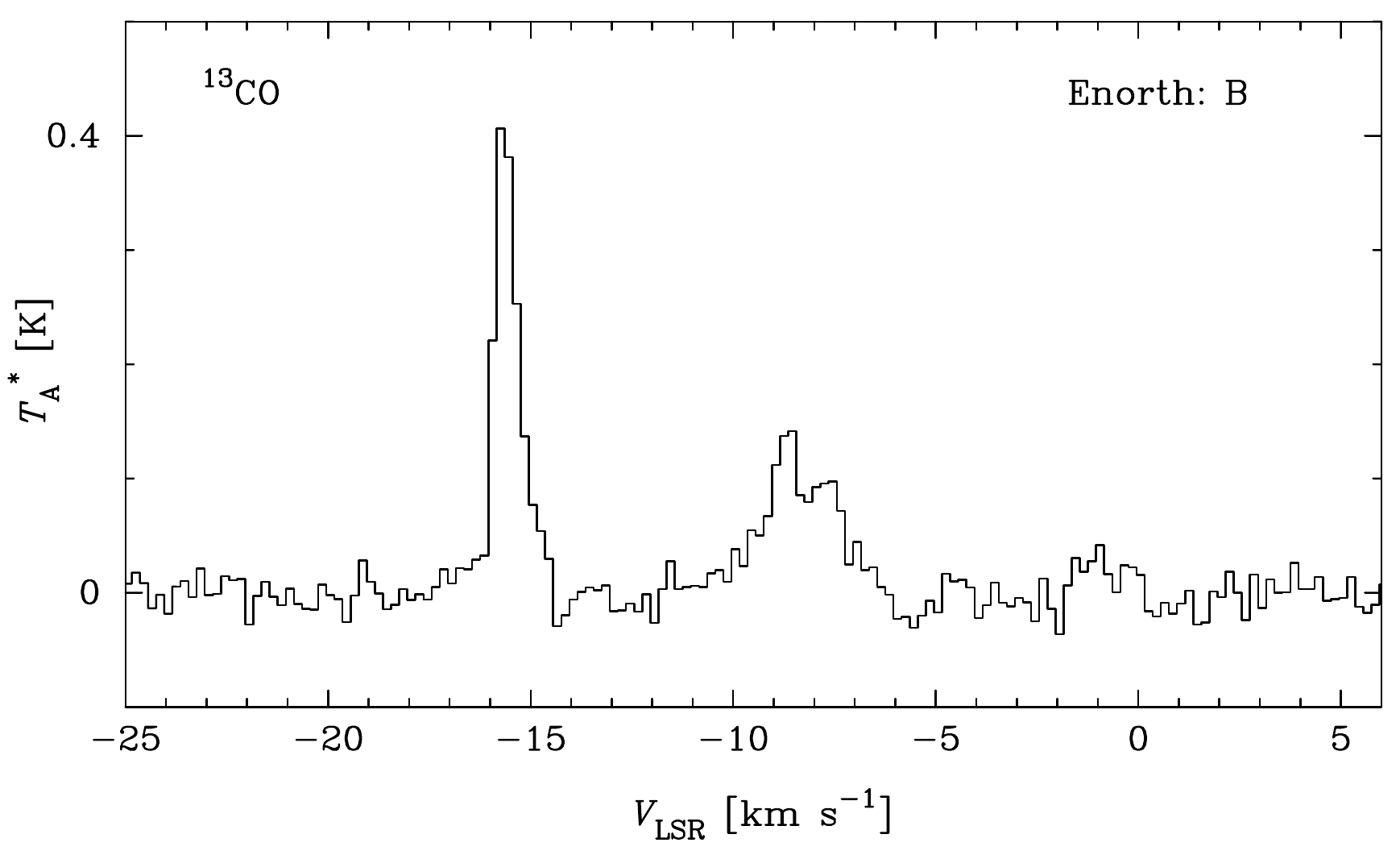}
\includegraphics[angle=00, width=6cm]{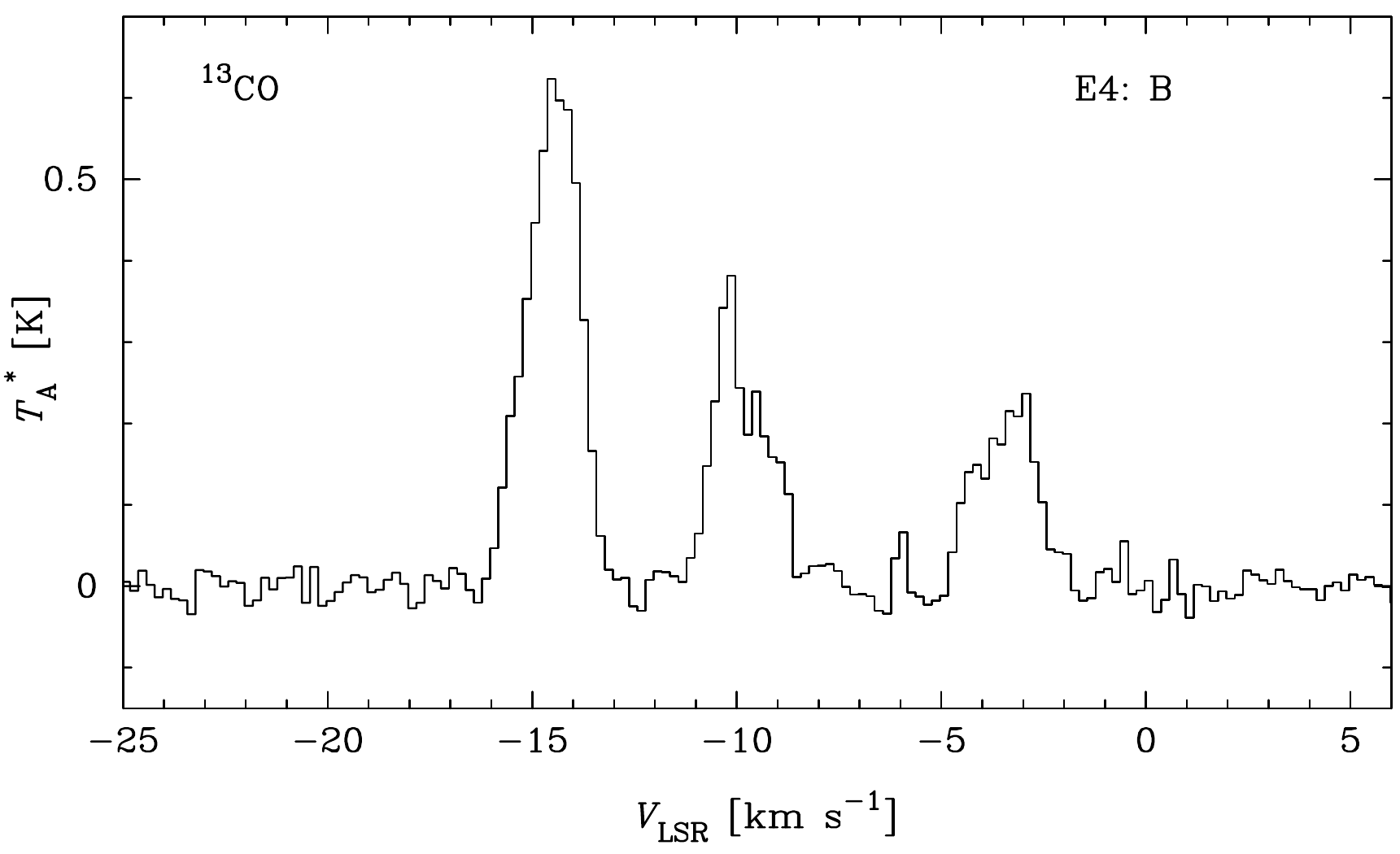}
\caption{As in Fig.~\ref{fig:COsp1}. Offsets are E1:B (0,0), E2:B (+300,+60), E3:A (-540,-480), Esouth:E (0,0), Enorth:B ((0,0), and E4:B (-360,-360).}
\label{fig:COsp2}
\end{figure*}

\end{appendix}

\end{document}